\documentclass[a4paper,11pt]{article}
\pdfoutput=1 

\usepackage{jheppub} 

\usepackage[T1]{fontenc} 
\usepackage[usenames,dvipsnames]{xcolor}
\usepackage{feynmp-auto}
\usepackage{cancel}
\usepackage{changepage}

\newcommand{\da}[1] {\left\langle #1 \right\rangle}
\newcommand{\ds}[1] {\left\lbrack #1 \bf \right\rbrack}

\usepackage{verbatim}

\usepackage{amstext} 
\usepackage{array}   
\newcolumntype{L}{>{$}l<{$}} 

\title{Causality, Unitarity and Symmetry in Effective Field Theory}

\author[a]{Timothy Trott}
\affiliation[a]{Department of Physics, University of California, \\
Santa Barbara, CA 93106, U.S.A.}

\emailAdd{ttrott@ucsb.edu}

\abstract{Sum rules in effective field theories, predicated upon causality, place restrictions on scattering amplitudes mediated by effective contact interactions. Through unitarity of the $S$-matrix, these imply that the size of higher dimensional corrections to transition amplitudes between different states is bounded by the strength of their contributions to elastic forward scattering processes. This places fundamental limits on the extent to which hypothetical symmetries can be broken by effective interactions. All analysis is for dimension $8$ operators in the forward limit. Included is a thorough derivation of all positivity bounds for a chiral fermion in $SU(2)$ and $SU(3)$ global symmetry representations resembling those of the Standard Model, general bounds on flavour violation, new bounds for interactions between particles of different spin, inclusion of loops of dimension $6$ operators and illustration of the resulting strengthening of positivity bounds over tree-level expectations, a catalogue of supersymmetric effective interactions up to mass dimension $8$ and $4$ legs and the demonstration that supersymmetry unifies the positivity theorems as well as the new bounds.}

\begin{document} 
\maketitle
\flushbottom

\section{Introduction}

Effective field theory (EFT) is a method to mock-up the long distance effects of high energy states with simplified and fake microphysics that nevertheless successfully approximates departures from otherwise universal behaviour. This is predicated on the approximate locality of the interactions involving heavy states (see e.g. \cite{Burgess:2007pt} for review). The heavier the state, the shorter its range. If the scales over which these fields can propagate is too small to be resolved, then the effects can be instead approximated by a series of local contact interactions. These are usually restricted by the symmetries, from which the possible interactions can be systematically identified and used to parameterise the impacts of the microphysics on long distance observables without knowing what it actually is. This provides a general strategy for accounting for the effects of unknown short-distance effects and identifying a classification scheme for the possible interactions that low energy particles may be involved with. 

However, demanding local contact interactions alone is not sufficient for consistency with causality (see e.g. \cite{Adams:2006sv}). The couplings parameterising the strength and phase of these interactions are restricted so that they cannot conspire to mediate macroscopic superluminal signal transmission. The focus of this work will be on exploring these consistency constraints on the space of effective interactions. The UV completion of these interactions will be assumed to be a conventional quantum field theory, obeying microcausality (analyticity of the $S$-matrix) and polynomially bounded energy dependence. For discussions of the relevance of this to various ideas of theories where these conditions could be modified (usually in relation to quantum gravity), see e.g. \cite{Dvali:2012zc}, \cite{Cooper:2013ffa},  \cite{Giddings:2009gj}, \cite{Keltner:2015xda}, \cite{Tokuda:2019nqb}. For applications of causality constraints to CFTs and holography, see e.g. \cite{Hartman:2015lfa}, \cite{Komargodski:2016gci}, \cite{Camanho:2014apa}, \cite{Caron-Huot:2017vep}, \cite{Afkhami-Jeddi:2018apj}, \cite{Kologlu:2019bco}, \cite{Caron-Huot:2020adz}, while for application to EFTs, the subject of discussion here, see \cite{Adams:2006sv}, \cite{Bellazzini:2014waa}, \cite{Bellazzini:2016xrt}, \cite{deRham:2017zjm}, \cite{deRham:2018qqo}, \cite{Komargodski:2011vj}, \cite{Bellazzini:2020cot}, \cite{Caron-Huot:2020cmc}, \cite{Tolley:2020gtv} for a sample of past studies (among many others). 

Unitarity in quantum mechanics is the statement of the consistency of time evolution with the probability interpretation of quantum mechanics - that is, the square magnitude of a transition amplitude between different states at different times have the interpretation of a probability. Of interest here will be asymptotic scattering states. Unitarity of the $S$-matrix is expressed through the optical theorem and the Cutting rules. For (near) forward scattering, this relates the residues or discontinuities over the singularities of the $S$-matrix to on-shell particle production. Together with its analytic causal properties, this enables the construction of dispersion relations that constrain the $S$-matrix for general (complex) momentum. For low energy scattering states, the $S$-matrix is calculable from an EFT. For high enough order interactions, the dispersion relations form a sum rule that determines the EFT couplings entirely from the (usually unknown) on-shell production rates of states in the UV. The (by now standard) dispersion relation between IR EFT-calculable processes and high energy production rates is reviewed below in Section \ref{sec:SumUni}, but see again \cite{Adams:2006sv} for more background.

For scattering processes in which the identity of the particles do not change, the sum rule implies that the IR contact interactions are equated with a positive sum of production rates in the UV. Because this is necessarily a positive number, the resulting constraints on the corresponding low energy Wilson coefficients have been called ``positivity theorems''. However, unitarity implies that the dispersion relation contains more information than simply positivity. 

As will be shown in Section \ref{sec:SumRule}, ``inelastic processes'' (processes in which the identity of the particles change) are bounded above by elastic ones according to the general constraint:
\begin{align}\label{InelasticContraint0}
    |M^{ijkl}|+|M^{ilkj}|\leq \sqrt{M^{ijij}M^{klkl}}+\sqrt{M^{ilil}M^{kjkj}}.
\end{align}
Here, $M^{ijkl}=\frac{d^2}{ds^2}A^{ijkl}(s)\Big|_{s=0}$, where $A^{ijkl}(s)$ is the forward amplitude describing the scattering process $i,j\rightarrow k,l$, where $i,j,k,l$ are any species of particle. Consequently, the extent to which inelastic effective interactions can violate hypothetical symmetries is limited. 

The term ``inelastic'' will be consistently (mis)used here to broadly refer to transitions in which the identity of the scattered particles change, rather than simply their masses. Here ``identity'' will usually reference quantum numbers with respect to a complete set of commuting observables, although this is, of course, a basis-dependent statement. In particular, (massless) spinning particles will usually be identified by helicity eigenstates (I will usually also misuse the word ``helicity'' to mean only the magnitude).

The remainder of the paper focuses both on applications of this result and its consequences, as well as general exploration of the structure of the causality constraints. This is mostly with an eye toward the Standard Model Effective Theory (SMEFT) \cite{Brivio:2017vri}, a general EFT parameterisation of the imprints of new, high energy microphysics on the Standard Model of particle physics (SM). The scope of these bounds are theoretically more far-reaching than the positivity constraints that have been largely the focus of previous attention. For example, elastic amplitudes that vanish in the forward limit can often be crossed into inelastic amplitudes that do not, thus failing to escape from the sum rule. See \cite{Remmen:2019cyz}, \cite{Remmen:2020vts}, \cite{Bi:2019phv}, \cite{Zhang:2018shp}, \cite{Yamashita:2020gtt} for previous applications of causality bounds to the SMEFT.

\subsection{Overview of results}

Section \ref{sec:SumRule} reviews the standard derivation of the dispersion relation between the twice differentiated forward scattering amplitude and the transition rates into on-shell states in the UV. This will be focused on forward scattering at dimension $8$ level. It is then shown that unitarity implies the general constraint (\ref{InelasticContraint0}) on inelastic processes. Bounds of this form have been previously identified in \cite{Yamashita:2020gtt} for the specific case of parity-symmetric weak boson interactions, although, to the author's knowledge, the general statement above is new. Section \ref{sec:loops} gives some general discussion about the inclusion of loops of lower dimensional operators in the dimension $8$ order contribution to the amplitude. In particular, unitarity implies the general expectation that the elastic dim-$8$ Wilson coefficients decrease with energy scale, which would therefore strengthen the positivity bounds that would otherwise be inferred at tree-level. The general results are illustrated in Section \ref{sec:ComplexScalarEG} by a simple example: a complex scalar field. It is shown that processes that would violate a (hypothetical) $U(1)$ charge must necessarily be bounded above by charge conserving processes. 

In Section \ref{sec:Internal}, causality constraints for EFTs with preserved internal symmetries are examined. Following \cite{Zhang:2020jyn}, the causality sum rule can be equivalently characterised as a convex cone in which UV completable IR amplitudes must lie. In simple examples where there are no degenerate states unrelated by symmetry, the cone is polyhedral and elementary convex geometry allows for a complete set of positivity bounds to be extracted, including those inaccessible from scattering of factorised states. 
Section \ref{sec:Program} reviews the convex geometry interpretation of \cite{Zhang:2020jyn} (similar ideas were suggested in \cite{Bellazzini:2014waa}), for which much of this work was inspired at understanding. 
Using this picture, I compute the complete set of bounds for a single fermion species in the symmetry representations of the SM (fundamentals of $SU(2)$ and $SU(3)$) in Section \ref{sec:SMFermions}, possibly also including exact flavour symmetry. Some of these bounds are altogether new. It is also illustrated how, with increasingly more symmetry representations, the structure of the positivity bounds becomes increasingly intricate - the convex cone describing the space of allowed amplitudes becomes increasingly multifaceted. In Section \ref{sec:Flavour}, I give the general bounds on flavour violation admissible under the general bound on inelastic amplitudes, elaborating upon the observations of limits on flavour violation made by \cite{Remmen:2020vts}. The results discussed in this section have direct application to the SMEFT.

Section \ref{sec:SpinBounds} discusses rotational symmetry and the treatment of spin and parity ($P$). Spin is discussed in generality in Section \ref{sec:RotSym}, which is then illustrated with the simple (and known) example of four photon scattering in Section \ref{sec:idparticles}. It is in particular shown how $P$ and helicity violation (electric-magnetic duality) are necessarily limited, explaining the observations made by \cite{Remmen:2019cyz} for vector boson scattering.
The results are then generalised to two distinct species with the same helicity in Section \ref{sec:FermionBounds}.

Constraints on EFTs with multiple particles of different helicities will be derived in Section \ref{Sec:MultiSpin}. These add to bounds derived previously for elastic processes involving states of various spin in \cite{Bellazzini:2014waa}, \cite{Remmen:2019cyz}, \cite{Bellazzini:2018paj}. In Section \ref{sec:SUSY}, I catalogue all possible supersymmetric effective contact interactions with mass dimension up to eight and at most four particles, identifying which types of interactions are embeddable into supersymmetric EFTs. This culminates in the demonstration that the standard simple positivity theorems of \cite{Adams:2006sv}, \cite{Bellazzini:2016xrt} and the new additions derived here unify under supersymmetry. Less supersymmetric inelastic interactions appear as amplitudes that must necessarily raise the lower bounds on the more supersymmetric elastic operators. In this sense supersymmetry must at least partially emerge from the positivity bounds. 

An appendix gives a list of elementary results for projectors for spin indices represented in $SO(2)$ form, as well as a presentation of the sum rules for the toy two fermion and multispin theories.

Note added: As this work was being completed, the work of \cite{Bellazzini:2020cot} appeared, which involves partial overlap with the general discussion of loops provided here. In particular, the general observation made here about strengthening of the constraints with RG flow from dim $6$ loops is another instance of the result for Goldstone bosons described in \cite{Bellazzini:2020cot}, albeit applied in different examples.

More notes added: Subsequent to release of this preprint, the following relevant works appeared: \cite{Huang:2020nqy} make some pertinent and interesting comments on positivity constraints with loops, problems with the forward limit and bounds on multiparticle theories, in addition to its main thesis of deriving new constraints on higher dimension operators away from the forward limit. \cite{Liu:2020fgu} incorporates supersymmetry in constraints of the form of \cite{Huang:2020nqy} on Yang-Mills operators, although their discussion has little overlap with that presented here. In \cite{Bonnefoy:2020yee}, positivity bounds were derived for three flavours of SM quarks under the assumption of minimal flavour violation but only from consideration of scattering of pure states. Finally, \cite{Li:2021cjv} makes advances on the issue of constraining theories with multiple species in which the space of couplings is a non-polyhedral cone. This includes some more thorough derivations of bounds for theories with more than two distinct species unrelated by symmetries that, in some instances, improves over bounds presented here (such as on the flavour-violating operators of right-handed electrons).

\section{Sum Rule and Unitarity}\label{sec:SumUni}
This section reviews the derivation of the standard dispersion relation between the IR scattering amplitude in an EFT and the full transition rates in the UV, along with the various caveats and assumptions that will be implicit throughout the remainder of the paper. This derivation can be found in numerous previous works e.g. \cite{Adams:2006sv}, \cite{Bellazzini:2014waa}. The discussion presented below is closest to \cite{Yamashita:2020gtt}, from which much of this thinking is inspired. The inelastic scattering constraint (\ref{InelasticContraint0}) will then be derived. The affects of loops will also be discussed.

The subsequent discussion will then turn to the process of extracting constraints on the EFT out of the sum rules. To do this, I will use the picture presented in \cite{Zhang:2020jyn} and \cite{Yamashita:2020gtt} of the space of UV completions as delineating a convex cone to which the space of all forward effective transition amplitudes in the IR must belong.

All discussion in this paper will specifically focus on constraints at mass dimension $8$ order in standard power counting arising from scattering in the forward limit.

\subsection{Causality sum rule and structure}\label{sec:SumRule}

The standard assumptions of analyticity of the scattering amplitude as a function of energy will be made. From the point of view of QFT, this is an expected consequence of microcausality \cite{Bros:1964iho}, \cite{GellMann:1954db}, \cite{Martin:1965jj}. See e.g. \cite{Simmons-Duffin:2019lll} for review. This has been somewhat established for correlators in massive theories and those obeying the Wightman axioms, like CFTs. If the theory has particles, these presumably extend to the $S$-matrix. I will assume that these results also hold for the simple massless theories described here by assuming that a massless limit can be taken that commutes with the ensuing derivations. Similarly, it will be assumed that the only singularities of the two-to-two $S$-matrix in the forward limit are poles and branch cuts close to the real energy axis directly associated with on-shell particle production in the $s$ or $u$-channels.

Call $A^{ijkl}(s)$ the forward scattering amplitude ($t=0$) for the process $i,j\rightarrow k,l$. Then, for some complex-valued energy $\sigma$, define $\overline{M}^{ijkl}(\sigma^2)=\frac{d^2}{ds^2}A^{ijkl}(s)|_{s=\sigma^2}=\frac{1}{\pi i}\oint \frac{A^{ijkl}(s)}{\left(s-\sigma^2\right)^3}ds$. The contour is over a small loop enclosing $\sigma^2$ and no other singularities. The small loop may then be deformed in the standard way into a contour enclosing and railing along the poles and branch cuts on the real axis closed-off by a semi-circular arc of some large enough radius. The Froissart bound \cite{Froissart:1961ux} implies that the integrand decays fast enough along the arc that this part of the integral may be ignored. The supplemental locality assumption of polynomially bounded energy growth is invoked here (see e.g. \cite{Giddings:2009gj}, \cite{Cooper:2013ffa},  \cite{Keltner:2015xda}, \cite{Tokuda:2019nqb} for commentary about the necessity of this assumption and possible consequences of its modification, usually in the context of gravity). 

Crossing can be used to relate the forward amplitude along the negative real axis to the $u$-channel amplitude $A^{ijkl}(s)=A^{i\bar{l}k\bar{j}}(4m^2-s)$ (where $\bar{X}$ denotes the antiparticle of $X$). I will assume for simplicity that each particle has mass $m$, although this will be ignored for most of what follows (assumed to be small compared to the characteristic energies of the observed scattering processes). Crossing relations with spinning particles have been explained in \cite{Bellazzini:2016xrt} for the forward limit specifically for the present context. Note that the amplitude $A\left(X,Y\rightarrow Z,W\right)$ is obtained through LSZ reduction from a correlator with fields ordered as $\langle 0|W Z X Y|0\rangle$, which will be of relevance in Section \ref{sec:SUSY}. The remaining contour integral over the branch cuts is
\begin{align}
    \overline{M}^{ijkl}(\sigma^2)=\frac{1}{\pi i}\int_{4m^2}^\infty\left(\frac{\text{Disc}A^{ijkl}(s)}{\left(s-\sigma^2\right)^3}+\frac{\text{Disc}A^{i\bar{l}k\bar{j}}(s)}{\left(s+\sigma^2-4m^2\right)^3} \right)ds+\text{residues at poles}.
\end{align}
The amplitudes obey the reality condition following from analyticity and unitarity \cite{Olive:1962jj}
\begin{align}\label{reality}
\left(A^{ijkl}(s)\right)^*=A^{klij}(s^*).
\end{align}
This is the $S$-matrix statement of requiring the Hamiltonian to be Hermitian. This relation expresses the fact that the discontinuity over the branch cut is related to the intermediate state on-shell production rate through unitarity of the $S$-matrix:
\begin{align}
    \text{Disc}A^{ijkl}(s)=A^{ijkl}(s+i\epsilon)-\left(A^{klij}(s+i\epsilon)\right)^*=i\sum_X\mathcal{M}^{ij\rightarrow X}(s+i\epsilon)\left(\mathcal{M}^{kl\rightarrow X}(s+i\epsilon)\right)^*,
\end{align}
where $\mathcal{M}^{ab\rightarrow X}$ is the amplitude for particles $a$ and $b$ to transition into intermediate state $X$ and $\epsilon\rightarrow 0^+$ is implicit.

Mostly for simplicity, the theories discussed here will all be massless. While the results above have been derived under the assumption of a mass gap, I will disregard this and assume that the particle masses in the sum rule can be freely taken to zero without consequence at this stage. In particular, this will assume that the singularity structure of the $S$-matrix in the forward limit is not affected. See \cite{Bellazzini:2016xrt} for a list of some other possible issues. The validity of this remains an open question for investigation and becomes increasingly less certain with increasing spin. These questions have received particular recent attention in the context of gravity, see e.g. \cite{Cheung:2014ega}, \cite{Tokuda:2020mlf}, \cite{Alberte:2020jsk}, although I am satisfied here with restricting to flat spacetime QFT with particle helicities $\leq 1$. 

For massive theories, it is natural to make a real insertion $\sigma^2$ below the mass threshold where the amplitude is analytic. This would represent a region of energies in which RG evolution switches off and the energy-dependence of the amplitude is relatively simple. For massless theories, the branch cuts cleave the entire complex $s$ space. It will be assumed that the dispersion relation for these theories can be reached by taking a mass-deformed theory, analytically continuing the insertion point $\sigma^2$ above a branch cut to some $\sigma^2+i\delta$ with $\delta\rightarrow 0^+$ (keeping $\sigma^2$ real) and then taking the massless limit (so that the cuts extend to the origin). Note that this procedure does not require the masses to be sent to zero exactly, but only that they be much smaller than the insertion point. The insertion point itself can then be taken small in the IR for simplicity: $\sigma^2/\Lambda^2\rightarrow 0$ for UV cut-off $\Lambda$. One advantage of doing this is that the sum rule becomes symmetric in $s$ and $u$-channel cuts.

So taking the massless limit and invoking unitarity, the sum rule becomes
\begin{align}\label{ProtoSumRule}
\overline{M}^{ijkl}(\sigma^2)&=\frac{1}{\pi}\int^\infty_0 \frac{1}{\left(s-\sigma^2-i\delta\right)^3}\sum_X\mathcal{M}^{ij\rightarrow X}\left(\mathcal{M}^{kl\rightarrow X}\right)^*ds\nonumber\\
&\qquad\qquad\qquad+\frac{1}{\pi}\int^\infty_0 \frac{1}{\left(s+\sigma^2+i\delta\right)^3}\sum_X\mathcal{M}^{i\bar{l}\rightarrow X}\left(\mathcal{M}^{k\bar{j}\rightarrow X}\right)^*ds.
\end{align}
The limit $\delta\rightarrow 0^+$ is implicit. The insertion point $\sigma\ll\Lambda$ is chosen to be a characteristic IR energy scale so that the LHS can be evaluated in the EFT. The sum is over all possible intermediate state $X$, which may be infinite and continuous. There may also be poles on the real axis - these will be implicitly included in the integral over the cuts. In the IR, these may be explicitly calculated anyway.

The integral on the RHS of (\ref{ProtoSumRule}) is over both the known IR and the unknown UV. As a relation between the IR and the UV, the calculable IR part of the integral really belongs on the LHS and contains significant information. Defining this new combined left-hand side by $M^{ijkl}(\sigma^2)$, the sum rule therefore becomes
\begin{align}\label{ProtoSumRule2}
\hspace{0cm}&M^{ijkl}(\sigma^2)=\overline{M}^{ijkl}(\sigma^2)\nonumber\\
&-\frac{1}{\pi}\int^{\lambda^2}_0 \left(\frac{1}{\left(s-\sigma^2-i\delta\right)^3}\sum_{X\in IR}\mathcal{M}^{ij\rightarrow X}\left(\mathcal{M}^{kl\rightarrow X}\right)^*+\frac{1}{\left(s+\sigma^2\right)^3}\sum_{X\in IR}\mathcal{M}^{i\bar{l}\rightarrow X}\left(\mathcal{M}^{k\bar{j}\rightarrow X}\right)^*\right)ds\nonumber\\
&=\frac{1}{\pi}\int^\infty_{\lambda^2}\left( \frac{1}{\left(s-\sigma^2-i\delta\right)^3}\sum_{X\in UV}\mathcal{M}^{ij\rightarrow X}\left(\mathcal{M}^{kl\rightarrow X}\right)^*+\frac{1}{\left(s+\sigma^2\right)^3}\sum_{X\in UV}\mathcal{M}^{i\bar{l}\rightarrow X}\left(\mathcal{M}^{k\bar{j}\rightarrow X}\right)^*\right)ds.
\end{align}
Here $\lambda$ is some high energy scale up to which the EFT is still reliable, which may be taken up to the cut-off $\Lambda$. If loops can be ignored in some approximation, then the IR integral over the branch cut can be ignored and there is no problem with choosing $\sigma\approx 0$, where RG evolution (by assumption) has ceased. Otherwise the amplitudes are to be evaluated above mass thresholds, as would be necessary for (approximately) massless particles, and $\sigma$ is chosen to lie over a branch cut. The IR integral is non-trivial and can be computed to the required level of accuracy in the energy expansion underpinning the EFT. Because of the pole in the integrand, if $\sigma$ is to be identified with a real energy scale, then a non-zero $\delta$ is required that must be sent to $0$ (it has been assumed that $\sigma^2$ is positive and lies above the $s$-channel cut in (\ref{ProtoSumRule2})). When evaluating the dispersion integral along the branch cut, this leaves behind a finite imaginary part that cancels against the imaginary part in the loop amplitude $\overline{M}^{ijkl}(\sigma^2)$. 
More generally, the loop amplitudes also include logarithmic RG-evolution from the renormalisation scale to the insertion point $\sigma$, while the dispersion integral accounts for further evolution from $\sigma$ to the scale up to which the integral is being evaluated. 

Taking the insertion point to the origin $\sigma^2\rightarrow 0$, the dispersion relation becomes
\begin{align}
M^{ijkl}(0)=\frac{1}{\pi}\int^\infty_{\lambda^2}\frac{1}{s^3}\sum_X\left(\mathcal{M}^{ij\rightarrow X}\left(\mathcal{M}^{kl\rightarrow X}\right)^*+\mathcal{M}^{i\bar{l}\rightarrow X}\left(\mathcal{M}^{k\bar{j}\rightarrow X}\right)^*\right)ds.
\end{align}
The amplitudes $\mathcal{M}^{ij\rightarrow X}(s)$ are vectors in a complex inner product space, where both the energy $s$ and the couplings to each intermediate state $X$ in the UV completion are the (infinite and continuous) components. To emphasize this, I will rewrite this full complex vector as $\boldsymbol{m}^{ij}$. The sum over states $X$ and the dispersion integral define an inner product in these variables (the accompanying multiplicative factors in the integrand are positive, so enable this interpretation). The sum rule can then be expressed as
\begin{align}\label{SumRule}
M^{ijkl}=\boldsymbol{m}^{kl}\cdot\boldsymbol{m}^{ij}+\boldsymbol{m}^{k\bar{j}}\cdot\boldsymbol{m}^{i\bar{l}},
\end{align}
omitting the specification that $\sigma =0$ from the notation for convenience. The sum rule (\ref{SumRule}) is the centerpiece of this work. All references to ``the sum rule'' refer to this equation, while the terms ``LHS'' and ``RHS'' will be used to refer to the left-hand side and right-hand side of this equation without qualification throughout. 

Note that it is not essential to evaluate the sum rule with $\sigma=0$ exactly. In this case, the $s$ and $u$-channel terms in (\ref{SumRule}) do not have identical coefficients, but differ only by subleading factors in $\sigma^2/\lambda^2$. Taking $\lambda$ close to $\Lambda$ and $\sigma\ll\Lambda$, these terms are already consistent with the truncation error of the low energy expansion. 

The space of couplings of the states in the UV to those in the EFT are parameterised by the vectors $\boldsymbol{m}^{ij}$. Organised in this way, it is possible to draw many immediate conclusions from the sum rule about the EFT directly from the combination of vectors in this expression. This will be illustrated in numerous examples below. The traditional positivity theorems for elastic scattering following from the optical theorem are obvious from (\ref{SumRule}) when $k=i$ and $l=j$, as each term is the norm of a complex vector, which must be positive if non-zero. The elastic amplitudes all have the form
\begin{align}
    M^{ijij}=|\boldsymbol{m}^{ij}|^2+|\boldsymbol{m}^{i\bar{j}}|^2.
\end{align}
However, inelastic processes are necessarily bounded from above by elastic processes as well. The Schwarz and triangle inequalities give upper bounds on the inelastic amplitudes:
\begin{align}
    |M^{ijkl}|= \left|\boldsymbol{m}^{kl}\cdot\boldsymbol{m}^{ij}+\boldsymbol{m}^{k\bar{j}}\cdot\boldsymbol{m}^{i\bar{l}}\right|\leq|\boldsymbol{m}^{kl}||\boldsymbol{m}^{ij}|+|\boldsymbol{m}^{k\bar{j}}||\boldsymbol{m}^{i\bar{l}}|.
\end{align}
A general upper bound can then be obtained as
\begin{align}\label{InelBound}
    |M^{ijkl}|+|M^{ilkj}|\leq \sqrt{M^{ijij}M^{klkl}}+\sqrt{M^{ilil}M^{kjkj}}.
\end{align}
This demonstrates the schematic pattern and is an entirely general result. When there are symmetries relating the states, a subset of the vectors $\boldsymbol{m}^{ij}$ are related and there are fewer independent vectors of UV couplings. In this case, stronger bounds may be possible after the states are classified into symmetry irreps, as will be discussed in Section \ref{sec:Internal}. In general, the bounds of the form (\ref{InelBound}) are necessary but not sufficient, but are being highlighted here both because they are simple and that they directly demonstrate the way in which unitarity fundamentally limits the size of inelastic transitions. Improvements remain an open problem, such as those explored in \cite{Yamashita:2020gtt}.

Recently, \cite{Zhang:2020jyn}, \cite{Yamashita:2020gtt} offered the interpretation of the space of points of the form $\{\boldsymbol{m}^{kl}\cdot\boldsymbol{m}^{ij}+\boldsymbol{m}^{k\bar{j}}\cdot\boldsymbol{m}^{i\bar{l}}\}$ as a convex cone - a convex hull generated by positive linear combinations of a subset of vectors. Note that it is not necessary that the vectors themselves be real-valued, nor that the states be self-conjugate. See \cite{Zhang:2020jyn}, \cite{Yamashita:2020gtt} for further details. Convex cones can be described in two equivalent ways: by a set of inequalities delineating hyperplanes (or facets) that bound the cone, or by a set of extremal rays (ERs) that determine the edges of the cone. Extremal rays are $1$d subspaces of single vectors that cannot be decomposed into a positive linear combination of any other set of linearly independent vectors in the cone. I will use the term ER ambiguously to mean either the subspace or a member vector. Any point in the cone can be expressed as a linear combination of extremal rays with positive coefficients, so the ERs generate the cone. 

The inequality representation is a manifest statement of the constraints on the space of forward amplitudes or, equivalently, the bounds on the space of Wilson coefficients allowed in the EFT. The problem at hand is to extract from the sum rule a complete set of such bounds. The ER representation provides an intermediate alternative that is straight-forward to determine directly from the sum rule. If the cone is polyhedral, as expected for theories in which all transitions between states are rigidly fixed by symmetries, then standard results from convex geometry may be applied to derive the inequalities describing the facets. This insight was applied to some simple examples by \cite{Zhang:2020jyn} to derive new constraints through rudimentary convex geometry that were inaccessible from considering only scattering amplitudes of factorised states. This will be applied to some more examples below in Section \ref{sec:Internal} with similar results.

It is worth discussing here the action of the discrete symmetries that the forward amplitudes. The kinematics of forward scattering preserves rotational invariance about the ``beam direction'' in the center of mass frame. The angular momentum of each state projected in this direction is a conserved charge that the external states are labelled by. This will be the subject of Section \ref{sec:SpinBounds}. Besides this however, there remains one further action of rotational invariance on the $S$-matrix. Rotations by $\pi$ perpendicular to the beam axis effectively interchanges (in the centre-of-mass frame) both particles one with two and three with four. This equates, up to a possible little group phase for inelastic amplitudes, the forward amplitudes $M^{ijkl}$ and $M^{jilk}$. This discrete rotation will be referred to as ``$Y$''. This symmetry is in addition to crossing, with which it can combine to produce $CPT$. In particular, $Y$-symmetry also acts on the vectors of UV couplings to imply that, in general,  $|\boldsymbol{m}^{ij}|=|\boldsymbol{m}^{ji}|$. In many cases, when transitions between the $Y$-rotated pairs of states are prohibited (such as when there is angular momentum about the beam axis), the vectors themselves lie in orthogonal subspaces that may be directly identified through $Y$. Crossing, the Hermitian analyticity condition (\ref{reality}) and $Y$ symmetries (as well as the emergent $CPT$) can generally act to simplify the structure of the sum rule. In particular, $\boldsymbol{m}^{ij}\cdot\boldsymbol{m}^{kl}=\boldsymbol{m}^{\bar{k}\bar{l}}\cdot\boldsymbol{m}^{\bar{i}\bar{j}}$. Other discrete symmetries may exist for a particular theory, such as parity, time-reversal and identical particle exchange symmetries. Examples of these will be given throughout this work, but their existence is theory-dependent.

\subsection{Loops}\label{sec:loops}

It is appropriate here to emphasise that, with no further assumptions beyond standard power counting, the dimension $8$ order scattering amplitudes receive contributions not only from terms with single insertions of dimension $8$ operators, but also from terms with multiple insertions of lower dimension operators that altogether give energy scaling of the same order. Double insertions of dimension $6$ operators are particularly common. Much of previous work on these constraints has neglected the latter terms and are naively restricted in applicability to UV completions that generate small lower dimension Wilson coefficients (usually justified by appealing to a weak coupling expansion). It is not obvious how these bounds would apply to theories saturating naive dimensional analysis \cite{Gavela:2016bzc}, which is characteristic of strongly coupled UV completions. One such example is chiral perturbation theory in the real world - see \cite{Manohar:2008tc}, \cite{Mateu:2008gv}, \cite{Wang:2020jxr} for the results of applications of causality constraints to this. Double insertions of dim-$6$ cubic vector operators were, however, considered in \cite{Zhang:2018shp} and \cite{Yamashita:2020gtt}, where it was interestingly observed that they enhanced the positivity constraints on the quartic vector operators. 

A significant general statement about loop corrections from $4$-point dimension $6$ operators can be likewise made. A loop of two dim-$6$ insertions produce UV divergent bubble integrals, which are proportional to logarithm of the Mandelstam variable corresponding to the partitioning of the legs on either side of the loop. The coefficient of the logarithm is determined by the unitarity cut across the appropriate channel. For elastic amplitudes, both the $s$ and $u$-channel cuts are positive in the forward limit by the optical theorem. The $t$-channel cuts vanish in the forward limit by conservation angular momentum. This is because the dim-$6$ effective interactions can only mediate scattering in, at most, the $j=1$ partial wave, implying that these terms cannot be proportional to more than one power of the crossed channel Mandelstam variable, here $s$ or $u$ (or combinations of spinor bilinears that effectively behave as square roots of these). There must therefore be an overall factor that vanishes in the forward limit. The positivity of the $s$ and $u$-channel cuts implies that the coefficient of the UV log generated by the loop must be positive. This means that the dim-$8$ contact coefficients for elastic processes are always decreasing with increasing energy scale under renormalisation group (RG) evolution.

As explained above, the IR segment of the dispersion integral effectively accounts for RG evolution of the coupling from renormalisation scale $\mu$ to cut-off $\Lambda$. For elastic scattering, the higher $\Lambda$ is pushed, the more negative the loop correction appearing on the LHS becomes. It is for this reason optimal to integrate the IR dispersion integral as far as the cut-off. It is therefore the smaller, RG-evolved high-scale coupling that is constrained to be positive. The naive tree-level bounds on the low energy coupling are therefore strengthened by these dim-$6$ loops. The low-energy contact interaction is therefore subject to a stronger lower bound that depends both on the size of the dim-$6$ operators and the size of the energy hierarchy. 

For bubble loop corrections to inelastic processes, another Schwarz-like bound can be placed on the cuts by noting that the sum over intermediate states (including phase space integration) is itself an inner product, so that:
\begin{align}
    \left|\sum_X\mathcal{M}^{ij\rightarrow X}\left(\mathcal{M}^{kl\rightarrow X}\right)^*\right|\leq\sqrt{\left(\sum_X\mathcal{M}^{ij\rightarrow X}\left(\mathcal{M}^{ij\rightarrow X}\right)^*\right)\left(\sum_X\mathcal{M}^{kl\rightarrow X}\left(\mathcal{M}^{kl\rightarrow X}\right)^*\right)}.
\end{align}
This bounds the size of the log coefficients for inelastic processes by those of elastic processes. In other words, the RG evolution of the corresponding tree operators is restricted by the elastic ones. This implies that the UV logs from loop corrections on the inelastic side of the constraints must be smaller than those on the elastic side. The RG evolution of the elastic amplitudes is therefore typically larger and determines how the constraints tighten with scale. 

These effects will be illustrated below in Section \ref{sec:Scalars} in a simple, concrete example. All calculations performed here will be with the $\overline{MS}$ renormalisation scheme. Lower-dimension operators can also contribute to the amplitudes in more ways than simple UV renormalisations. A more thorough examination of these effects will be left for the future. 

It will be likewise assumed that all marginal or renormalisable couplings are perturbative and that the loop corrections that they mediate are subdominant at leading order in the energy expansion of the EFT. These corrections may nevertheless be included in a similar way to the loops discussed above. If the energy hierarchy is large enough, the logarithms associated with these corrections become large and this is no longer justified. It would be interesting to also investigate how the RG flow would interact with the simplified conclusion derived above. Note that a perturbative treatment would apply under these conditions to the relevant couplings in the SM with the exception of the strong gauge coupling, which is non-perturbative at energies below $\sim 1$ GeV. Extrapolating amplitudes in perturbative QCD to low energies is therefore unclear. A possible way of dodging the problem may be to modify the integration contour to cut-off the dispersion integral in the IR at some energy $s=r$ and then integrate over a semi-circular arc to the opposite branch cut. As long as $r\gg \Lambda_{QCD}$, then this should be computable within the perturbative theory, as long as the analytic continuation is still valid. If $r\ll\sigma^2\ll\Lambda^2$, then this will also have only a small effect on the results derived under the assumptions above. Of course, this different contour choice does nothing to address the question of the validity of the foundational arguments underpinning the sum rule to Yang-Mills gauge theory, where the perturbative $S$-matrix must be presumably matched onto an inclusive IR observable.

\subsection{Constraints}\label{sec:Program}

The sum rule in the form (\ref{SumRule}) and the convex cone picture of \cite{Zhang:2020jyn} yields a program for systematically extracting information in the sum rule into constraints on Wilson coefficients in an effective action that follows three stages:

\begin{enumerate}
    \item Write down the sum rule and find the (potential) ERs. 
    \item Convert into inequalities among amplitudes.
    \item Convert into inequalities among Wilson coefficients.
\end{enumerate}
For simple enough theories, inequalities may be deduced directly from inspection of the sum rule without recourse to the convex cone picture. Invocation of convex geometry is most useful when the number of ERs is greater than the dimension of the space of independent amplitudes. As will be elaborated upon much more in Section \ref{sec:Internal}, this typically occurs when many of the states are related by symmetries. 
While it is simple enough to outline the strategy above, much of the challenge lies in step 2 which itself thus far lacks a general procedure, although direct application of (\ref{InelBound}) is often substantial.

\subsubsection{Sum rule and extremal rays}
If sufficiently simple, constraints on the EFT can be deduced from the sum rule directly from inspection by expressing it in the form of (\ref{SumRule}), similarly to the way that the standard postivity results from elastic scattering and the inelastic bound (\ref{InelBound}) were derived. Examples of this will be given in the sections below. However, this is not always so simple when symmetries are present that impose further structure over the $S$-matrix.

Candidate ERs can be constructed by finding the ERs of the cone generated by only the $s$-channel term in (\ref{SumRule}), that is, the cone of positive semi-definite (PSD) matrices. Following \cite{Zhang:2020jyn}, these will be referred to as potential ERs (PERs). Once the $u$-channel term is added, all ERs must be PERs of the $s$-channel cone, but the converse does not necessarily hold and some PERs may be redundant (lie within the interior).

In the simple case where the symmetries of the theory are stringent enough to restrict $S$-matrix transitions to unique (irreps of) initial and final state species, then there is only a single independent vector in (\ref{SumRule}) that parameterises each transition and the magnitude squared of a single component of this vector, by itself, represents a full PSD matrix and defines a PER. Again, see Section \ref{sec:GroupTheory} below for further details and explanation. This will be the situation discussed further below in Section \ref{sec:SMFermions}. However, in the presence of multiple ``degenerate'' states (irreps) between which ``inelastic'' transitions are permitted, each PSD matrix consists of multiple parameters. Each parameter is a complex number that can be interpreted as a coupling of the IR states to a particular UV state with a specific set of quantum numbers. As rays, these are only of interest up to an overall scale. For example, for a theory with two degenerate irreps of scattering states under some symmetry (or, equivalently, distinct states with transition amplitudes permitted by symmetries), a PER has the form 
\begin{align}
    \begin{pmatrix}
    1 & r \\
    r^* & |r|^2
    \end{pmatrix}
\end{align}
for some unknown $r\in\mathbb{C}$. The undetermined components effectively parameterise a continuous family of (P)ERs that generate curved facets. A simple example of this will be illustrated in Section \ref{sec:idparticles}. Again see \cite{Yamashita:2020gtt} for more details and discussion of application to SM electroweak bosons.

\subsubsection{From rays to amplitudes}
While the cone is fully determined as the convex hull of the rays, it is still required to convert the description into a set of inequalities on the amplitudes. For many of the simple examples described here, this step is relatively easy given the structure of the sum rule. However, when the shape of the cone becomes more intricate, as typically happens when the number of edges is larger than the dimension of the ambient space of amplitudes, then there is not a simple correspondence between coordinates/amplitudes and ERs. If the cone is polyhedral, the algorithm of vertex enumeration from convex geometry may be applied. This will be illustrated in the examples in Section \ref{sec:SMFermions} below. However, whenever degenerate states exist (two-particle states with the same quantum numbers, which are typically pervasive amongst theories), the cone is non-polyhedral. A systematic method for determining the shape of the cone, and hence the causality bounds, remains a problem for further work.

\subsubsection{From amplitudes to effective operators}
This step is well-known and not new. Given the effective action for the EFT, the standard Dyson series expansion can be performed to obtain the relevant scattering amplitudes at the relevant order of precision - here $\sim s^2/\Lambda^4$ for typical centre of mass energy scale $\sqrt{s}$. Their derivatives $\frac{d^2}{ds^2}A(s)$ that appear in the sum rule are then functions of the Wilson coefficients in the action. 

The present work will include some exploration of lower dimension operators in the sum rule, mostly focused on dimension $6$, and discuss how they modify constraints previously limited to dimension $8$ Wilson coefficients. It is at this third stage in the program where this issue becomes relevant. However, an interpretation of the constraints directly on the structure of the $S$-matrix is unaffected.

It is interesting to wonder whether this step can be made altogether redundant. In such a formulation, the effective action would be redundant and the $S$-matrix may be perturbatively constructed directly from its singularity structure out of a set of contact interactions consistent with Lorentz invariance. The strength of the contact interactions would be an equivalent parameterisation to the Wilson coefficients. This program would require both a systematic understanding of the all-order singularity structure of the $S$-matrix (i.e. causality and locality) and a systematic method for actually performing this reconstructing in order to be a complete replacement, although for simple enough theories at low enough order and few enough legs (which cover all applications considered here), this is currently feasible.

\section{Bounds on Inelastic Transitions}\label{sec:ComplexScalarEG}

This section presents a simple example to concretely illustrate the general discussion presented above. However, the constraints presented here are also new and demonstrate the general way in which these bounds fundamentally limit the extent of symmetry violation by effective interactions.

\subsection{Multiple scalars}\label{sec:Scalars}
For a theory of a single scalar field $\phi$, the positivity of the coefficient of the $(\partial\phi)^4$ operator is well known. This would be the leading irrelevant operator if the scalar was a Goldstone boson. Now consider a more general EFT of a complex scalar with effective interaction Lagrangian density
\begin{align}
    \mathcal{L}_{EFT_6}&=\frac{c_6}{\Lambda^2}\phi\phi\left(\partial\phi^\dagger\cdot\partial\phi^\dagger\right)
\end{align}
and 
\begin{align}
    \mathcal{L}_{EFT_8}=\frac{c_8}{\Lambda^4}\left(\partial\phi\cdot\partial\phi\right)\left(\partial\phi^\dagger\cdot\partial\phi^\dagger\right)+\frac{\tilde{c}_8}{\Lambda^4}\left(\partial\phi\cdot\partial\phi^\dagger\right)\left(\partial\phi\cdot\partial\phi^\dagger\right)\nonumber\\
    +\frac{d_8}{2\Lambda^4}\left(\partial\phi\cdot\partial\phi\right)^2+\frac{2\tilde{d}_8}{\Lambda^4}\left(\partial\phi\cdot\partial\phi\right)\left(\partial\phi\cdot\partial\phi^\dagger\right)+\text{conj.}.
\end{align}
Here $c_6$, $c_8$ and $\tilde{c}_8$ are real while $d_8$ and $\tilde{d_8}$ are complex. Complex scalars are usually associated with $U(1)$ symmetries, but this will not be assumed here. Only the existence of  charge conjugation $C$ will be assumed to relate the two real scalar states. Nevertheless, at dim-$6$ level, the only possible $4$-point operator is charge conserving (this would not be true if there were more species). See \cite{Andriolo:2020lul} for an analogous recent analysis of a two scalar system (axion and dilaton) in which both degrees of freedom are totally unrelated by symmetry (the bound is exactly that expected from (\ref{InelBound})).

Note that marginal and relevant operators could also be included - it will be assumed that these are small perturbations such that they can be neglected from the leading contributions at each order the EFT expansion. Operators composed of more than four scalars $\phi$, such as $\phi^6$, have also been neglected for simplicity, but would ordinarily be considered at the same order. This would all be justified if the scalar was a Goldstone boson, in addition to ruling-out all possible dimension $6$ operators and higher point dimension $8$ operators so that $\mathcal{L}_{EFT_8}$ would give a complete leading order description of the interactions. However, I choose to include $\mathcal{L}_{EFT_6}$ here to provide a simple illustration of the inclusion of loops.

\begin{table}
\centering
\begin{adjustwidth}{-1.5cm}{0cm}
\begin{tabular}{ |p{1cm}||p{3.8cm}|p{3.8cm}|p{3.8cm}|p{3.8cm}| }
 \hline
  \, & $\phi\phi$ & $\phi\overline{\phi}$ & $\overline{\phi}\phi$ & $\overline{\phi}\overline{\phi}$\\
 \hline
 \hline
 $\phi\phi$   & $|\boldsymbol{m}^{\phi\phi}|^2+|\boldsymbol{m}^{\phi\bar{\phi}}|^2$    & $2\boldsymbol{m}^{\phi\bar{\phi}}\cdot\boldsymbol{m}^{\phi\phi}$ & $\boldsymbol{m}^{\bar{\phi}\phi}\cdot\boldsymbol{m}^{\phi\phi}+\boldsymbol{m}^{\bar{\phi}\bar{\phi}}\cdot\boldsymbol{m}^{\phi\bar{\phi}}$  & $2\boldsymbol{m}^{\bar{\phi}\bar{\phi}}\cdot\boldsymbol{m}^{\phi\phi}$\\
 $\phi\overline{\phi}$ & $2\boldsymbol{m}^{\phi\phi}\cdot\boldsymbol{m}^{\phi\bar{\phi}}$ &  $|\boldsymbol{m}^{\phi\bar{\phi}}|^2+|\boldsymbol{m}^{\phi\phi}|^2$  & $2\boldsymbol{m}^{\bar{\phi}\phi}\cdot\boldsymbol{m}^{\phi\bar{\phi}}$  & $\boldsymbol{m}^{\bar{\phi}\bar{\phi}}\cdot\boldsymbol{m}^{\phi\bar{\phi}}+\boldsymbol{m}^{\bar{\phi}\phi}\cdot\boldsymbol{m}^{\phi\phi}$\\
 $\overline{\phi}\phi$ & $\boldsymbol{m}^{\phi\phi}\cdot\boldsymbol{m}^{\bar{\phi}\phi}+\boldsymbol{m}^{\phi\bar{\phi}}\cdot\boldsymbol{m}^{\bar{\phi}\bar{\phi}}$ & $2\boldsymbol{m}^{\phi\bar{\phi}}\cdot\boldsymbol{m}^{\bar{\phi}\phi}$ &  $|\boldsymbol{m}^{\bar{\phi}\phi}|^2+|\boldsymbol{m}^{\bar{\phi}\bar{\phi}}|^2$ & $2\boldsymbol{m}^{\bar{\phi}\bar{\phi}}\cdot\boldsymbol{m}^{\bar{\phi}\phi}$\\
 $\overline{\phi}\overline{\phi}$    & $2\boldsymbol{m}^{\phi\phi}\cdot\boldsymbol{m}^{\bar{\phi}\bar{\phi}}$ & $\boldsymbol{m}^{\phi\bar{\phi}}\cdot\boldsymbol{m}^{\bar{\phi}\bar{\phi}}+\boldsymbol{m}^{\phi\phi}\cdot\boldsymbol{m}^{\bar{\phi}\phi}$ & $2\boldsymbol{m}^{\bar{\phi}\phi}\cdot\boldsymbol{m}^{\bar{\phi}\bar{\phi}}$ & $|\boldsymbol{m}^{\bar{\phi}\bar{\phi}}|^2+|\boldsymbol{m}^{\bar{\phi}\phi}|^2$\\
 \hline
\end{tabular}
\end{adjustwidth}
\caption{\label{Tab:CScalar} Sum rule for complex scalar theory.}
\end{table}

Firstly to analyse the structure of the constraints on the $S$-matrix entries. 
The sum rule can be organised into a matrix of incoming and outgoing states. This is given in Table \ref{Tab:CScalar}.
Here, there are four complex vectors with components corresponding to the amplitudes $m^{\phi\phi}_i=\mathcal{M}^{\phi\phi\rightarrow X_i}(s_i)$, $m^{\phi\bar{\phi}}_i=\mathcal{M}^{\phi\overline{\phi}\rightarrow X_i}(s_i)$, $m^{\bar{\phi}\phi}_i=\mathcal{M}^{\overline{\phi}\phi\rightarrow X_i}(s_i)$ and $m^{\bar{\phi}\bar{\phi}}_i=\mathcal{M}^{\overline{\phi}\overline{\phi}\rightarrow X_i}(s_i)$, where each entry in the vector corresponds to a particular state $i$ in the UV completion up to an unimportant overall positive scalar coefficient. $CPT$ implies that all of the elastic amplitudes are equal so that $|\boldsymbol{m}^{\phi\phi}|^2+|\boldsymbol{m}^{\phi\bar{\phi}}|^2=|\boldsymbol{m}^{\bar{\phi}\phi}|^2+|\boldsymbol{m}^{\bar{\phi}\bar{\phi}}|^2$ and, up to an irrelevant phase, $\boldsymbol{m}^{\phi\bar{\phi}}\cdot\boldsymbol{m}^{\phi\phi}=\boldsymbol{m}^{\bar{\phi}\bar{\phi}}\cdot\boldsymbol{m}^{\bar{\phi}\phi}$. Then $Y$ symmetry by itself also implies that $|\boldsymbol{m}^{\phi\bar{\phi}}|=|\boldsymbol{m}^{\bar{\phi}\phi}|$ and $|\boldsymbol{m}^{\phi\phi}|=|\boldsymbol{m}^{\bar{\phi}\bar{\phi}}|$, while $\boldsymbol{m}^{\phi\bar{\phi}}\cdot\boldsymbol{m}^{\phi\phi}+\boldsymbol{m}^{\bar{\phi}\bar{\phi}}\cdot\boldsymbol{m}^{\phi\bar{\phi}}=2\boldsymbol{m}^{\phi\bar{\phi}}\cdot\boldsymbol{m}^{\phi\phi}$. This simplifies the matrix, in particular equating each single-charge violating amplitude in the upper triangle.

The standard positivity theorems on elastic forward scattering \cite{Adams:2006sv} are immediately clear from the diagonal entries in this table. However, there is clearly more information. These constraints can be extracted by applying the Schwarz and triangle inequalities. For example, $|2\boldsymbol{m}^{\phi\phi}\cdot\boldsymbol{m}^{\phi\bar{\phi}}|\leq 2|\boldsymbol{m}^{\phi\phi}||\boldsymbol{m}^{\phi\bar{\phi}}|\leq |\boldsymbol{m}^{\phi\phi}|^2+|\boldsymbol{m}^{\phi\bar{\phi}}|^2$, which is the statement on the LHS that
\begin{align}\label{ChargeVolBound}
    |M^{\phi\phi\phi\overline{\phi}}|\leq M^{\phi\overline{\phi}\phi\overline{\phi}},
\end{align}
that is, the single-charge violating amplitude must be smaller than the charge conserving one. Likewise, 
\begin{align}
2|\boldsymbol{m}^{\phi\phi}\cdot\boldsymbol{m}^{\bar{\phi}\bar{\phi}}|+2|\boldsymbol{m}^{\phi\bar{\phi}}\cdot\boldsymbol{m}^{\bar{\phi}\bar{\phi}}|&\leq 2|\boldsymbol{m}^{\phi\phi}||\boldsymbol{m}^{\bar{\phi}\bar{\phi}}|+2|\boldsymbol{m}^{\phi\bar{\phi}}||\boldsymbol{m}^{\bar{\phi}\bar{\phi}}|\nonumber\\
&\leq 2\sqrt{\left(|\boldsymbol{m}^{\phi\phi}|^2+|\boldsymbol{m}^{\phi\bar{\phi}}|^2\right)\left(|\boldsymbol{m}^{\bar{\phi}\bar{\phi}}|^2+|\boldsymbol{m}^{\bar{\phi}\bar{\phi}}|^2\right)}
\end{align}
implies that 
\begin{align}\label{ScalarBound2}
|M^{\phi\phi\overline{\phi}\overline{\phi}}|+|M^{\phi\overline{\phi}\overline{\phi}\phi}|\leq 2M^{\phi\overline{\phi}\phi\overline{\phi}}.
\end{align}

These statements can then be converted into constraints on the Wilson coefficients. At tree-level, these bounds may be directly translated into the statement that the dim-$8$ charge conserving Wilson coefficients must be larger than the charge violating ones. This is, however, also an appropriate place to illustrate the inclusion of a loop process in the sum rule so that the affect of the dim-$6$ operators can be accounted for.

The relevant amplitudes at dim-$8$ order are 
\begin{align}
    A\left(\phi,\overline{\phi}\rightarrow\phi,\overline{\phi}\right)&=\frac{c_8}{\Lambda^4}u^2+\frac{\tilde{c}_8}{2\Lambda^4}\left(s^2+t^2\right)+\frac{4}{(4\pi)^2}\frac{c_6^2}{\Lambda^4}\Bigg(\frac{2}{3}u^2+\int_0^1x(1-x)\Bigg(3u^2\log\left(\frac{\mu^2}{-x(1-x)u}\right)\nonumber\\
    &\quad+t(t-u)\log\left(\frac{\mu^2}{-x(1-x)t}\right)+s(s-u)\log\left(\frac{\mu^2}{-x(1-x)s}\right)\Bigg)dx\Bigg)\\
    A\left(\phi,\phi\rightarrow\phi,\overline{\phi}\right)&=\frac{\tilde{d}_8}{\Lambda^4}\left(s^2+t^2+u^2\right)\\
    A\left(\phi,\phi\rightarrow\overline{\phi},\overline{\phi}\right)&=\frac{d_8}{\Lambda^4}\left(s^2+t^2+u^2\right)\\
    A\left(\phi,\overline{\phi}\rightarrow\overline{\phi},\phi\right)&=A\left(\phi,\overline{\phi}\rightarrow\phi,\overline{\phi}\right)|_{u\mapsto t, t\mapsto s, s\mapsto u}.
\end{align}
As usual, $\mu$ is the renormalisation scale. The couplings are implicitly functions of this.

Taking the forward limit and differentiating give the entries for the LHS of the sum rule. Because of the singularities, I will take the insertion at $s=\sigma^2+i\delta$ for some $\sigma^2>0$ and $\delta\rightarrow 0^+$, as explained in Section \ref{sec:SumRule} above:
\begin{align}
    \overline{M}(\phi,\overline{\phi}\rightarrow\phi,\overline{\phi})&=\frac{2c_8}{\Lambda^4}+\frac{\tilde{c}_8}{\Lambda^4}+\frac{4}{(4\pi)^2}\frac{c_6^2}{\Lambda^4}\left(\frac{29}{18}+\frac{2\pi i}{3}+\frac{5}{3}\log\left(\frac{\mu^2}{\sigma^2}\right)\right)\\
    \overline{M}(\phi,\phi\rightarrow\phi,\overline{\phi})&=\frac{4\tilde{d}_8}{\Lambda^4}\\
    \overline{M}(\phi,\phi\rightarrow\overline{\phi},\overline{\phi})&=\frac{4d_8}{\Lambda^4}\\
    \overline{M}(\phi,\overline{\phi}\rightarrow\overline{\phi},\phi)&=\frac{2\tilde{c}_8}{\Lambda^4}+\frac{4}{(4\pi)^2}\frac{c_6^2}{\Lambda^4}\left(\frac{1}{9}+\frac{\pi i}{3}+\frac{2}{3}\log\left(\frac{\mu^2}{\sigma^2}\right)\right).
\end{align}
The IR part of the dispersion integral needs to be added to this to obtain the full LHS. This cancels the imaginary part of the amplitudes above (corresponding to above threshold production of the light states in the EFT), as well as the logarithmic dependence on $\sigma^2$, which can be taken arbitrarily soft, leaving behind a scheme-dependent correction to the coupling representing its RG evolution from $\mu$ to the cut-off. The terms from the dispersion integral relevant for each loop amplitude are:
\begin{align}
    &\frac{2}{\pi}\int_0^{\Lambda^2}s\left(\frac{\sigma^{(6)}\left(\phi,\overline{\phi}\rightarrow\phi,\overline{\phi}\right)}{\left(s-\sigma^2-i0^+\right)^3}+\frac{\sigma^{(6)}\left(\phi,\phi\rightarrow\phi,\phi\right)}{\left(s+\sigma^2\right)^3}\right)ds\approx\frac{5}{12\pi^2}\frac{c_6^2}{\Lambda^4}\left(\frac{-3}{2}+\frac{2\pi i}{5}+\log\left(\frac{\Lambda^2}{\sigma^2}\right)\right)\label{ElasticDispInt}\\
    \frac{1}{\pi}&\int_0^{\Lambda^2}\left(\frac{1}{\left(s-\sigma^2-i0^+\right)^3}+\frac{1}{\left(s+\sigma^2\right)^3}\right)\int A^{(6)}\left(\phi,\overline{\phi}\rightarrow\phi,\overline{\phi}\right)\left(A^{(6)}\left(\overline{\phi},\phi\rightarrow\phi,\overline{\phi}\right)\right)^*d\Pi ds\nonumber\\
    &\qquad\qquad\qquad\qquad\qquad\qquad\qquad\qquad\approx\frac{1}{6\pi^2}\frac{c_6^2}{\Lambda^4}\left(\frac{-3}{2}+\frac{\pi i}{2}+\log\left(\frac{\Lambda^2}{\sigma^2}\right)\right)\label{ElasticDispInt2}
\end{align}
where $\sigma^{(6)}$ denotes cross section determined from the dim-$6$ tree amplitudes $A^{(6)}$ and $\Pi$ is the Lorentz-invariant phase space of the intermediate states being integrates over. The conventional optical theorem has been invoked in the statement of (\ref{ElasticDispInt}) because the relevant processes are elastic. The constraints (\ref{ChargeVolBound}) and (\ref{ScalarBound2}) therefore translate into bounds on Wilson coefficients:
\begin{align}
    4|\tilde{d}_8|\leq 2c_8+\tilde{c}_8+\frac{c_6^2}{(4\pi)^2}\left(\frac{148}{9}+\frac{20}{3}\log\left(\frac{\mu^2}{\Lambda^2}\right)\right)\\
    2|d_8|+\left|\tilde{c}_8+\frac{c_6^2}{(4\pi)^2}\left(\frac{20}{9}+\frac{4}{3}\log\left(\frac{\mu^2}{\Lambda^2}\right)\right)\right|\leq 2c_8+\tilde{c}_8+\frac{c_6^2}{(4\pi)^2}\left(\frac{148}{9}+\frac{20}{3}\log\left(\frac{\mu^2}{\Lambda^2}\right)\right).
\end{align}

The above example also makes explicit the issues of RG-scale dependence described in Section \ref{sec:SumRule}. The terms proportional to $\log\left(\frac{\mu^2}{\Lambda^2}\right)$ represent RG-evolution of the dim-$8$ couplings from the renormalisation scale $\mu$ to the cut-off $\Lambda$. The upper limit of the IR segment of the dispersion integral could have instead been chosen to be some $\lambda<\Lambda$. In this case, the above calculation would be mostly unchanged, but with $\Lambda\mapsto\lambda$ and the addition of terms $\mathcal{O}\left(\frac{\sigma^2}{\lambda^2}\right)$ on the RHS of (\ref{ElasticDispInt}) and (\ref{ElasticDispInt2}). These latter terms were dropped with $\lambda =\Lambda$ because they are higher order in the energy expansion organising the EFT, but must be retained for smaller $\lambda$. They are nevertheless eliminated by taking $\sigma\rightarrow 0$. As a result, the constraints differ only in the replacement of the cut-off $\Lambda$ by the lower energy $\lambda$ in the logarithms, representing RG-evolution to the scale $\lambda$ instead. While the bounds must hold for all $\lambda$ and therefore represent constraints on the entire flow, they are typically optimised by taking $\lambda\rightarrow\Lambda$ because of the positive sign of the $c_6^2\log$ contribution. This is the reason for integrating all the way to the cut-off. Interestingly, if the sign of the coefficient of the logarithm were instead negative, then taking $\lambda$ arbitrarily small would place arbitrarily strong lower bounds on the dim-$8$ coefficients mediating elastic scattering, effectively ruling them out. That this cannot happen is consequence of unitarity. 

However, it is also of note that the logarithmic term is increasingly negative with higher cut-off. The division between the dim-$8$ contact coefficients and the rational terms proportional to $c_6^2$ is renormalisation scheme-dependent and the sum of both terms should be the object of comparison with the pure tree-level bounds. The dim-$6$ loop contribution should be entirely attributed to the logarithm and this contributes negatively to the LHS of the sum rule, strengthening the bound relative to the tree-level expectation. Stated equivalently, the contact coefficients decrease with increasing RG scale to the extent that their IR values can be (at least partially) cancelled. For a given coupling at scale $\mu$, the constraints fundamentally limit the extent that the cut-off can be extended, or alternatively, for a given cut-off, improve the positivity bounds on the dim-$8$ contact interaction strength in the IR. The mere presence of dim-$6$ operators therefore enhances the lower bound on the size of dim-$8$ operators mediating elastic scattering.

The organisation of the sum rule presented here can be further applied to more complicated theories with multiple species, of which many examples will follow.

\section{Bounds with Internal Symmetries - Flavour and Colour}\label{sec:Internal}
The presentation of the sum rule illustrated in the previous sections demonstrates the nature of the new bounds for inelastic processes between distinct particles. The next level of sophistication to discuss is for theories with multiple states related by symmetries. Positivity bounds in theories with global symmetries have been discussed previously in \cite{Bellazzini:2014waa}. The present discussion will further examine the new positivity bounds suggested in \cite{Zhang:2020jyn} for theories of particles belonging to non-trivial representations of multiple symmetry groups. While the action of the symmetries on the states factorise, because all sets of indices are crossed simultaneously between the $s$-channel and $u$-channel terms in (\ref{SumRule}), they become effectively entangled across different entries in the full matrix of sum rules. This raises the possibility of new constraints that cannot be accessed by considering only elastic scattering of factorised superpositions of states.

The bounds are determined from the ERs in the convex cone picture. Each ER is itself identified with a particular irrep under the global symmetry that a pair of external scattering states can couple to. Thus if a particular representation $R$ lies in the Clebsch-Gordan decomposition of the incoming states in either the $s$-channel or the $u$-channel, it will yield an ER in the RHS of the sum rule. Stage one of the procedure enumerated in Section \ref{sec:Program} thus reduces to a decomposition of the theory's amplitudes into a set of partial amplitudes describing symmetry preserving transitions. This will be elaborated upon more precisely below. 

The goal of this section is to analyse theories of global symmetries with the features just described and discover new bounds. This will provide a further educational illustration of the structure of the sum rule constraints, the way in which symmetries are managed and the convex cone of UV completions. More importantly however, the cases considered here will directly apply to the fermionic operators with the global symmetries of the SM. The important special case of rotational symmetry will be deferred to the next section in order to avoid distraction from the goal. However, for the theories of (hyper)charged chiral fermions considered here, such a treatment for spin is not necessary and each left-handed particle and right-handed antiparticle may be treated as independent states that have amplitudes related by only $CPT$, similarly to the scalars in the previous section (although hypercharge conservation will be assumed here). Note that the following method for the accounting of symmetries and the algorithm used to derive the positivity bounds are not intended to represent an application of the simple inelastic bounds derived earlier in (\ref{InelBound}) and could include more information.

\subsection{Background group theory}\label{sec:GroupTheory}
A $2\rightarrow 2$ scattering amplitude can be decomposed into partial amplitudes corresponding to symmetry preserving transitions between particular irreps. Assume that the incoming and outgoing particles transform under representations of some symmetry group. Generally, the initial states may be decomposed into irreps with Clebsch-Gordan coefficients defined here as
\begin{align}
    C^{ab}_{R_\xi\iota}=\langle R_\xi;\iota|\left(|r1;a\rangle|r2;b\rangle\right)
\end{align}
to give 
\begin{align}
    |r1;a\rangle|r2;b\rangle=\sum_{R_\xi,\iota}C^{ab}_{R_\xi\iota}|R_\xi; \iota\rangle,
\end{align}
where $r_1$ and $r_2$ label the representations of the individual particles, $a$ and $b$ their components, while $R$ and $\iota$ index the product irreps and components. The index $\xi$ counts degenerate representations that may arise. 

Projection tensors can be defined as
\begin{align}
    P_{R_{\xi\xi'}}^{abcd}=\sum_{\iota}C^{ab}_{R_\xi\iota}\left(C^{cd}_{R_{\xi'}\iota}\right)^*.
\end{align}
These obey orthogonality conditions
\begin{align}
    \frac{1}{\text{dim} R}\sum_{a,b,c,d}\left(P_{R_{\xi\xi'}}^{abcd}\right)^*\left(P_{R'_{\eta\eta'}}^{abcd}\right)=\delta_{RR'}\delta_{\xi\eta}\delta_{\xi'\eta'}.
\end{align}

The final states may be likewise decomposed. The Wigner-Eckart theorem then implies that the resulting transition amplitude is diagonal in representation and components, although transitions between distinct but degenerate representations are permitted. This will be especially important when spin is discussed later. The full amplitude decomposes as 
\begin{align}
    A^{abcd}&=\,_{\text{out}}\left(\langle r4;d|\langle r3;c|\right)\left(|r1;a\rangle|r2;b\rangle\right)_{\text{in}}\nonumber\\
    &=\sum_{R,\xi,\xi'} P_{R_{\xi\xi'}}^{abcd}A_{R_{\xi\xi'}},\label{WEtheorem}
\end{align}
where the partial amplitudes are defined as 
\begin{align}\label{PartialAmp}
    A_{R_{\xi\xi'}} = \,_{\text{out}}\langle R_{\xi'};\iota|R_\xi;\iota\rangle_{\text{in}}
\end{align}
and may be extracted from the full amplitude by the action of projectors (note that the RHS of (\ref{PartialAmp}) is independent of the choice of the component $\iota$ and no sum is implied).

The projection operators thus encode all of the symmetry relations between amplitudes of different states. The irreps are the states that block-diagonalise the $S$-matrix. Each term in the $s$-channel of the sum rule (the first term in (\ref{SumRule})) can be decomposed into irreps into the form 
\begin{align}
    M_s=\sum_{R_s,\xi,\xi'}P_{R_{s{\xi\xi'}}}^{abcd}\,\boldsymbol{m}_{R_{s\xi'}}\cdot\boldsymbol{m}_{R_{s\xi}},
\end{align}
where the $s$ subscript on the irrep label $R$ has been used to emphasise applicability to the $s$-channel decomposition. The u-channel term in (\ref{SumRule}) can likewise be decomposed with particles $b$ and $d$ exchanged with each others' antiparticles. The irreps in this case may be entirely different. However, by the Wigner-Eckart theorem, it must be possible for the $u$-channel projectors to be decomposed as linear combinations of the $s$-channel ones so that the amplitude takes the form (\ref{WEtheorem}) given entirely in terms of $s$-channel projectors. 
\begin{align}
    P_{R_{u{\rho\rho'}}}^{a\bar{d}c\bar{b}}=\sum_{R_s,\xi,\xi'}c_{R_{u{\rho\rho'}}R_{s{\xi\xi'}}}P_{R_{s{\xi\xi'}}}^{abcd}.
\end{align}
The numerical constants $c_{R_{u{\rho\rho'}}R_{s{\xi\xi'}}}$ of this decomposition are entirely determined by the Clebsch-Gordan coefficients of the group. These will be presented below for various examples relevant to the SM. As a result of this decomposition, the sum rule (\ref{SumRule}) in the presence of global symmetries may be expressed as 
\begin{align}\label{SymmetrySumRule}
    M^{abcd}=P_{R_{\xi\xi'}}^{abcd}\,\boldsymbol{m}_{R_{\xi'}}\cdot\boldsymbol{m}_{R_\xi}+\sum_{R_s,\xi,\xi'}c_{R_{u{\rho\rho'}}R_{s{\xi\xi'}}}P_{R_{{\xi\xi'}}}^{abcd}\,\boldsymbol{m}_{R_{u\rho'}}\cdot\boldsymbol{m}_{R_{u\rho}}
\end{align}
The $s$ label on the irreps of the $s$-channel has now been dropped. If there are no degeneracies, then each term in (\ref{SymmetrySumRule}) is of the form $|\boldsymbol{m}_R|^2$ and can be identified with a PER.

The special case of $SU(3)$ will be used in examples below. Projectors for the irreps that arise in combining fundamental and antifundamental representations will be necessary. The Clebsch-Gordan coefficients may be easily inferred from the exchange symmetry structure of the representations in tensor form. The projectors onto each irrep appearing in the products $\mathbf{3}\otimes \mathbf{3}$ are then determined as
\begin{align}
    P_{\overline{\mathbf{3}}\,\,\,cd}^{ab}&=\frac{1}{2}\left(\delta^a_c\delta^b_d-\delta^a_d\delta^b_c\right)\\
    P_{\mathbf{6}\,\,\,cd}^{ab}&=\frac{1}{2}\left(\delta^a_c\delta^b_d+\delta^a_d\delta^b_c\right).
\end{align}
For $\mathbf{3}\otimes \overline{\mathbf{3}}$ transitioning into $\mathbf{3}\otimes \overline{\mathbf{3}}$, they are
\begin{align}
    P_{\mathbf{1}\,bc}^{a\,\,\,\,\,d}&=\frac{1}{3}\delta^a_b\delta^d_c\\
    P_{\mathbf{8}\,bc}^{a\,\,\,\,\,d}&=\delta^a_c\delta^d_b-\frac{1}{3}\delta^a_b\delta^d_c,
    \end{align}
while for $\mathbf{3}\otimes \overline{\mathbf{3}}$ transitioning into $\overline{\mathbf{3}}\otimes\mathbf{3}$, they are 
\begin{align}
    P_{\mathbf{1}\,b\,\,d}^{a\,\,c}&=\frac{1}{3}\delta^a_b\delta^c_d\\
    P_{\mathbf{8}\,b\,\,d}^{a\,\,c}&=\delta^a_d\delta^c_b-\frac{1}{3}\delta^a_b\delta^c_d.
\end{align}
Raised indices indicate fundamental, lowered are antifundamental. The cases where the representations are conjugate are identical, with index heights reversed.

It will also be necessary to decompose the projectors in the $u$-channel into projectors for the $s$-channel, which is traditionally referred to as ``finding the crossing matrix''. For the projectors above, these are 
\begin{align}
    P_{\mathbf{1}\,dc}^{a\,\,\,\,\,b}&=\frac{1}{3}\left(P_{\mathbf{6}\,\,\,cd}^{ab}-P_{\overline{\mathbf{3}}\,\,\,cd}^{ab}\right)\\
    P_{\mathbf{8}\,dc}^{a\,\,\,\,\,b}&=\frac{2}{3}\left(P_{\mathbf{6}\,\,\,cd}^{ab}+2P_{\overline{\mathbf{3}}\,\,\,cd}^{ab}\right)\\
    P_{\mathbf{1}\,d\,\,b}^{a\,\,c}&=\frac{1}{3}\left(P_{\mathbf{1}\,b\,\,d}^{a\,\,c}+P_{\mathbf{8}\,b\,\,d}^{a\,\,c}\right)\\
    P_{\mathbf{8}\,d\,\,b}^{a\,\,c}&=\frac{1}{3}\left(8P_{\mathbf{1}\,b\,\,d}^{a\,\,c}-P_{\mathbf{8}\,b\,\,d}^{a\,\,c}\right)\\
    P_{\overline{\mathbf{3}}\,\,\,cb}^{ad}&=\frac{1}{2}\left(P_{\mathbf{8}\,bc}^{a\,\,\,\,\,d}-2P_{\mathbf{1}\,bc}^{a\,\,\,\,\,d}\right)\\
    P_{\mathbf{6}\,\,\,cb}^{ad}&=\frac{1}{2}\left(P_{\mathbf{8}\,bc}^{a\,\,\,\,\,d}+4P_{\mathbf{1}\,bc}^{a\,\,\,\,\,d}\right).
\end{align}

Also of use will be the projectors for  $SU(2)$. For the product $\mathbf{2}\otimes \mathbf{2}$, they are \begin{align}
    P_{\mathbf{1}\,\,\,cd}^{ab}&=-\frac{1}{2}\epsilon^{ab}\epsilon_{cd}\\
    P_{\mathbf{3}\,\,\,cd}^{ab}&=\frac{1}{2}\left(\delta^{a}_c\delta^b_{d}+\delta^{a}_d\delta^{b}_c\right).
\end{align}
The indices can be raised and lowered by $\epsilon$ tensors in order to relate these to the projectors appearing in amplitudes involving the conjugate representations. An additional factor of $-1$ must be included for each index either raised or lowered in this way (because, for a state $\psi^a$, defining $\psi_a=\epsilon_{ab}\psi^b$ and $\psi^{\dagger a}=\epsilon^{ab}\psi^\dagger_b$, then $(\psi_a)^\dagger=-\psi^{\dagger a}$). The $u$-channel projectors decompose as
\begin{align}
    P_{\mathbf{1}\,dc}^{a\,\,\,\,\,b}&=-\frac{1}{2}\left(P_{\mathbf{1}\,\,\,cd}^{ab}-P_{\mathbf{3}\,\,\,cd}^{ab}\right)\\
    P_{\mathbf{3}\,dc}^{a\,\,\,\,\,b}&=\frac{1}{2}\left(3P_{\mathbf{1}\,\,\,cd}^{ab}+P_{\mathbf{3}\,\,\,cd}^{ab}\right).
\end{align}

For the product $\mathbf{3}\otimes \mathbf{3}$, they are 
\begin{align}
    P_{\mathbf{1}}^{abcd}&=\frac{1}{2}\delta^{ab}\delta^{cd}\\
    P_{\mathbf{3}}^{abcd}&=\frac{1}{2}\left(\delta^{ac}\delta^{bd}-\delta^{ad}\delta^{bc}\right)\\
    P_{\mathbf{5}}^{abcd}&=\frac{1}{2}\left(\delta^{ac}\delta^{bd}+\delta^{ad}\delta^{bc}-\delta^{ab}\delta^{cd}\right).
\end{align}
The indices here label components of the $\mathbf{3}$ representation, rather than fundamental. The $u$-channel projectors decompose as 
\begin{align}
    P_{\mathbf{1}}^{adcb}&=\frac{1}{3}\left(P_{\mathbf{5}}^{abcd}-P_{\mathbf{3}}^{abcd}+P_{\mathbf{1}}^{abcd}\right)\\
    P_{\mathbf{3}}^{adcb}&=\frac{1}{2}\left(P_{\mathbf{5}}^{abcd}+P_{\mathbf{3}}^{abcd}-2P_{\mathbf{1}}^{abcd}\right)\\
    P_{\mathbf{5}}^{adcb}&=\frac{1}{6}\left(P_{\mathbf{5}}^{abcd}+5P_{\mathbf{3}}^{abcd}+10P_{\mathbf{1}}^{abcd}\right).
\end{align}

\subsection{Standard Model fermions}\label{sec:SMFermions}
As the SM is a theory of chiral fermions, EFTs of these states will be the focus of this section. The isospin, colour and flavour representations of the SM fermions will be systematically considered, with hypercharge conservation imposed. Helicity and hypercharge are not independent quantum numbers, so both sets of representations are equivalent. However, in the product representations of two such states, the non-zero charged irreps correspond to the rotational singlets, while the charge singlets constitute the non-trivial angular momentum irreps (see Section \ref{sec:RotSym} below for more explanation). This ensures that there are no transitions between distinct degenerate irreps and that hypercharge conservation is otherwise sufficient to account for both of these symmetries. 

I begin with a theory of hypercharged chiral fermions in the fundamental representation of $SU(3)$. These results would apply to a single flavour of right-handed down or up quarks, with the $SU(3)$ symmetry being interpreted as colour, or to right-handed leptons in which the full $SU(3)$ flavour symmetry is preserved in the UV. I will use notation describing the former. The terms in the sum rule are determined by finding projectors for the irreps of the external legs, beginning with the $s$-channel and then crossing to the $u$-channel, in a similar way to that illustrated in the previous section. Again parameterising the couplings to UV states by complex vectors, the relevant amplitudes are of the form
\begin{align}
    M\left(q_R,q_R\rightarrow q_R,q_R\right)&=M_{\overline{\mathbf{3}}}P_{\overline{\mathbf{3}}}+M_{\mathbf{6}}P_{\mathbf{6}}\nonumber\\
    &=\left(|\boldsymbol{m}_{\overline{\mathbf{3}}}|^2-\frac{1}{3}|\boldsymbol{m}_{\mathbf{1}}|^2+\frac{4}{3}|\boldsymbol{m}_{\mathbf{8}}|^2\right)P_{\overline{\mathbf{3}}}+\left(|\boldsymbol{m}_{\mathbf{6}}|^2+\frac{1}{3}|\boldsymbol{m}_{\mathbf{1}}|^2+\frac{2}{3}|\boldsymbol{m}_{\mathbf{8}}|^2\right)P_{\mathbf{6}}\\
    M\left(q_R,\overline{q}_L\rightarrow q_R,\overline{q}_L\right)&=M_{\mathbf{1}}P_{\mathbf{1}}+M_{\mathbf{8}}P_{\mathbf{8}}\nonumber\\
    &=\left(|\boldsymbol{m}_{\mathbf{1}}|^2-|\boldsymbol{m}_{\overline{\mathbf{3}}}|^2+2|\boldsymbol{m}_{\mathbf{6}}|^2\right)P_{\mathbf{1}}+\left(|\boldsymbol{m}_{\mathbf{8}}|^2+\frac{1}{2}|\boldsymbol{m}_{\overline{\mathbf{3}}}|^2+\frac{1}{2}|\boldsymbol{m}_{\mathbf{6}}|^2\right)P_{\mathbf{8}}.
\end{align}
The others are either related by $CPT$ or are zero. The partial amplitudes for only one of these transitions are independent - the other channel is determined by crossing. It is easy to see that the parameters $\boldsymbol{m}_{\mathbf{6}}$ and $\boldsymbol{m}_{\mathbf{8}}$ are redundant. The remaining terms clearly span a $2d$ cone and can be converted into inequalities
\begin{align}
    M_{\overline{\mathbf{3}}}+M_{\mathbf{6}}>0\\
    M_{\mathbf{6}}>0.
\end{align}
These bounds correspond to those found in \cite{Remmen:2020vts} and (unsurprisingly) contain no new information.

Now to advance to fermions in the fundamental representation of $SU(2)\otimes SU(3)$, for example, left-handed leptons with flavour symmetry or a single flavour of left-handed quark. Adopting the latter interpretation, the independent, non-zero amplitudes are 
\begin{align}
    M\left(Q_L,Q_L\rightarrow Q_L,Q_L\right)&=\left(|\boldsymbol{m}_{(\mathbf{1},\overline{\mathbf{3}})}|^2+\frac{1}{6}|\boldsymbol{m}_{(\mathbf{1},\mathbf{1})}|^2-\frac{2}{3}|\boldsymbol{m}_{(\mathbf{1},\mathbf{8})}|^2-\frac{1}{2}|\boldsymbol{m}_{(\mathbf{3},\mathbf{1})}|^2+2|\boldsymbol{m}_{(\mathbf{3},\mathbf{8})}|^2\right)P_{\mathbf{1}}P_{\overline{\mathbf{3}}}\nonumber\\
    &+\left(|\boldsymbol{m}_{(\mathbf{1},\mathbf{6})}|^2-\frac{1}{6}|\boldsymbol{m}_{(\mathbf{1},\mathbf{1})}|^2-\frac{1}{3}|\boldsymbol{m}_{(\mathbf{1},\mathbf{8})}|^2+\frac{1}{2}|\boldsymbol{m}_{(\mathbf{3},\mathbf{1})}|^2+|\boldsymbol{m}_{(\mathbf{3},\mathbf{8})}|^2\right)P_{\mathbf{1}}P_{\mathbf{6}}\nonumber\\
    &+\left(|\boldsymbol{m}_{(\mathbf{3},\overline{\mathbf{3}})}|^2-\frac{1}{6}|\boldsymbol{m}_{(\mathbf{1},\mathbf{1})}|^2+\frac{2}{3}|\boldsymbol{m}_{(\mathbf{1},\mathbf{8})}|^2-\frac{1}{6}|\boldsymbol{m}_{(\mathbf{3},\mathbf{1})}|^2+\frac{2}{3}|\boldsymbol{m}_{(\mathbf{3},\mathbf{8})}|^2\right)P_{\mathbf{3}}P_{\overline{\mathbf{3}}}\nonumber\\
    &+\left(|\boldsymbol{m}_{(\mathbf{3},\mathbf{6})}|^2+\frac{1}{6}|\boldsymbol{m}_{(\mathbf{1},\mathbf{1})}|^2+\frac{1}{3}|\boldsymbol{m}_{(\mathbf{1},\mathbf{8})}|^2+\frac{1}{6}|\boldsymbol{m}_{(\mathbf{3},\mathbf{1})}|^2+\frac{1}{3}|\boldsymbol{m}_{(\mathbf{3},\mathbf{8})}|^2\right)P_{\mathbf{3}}P_{\mathbf{6}}\\
    M\left(Q_L,\overline{Q}_R\rightarrow Q_L,\overline{Q}_R\right)&=\left(|\boldsymbol{m}_{(\mathbf{1},\mathbf{1})}|^2+\frac{1}{2}|\boldsymbol{m}_{(\mathbf{1},\overline{\mathbf{3}})}|^2-|\boldsymbol{m}_{(\mathbf{1},\mathbf{6})}|^2-\frac{3}{2}|\boldsymbol{m}_{(\mathbf{3},\overline{\mathbf{3}})}|^2+3|\boldsymbol{m}_{(\mathbf{3},\mathbf{6})}|^2\right)P_{\mathbf{1}}P_{\mathbf{1}}\nonumber\\
    &+\left(|\boldsymbol{m}_{(\mathbf{1},\mathbf{8})}|^2-\frac{1}{4}|\boldsymbol{m}_{(\mathbf{1},\overline{\mathbf{3}})}|^2-\frac{1}{4}|\boldsymbol{m}_{(\mathbf{1},\mathbf{6})}|^2+\frac{3}{4}|\boldsymbol{m}_{(\mathbf{3},\overline{\mathbf{3}})}|^2+\frac{3}{4}|\boldsymbol{m}_{(\mathbf{3},\mathbf{6})}|^2\right)P_{\mathbf{1}}P_{\mathbf{8}}\nonumber\\
    &+\left(|\boldsymbol{m}_{(\mathbf{3},\mathbf{1})}|^2-\frac{1}{2}|\boldsymbol{m}_{(\mathbf{1},\overline{\mathbf{3}})}|^2+|\boldsymbol{m}_{(\mathbf{1},\mathbf{6})}|^2-\frac{1}{2}|\boldsymbol{m}_{(\mathbf{3},\overline{\mathbf{3}})}|^2+|\boldsymbol{m}_{(\mathbf{3},\mathbf{6})}|^2\right)P_{\mathbf{3}}P_{\mathbf{1}}\nonumber\\
    &+\left(|\boldsymbol{m}_{(\mathbf{3},\mathbf{8})}|^2+\frac{1}{4}|\boldsymbol{m}_{(\mathbf{1},\overline{\mathbf{3}})}|^2+\frac{1}{4}|\boldsymbol{m}_{(\mathbf{1},\mathbf{6})}|^2+\frac{1}{4}|\boldsymbol{m}_{(\mathbf{3},\overline{\mathbf{3}})}|^2+\frac{1}{4}|\boldsymbol{m}_{(\mathbf{3},\mathbf{6})}|^2\right)P_{\mathbf{3}}P_{\mathbf{8}}.
\end{align}
Choosing the coordinates to be the partial amplitudes $(M_{(\mathbf{1},\overline{\mathbf{3}})},M_{(\mathbf{1},\mathbf{6})},M_{(\mathbf{3},\overline{\mathbf{3}})},M_{(\mathbf{3},\mathbf{6})})$, the extremal rays can be read-off the sum rule and are (up to an arbitrary scale):
\begin{align}
\{(1,0,0,0), (0,1,0,0), (0,0,1,0), (1,-1,-1,1), (-2,-1,2,1), (-3,3,-1,1)\}. 
\end{align}
Note that the terms parameterised by both the vectors $\boldsymbol{m}_{(\mathbf{3},\mathbf{6})}$ and $\boldsymbol{m}_{(\mathbf{3},\mathbf{8})}$ are redundant, so have been excluded. The standard techniques of vertex enumeration may be directly applied to this system (see e.g. \cite{Fukuda:2016xxx}) in order to convert the extremal rays into linear inequalities among coordinates. The present example is readily computed by hand. However, the following examples rapidly grow in complexity. I use lrs \cite{Avis:2000xxx} as a cross-check here and to compute the more complicated examples to follow. The resulting constraints are 
\begin{align}
    M_{(\mathbf{1},\mathbf{6})}+M_{(\mathbf{3},\mathbf{6})}&>0\\
    M_{(\mathbf{3},\mathbf{6})}&>0\\
    M_{(\mathbf{1},\overline{\mathbf{3}})}+M_{(\mathbf{1},\mathbf{6})}+M_{(\mathbf{3},\overline{\mathbf{3}})}+M_{(\mathbf{3},\mathbf{6})}&>0\\
    M_{(\mathbf{3},\overline{\mathbf{3}})}+M_{(\mathbf{3},\mathbf{6})}&>0\\
    4M_{(\mathbf{1},\overline{\mathbf{3}})}+M_{(\mathbf{1},\mathbf{6})}+9M_{(\mathbf{3},\mathbf{6})}&>0\\
    M_{(\mathbf{1},\overline{\mathbf{3}})}+3M_{(\mathbf{3},\mathbf{6})}&>0.
\end{align}

These bounds can be compared to those given in \cite{Remmen:2020vts} derived specifically from elastic forward scattering. At tree-level, each partial amplitude corresponds to a particular dim-$8$ operator of the form $\left(\psi D\psi\right)\cdot\left(\psi^\dagger D\psi^\dagger\right)$ for chiral fermionic operator $\psi$. The correspondence is determined by decomposing the bilinears $\psi D\psi$ into the irreps of $SU(2)$ and $SU(3)$ given above - each different operator corresponds to one of the four such representations and thus corresponds to one of the four such partial amplitudes. The irrep into which $\psi D\psi$ is decomposed can only be contracted into a singlet with its conjugate, thus fully determining the operator. From the correspondence between operators and partial amplitudes it is sufficient to see that this set of six irreducible bounds contains more information than the four linear inequalities stated in \cite{Remmen:2020vts}. In particular, the last two are new. As argued in \cite{Zhang:2020jyn}, this is because the crossing between $s$ and $u$-channel terms in (\ref{SumRule}) involves simultaneously exchanging all degrees of freedom associated to the states, effectively entangling them.

This example illustrates the insight of the convex cone picture in cases where the space of allowed couplings is bounded by more faces than the dimension of the ambient space. In simple examples like the pure $SU(3)$ case described above, the number of linear positivity constraints are comparable to the dimension of the space being bounded, so bounds can be derived almost by direct inspection of the sum rule and some simple geometry. However, when the number of vectors parameterising the RHS of the sum rule exceeds the number of independent partial amplitudes, the space of allowed couplings becomes a multi-faceted polyhedral cone and the tools of convex geometry must be invoked. 

Next are fermions in the fundamental representation of $SU(3)\otimes SU(3)$, such as right-handed quarks with flavour symmetry. This is similar to the above example. The forward elastic amplitude $M\left(q_R,q_R\rightarrow q_R,q_R\right)$ can be decomposed into four independent partial amplitudes $\{M_{(\overline{\mathbf{3}},\overline{\mathbf{3}})},M_{(\overline{\mathbf{3}},\mathbf{6})},M_{(\mathbf{6},\overline{\mathbf{3}})},M_{(\mathbf{6},\mathbf{6})}\}$. The positivity constraints can be deduced by the same procedure to be 
\begin{align}
    M_{(\overline{\mathbf{3}},\mathbf{6})}+M_{(\mathbf{6},\mathbf{6})}&>0\\
    M_{(\mathbf{6},\mathbf{6})}&>0\\
     M_{(\overline{\mathbf{3}},\overline{\mathbf{3}})}+M_{(\overline{\mathbf{3}},\mathbf{6})}+M_{(\mathbf{6},\overline{\mathbf{3}})}+M_{(\mathbf{6},\mathbf{6})}&>0\\
    M_{(\mathbf{6},\overline{\mathbf{3}})}+M_{(\mathbf{6},\mathbf{6})}&>0\\
    M_{(\overline{\mathbf{3}},\overline{\mathbf{3}})}+2M_{(\mathbf{6},\mathbf{6})}&>0.
\end{align}
As the $SU(2)$ and $SU(3)$ projectors for these fundamental representations have the same tensor form, the first four bounds are analogous to those derived from forward scattering in the $Q_L$ case above. The last bound again corresponds to positivity of an entangled amplitude and differs from the example above because of the different crossing relations.

The final case considered here will be fermions in the fundamental representation of $SU(2)\otimes SU(3)\otimes SU(3)$, corresponding to states with isospin, colour and flavour, such as left-handed quarks. This is substantially more complicated than the previous two examples. The independent non-zero partial amplitudes are those of 
$M\left(Q_L,Q_L\rightarrow Q_L,Q_L\right)=\sum_{I,a,i}M_{(\boldsymbol{I},\boldsymbol{a},\boldsymbol{i})}P_{(\boldsymbol{I},\boldsymbol{a},\boldsymbol{i})}$, where $(\boldsymbol{I},\boldsymbol{a},\boldsymbol{i})$ indexes $SU(2)\otimes SU(3)\otimes SU(3)$ representations. This is an $8$-dimensional space, with a general vector of partial amplitudes denoted by $\boldsymbol{M}=(M_{(\mathbf{1},\overline{\mathbf{3}},\overline{\mathbf{3}})},M_{(\mathbf{3},\overline{\mathbf{3}},\overline{\mathbf{3}})},M_{(\mathbf{1},\mathbf{6},\overline{\mathbf{3}})},M_{(\mathbf{3},\mathbf{6},\overline{\mathbf{3}})},M_{(\mathbf{1},\overline{\mathbf{3}},\mathbf{6})},M_{(\mathbf{3},\overline{\mathbf{3}},\mathbf{6})},M_{(\mathbf{1},\mathbf{6},\mathbf{6})},M_{(\mathbf{3},\mathbf{6},\mathbf{6})})$. Repeating the procedure as above, there are $16$ PERs (for each partial amplitude in this and the crossed channel), of which two are redundant, leaving $14$ ERs. These may be converted into $44$ positivity bounds: 
\begin{equation*}
\begin{aligned}[c]
    \begin{matrix}
    0 & 1 & 0 & 1 & 0 & 1 & 0 & 1 \\
    1 & 1 & 1 & 1 & 1 & 1 & 1 & 1 \\
    1 & 3 & 2 & 0 & 2 & 0 & 0 & 6 \\
    5 & 9 & 8 & 0 & 8 & 0 & 2 & 18 \\
    11 & 15 & 16 & 0 & 16 & 0 & 0 & 48 \\
    1 & 0 & 1 & 0 & 0 & 3 & 0 & 3 \\
    4 & 0 & 4 & 0 & 1 & 9 & 1 & 9 \\
    8 & 0 & 8 & 0 & 3 & 15 & 0 & 24 \\
    1 & 1 & 0 & 0 & 0 & 0 & 2 & 2 \\
    8 & 0 & 4 & 0 & 0 & 12 & 5 & 21 \\
    0 & 1 & 0 & 0 & 0 & 0 & 0 & 6 \\
    7 & 0 & 2 & 0 & 2 & 0 & 0 & 36 \\
    2 & 0 & 0 & 0 & 0 & 0 & 1 & 9 \\
    4 & 0 & 0 & 0 & 0 & 4 & 5 & 13 \\
    8 & 0 & 0 & 12 & 4 & 0 & 5 & 21 \\
    3 & 0 & 0 & 5 & 0 & 5 & 0 & 13 \\
    8 & 0 & 0 & 4 & 0 & 4 & 7 & 23 \\
    4 & 0 & 0 & 4 & 0 & 0 & 5 & 13 \\
    0 & 1 & 0 & 0 & 0 & 0 & 0 & 2 \\
    0 & 0 & 0 & 0 & 1 & 1 & 1 & 1 \\
    0 & 0 & 0 & 0 & 0 & 1 & 0 & 1 \\
    0 & 3 & 1 & 0 & 2 & 1 & 0 & 7 \\
    \end{matrix}   
\end{aligned}
\qquad\qquad\qquad
\begin{aligned}[c]
    \begin{matrix}    
    1 & 0 & 0 & 3 & 1 & 0 & 0 & 3 \\
    4 & 0 & 1 & 9 & 4 & 0 & 1 & 9 \\
    8 & 0 & 3 & 15 & 8 & 0 & 0 & 24 \\
    0 & 0 & 0 & 0 & 4 & 0 & 1 & 9 \\
    4 & 0 & 0 & 8 & 4 & 0 & 3 & 11 \\
    0 & 0 & 0 & 0 & 1 & 0 & 0 & 3 \\
    1 & 0 & 1 & 0 & 6 & 0 & 0 & 18 \\
    0 & 3 & 1 & 0 & 5 & 0 & 0 & 15 \\
    0 & 1 & 0 & 1 & 3 & 0 & 0 & 9 \\
    0 & 0 & 1 & 1 & 0 & 0 & 1 & 1 \\
    4 & 0 & 4 & 0 & 0 & 8 & 3 & 11 \\
    0 & 0 & 4 & 0 & 0 & 0 & 1 & 9 \\
    0 & 0 & 1 & 0 & 0 & 0 & 0 & 3 \\
    1 & 0 & 6 & 0 & 1 & 0 & 0 & 18 \\
    0 & 1 & 3 & 0 & 0 & 1 & 0 & 9 \\
    0 & 3 & 5 & 0 & 1 & 0 & 0 & 15 \\
    0 & 0 & 0 & 0 & 0 & 0 & 0 & 1 \\
    0 & 0 & 0 & 0 & 0 & 0 & 1 & 1 \\
    0 & 4 & 8 & 0 & 8 & 0 & 7 & 23 \\
    0 & 0 & 0 & 1 & 0 & 0 & 0 & 1 \\
    0 & 3 & 2 & 1 & 1 & 0 & 0 & 7 \\
    0 & 5 & 3 & 0 & 3 & 0 & 0 & 13
    \end{matrix}
\end{aligned}\addtocounter{equation}{1}\tag{\theequation}
\end{equation*}
where each row $A_i$ corresponds to an inequality $A_i\cdot \boldsymbol{M}>0$.

Operators mixing the different types of fermions together can also be considered, but these will be deferred to a more systematic analysis for now. The cases just described are especially simple because the UV cone is polyhedral. It is in this sense that the bounds derived here are well-described as generalisations of ``positivity'' constraints - they correspond to identifying the positive combinations of partial amplitudes implied by the optical theorem, which geometrises into the problem of finding the cone generated by a finite set of ERs. Examples of non-polyhedral cones, in which a section of the cone is described by smooth curved surface, will be discussed below. The parameters of the section correspond to the possible $S$-matrix transitions between distinct states that are permitted by the symmetries. No such transitions exist for the simple cases just discussed, but introducing new particles will usually spoil this.

To close this section, I emphasise that, for the restricted theories of a single species of fermion in the representations considered here, the listed constraints are complete. There are no further implications of the sum rule as stated in (\ref{SumRule}) for the structure of the EFT.

\subsection{Flavour violation}\label{sec:Flavour}
The inelastic bounds can also be adapted to multiple flavours of particles unrelated by symmetries, of particular relevance to the SM. These were discussed in \cite{Remmen:2020vts}, where it was observed that flavour-violating fermion operators were bounded above by the flavour-conserving ones. I here show that this is a consequence of general inelastic unitarity bounds and derive general statements for the simplest cases.

Begin with right-handed leptons, the simplest states without non-Abelian symmetries, and allow for any number of flavours. As explained in Section \ref{sec:SMFermions}, both angular momentum and hypercharge conservation ensure that, even with multiple flavours, amplitudes of the form $e_{Ri},e_{Rj}\rightarrow e_{Rk},e_{Rl}$ contain all possibly independent interactions (other processes being related by $CPT$ or crossing). Labelling the UV coupling vectors as $\mathcal{M}^{ij\rightarrow X}=\boldsymbol{\alpha}^{ij}$ and $\mathcal{M}^{i\bar{j}\rightarrow X}=\boldsymbol{\beta}^{ij}$, then all elastic and inelastic amplitudes have the respective forms 
\begin{align}
    M^{ijij}=|\boldsymbol{\alpha}^{ij}|^2+|\boldsymbol{\beta}^{ij}|^2\\
    M^{ijkl}=\boldsymbol{\alpha}^{kl}\cdot\boldsymbol{\alpha}^{ij}+\boldsymbol{\beta}^{kl}\cdot\boldsymbol{\beta}^{il}.
\end{align}
This implies the existence of general bounds on flavour-violating transitions in any particular flavour basis
\begin{align}\label{FlavInelBound}
    |M^{ijkl}|+|M^{ilkj}|\leq\sqrt{M^{ijij}M^{klkl}}+\sqrt{M^{ilil}M^{kjkj}}.
\end{align}
This result similarly holds for elastic scattering of right-handed leptons off any other species of fermion in the SM, as well as left-handed leptons off right-handed quarks, where the decompositions of the non-Abelian symmetries are trivial. 

However, stronger statements can be made when the particles are both right-handed leptons. This is because identical-particle exchange symmetries requires that the amplitudes be symmetric under exchange of the flavour labels of the incoming or outgoing electrons (the kinematical factors already being antisymmetric). The amplitudes must obey the relations $M^{ijkl}=M^{jikl}=M^{ijlk}$. For example, restricting to two flavours, as discussed in \cite{Remmen:2020vts}, this implies that the bounds can be simplified to the set
\begin{align}
   |M^{1122}|\leq\sqrt{M^{1111}M^{2222}}\\
   |M^{1112}|\leq\sqrt{M^{1111}M^{1212}}\\
   |M^{2221}|\leq\sqrt{M^{1212}M^{2222}}.
\end{align}
All other amplitudes are related by exchange symmetries or the usual ones explained in Section \ref{sec:SumRule}. With more flavours, the exchange symmetries will also lead to simpler bounds than (\ref{FlavInelBound}). In any case however, the bounds (\ref{FlavInelBound}) are generally necessary and not sufficient. A more thorough exploration of the flavoured bounds, including internal symmetries (that will be incorporated to produce the analogous results to (\ref{FlavInelBound}) next) will be left to further work. 

With left-handed leptons instead, the amplitudes must be decomposed into $SU(2)$ irreps. The partial amplitudes for the process $L_{Li},L_{Lj}\rightarrow L_{Lk},L_{Ll}$ are, using the projectors from Section \ref{sec:GroupTheory},
\begin{align}
    M_{\mathbf{1}}^{ijkl}=\boldsymbol{m}_{\mathbf{1}}^{kl}\cdot\boldsymbol{m}^{ij}_{\mathbf{1}}-\frac{1}{2}\boldsymbol{m}_{\mathbf{1u}}^{kj}\cdot\boldsymbol{m}_{\mathbf{1u}}^{il}+\frac{3}{2}\boldsymbol{m}_{\mathbf{3u}}^{kj}\cdot\boldsymbol{m}_{\mathbf{3u}}^{il}\\
    M_{\mathbf{3}}^{ijkl}=\boldsymbol{m}_{\mathbf{3}}^{kl}\cdot\boldsymbol{m}_{\mathbf{3}}^{ij}+\frac{1}{2}\boldsymbol{m}_{\mathbf{1u}}^{kj}\cdot\boldsymbol{m}_{\mathbf{1u}}^{il}+\frac{1}{2}\boldsymbol{m}_{\mathbf{3u}}^{kj}\cdot\boldsymbol{m}_{\mathbf{3u}}^{il}.
\end{align}
Here, the subscript $u$ has been introduced to distinguish the UV coupling vectors in the $u$-channel term from the $s$-channel term. The bounds become
\begin{align}
    |M_{\mathbf{3}}^{ijkl}|+|M_{\mathbf{3}}^{ilkj}|&\leq \sqrt{M^{ijij}_{\mathbf{3}}M^{klkl}_{\mathbf{3}}}+\sqrt{M^{ilil}_{\mathbf{3}}M^{kjkj}_{\mathbf{3}}}\\
    |M_{\mathbf{3}}^{ijkl}+\frac{1}{3}M_{\mathbf{1}}^{ijkl}|+|M_{\mathbf{3}}^{ilkj}+\frac{1}{3}M_{\mathbf{1}}^{ilkj}|&\leq \sqrt{\left(M_{\mathbf{3}}^{ijij}+M_{\mathbf{1}}^{ijij}\right)\left(M_{\mathbf{3}}^{klkl}+M_{\mathbf{1}}^{klkl}\right)}\nonumber\\
    &+\sqrt{\left(M_{\mathbf{3}}^{ilil}+M_{\mathbf{1}}^{ilil}\right)\left(M_{\mathbf{3}}^{kjkj}+M_{\mathbf{1}}^{kjkj}\right)}.
\end{align}
The factors in the square roots on right-hand sides of these inequalities are positive, by analogous derivations to the $SU(3)$ case given at the beginning of Section \ref{sec:SMFermions}. Because they have the same non-trivial non-Abelian symmetry structure, bounds for elastic scattering of left-handed leptons off left-handed quarks have the same form.

For right-handed quarks, analogous results can be derived but with $SU(3)$ partial amplitudes instead of $SU(2)$. In this case, the partial amplitudes have the form:
\begin{align}
    M_{\mathbf{6}}^{ijkl}=\boldsymbol{m}_{\mathbf{6}}^{kl}\cdot\boldsymbol{m}_{\mathbf{6}}^{ij}+\frac{1}{3}\boldsymbol{m}_{\mathbf{1}}^{kj}\cdot\boldsymbol{m}_{\mathbf{1}}^{il}+\frac{2}{3}\boldsymbol{m}_{\mathbf{8}}^{kj}\cdot\boldsymbol{m}_{\mathbf{8}}^{il}\\
    M_{\overline{\mathbf{3}}}^{ijkl}=\boldsymbol{m}_{\overline{\mathbf{3}}}^{kl}\cdot\boldsymbol{m}_{\overline{\mathbf{3}}}^{ij}-\frac{1}{3}\boldsymbol{m}_{\mathbf{1}}^{kj}\cdot\boldsymbol{m}_{\mathbf{1}}^{il}+\frac{4}{3}\boldsymbol{m}_{\mathbf{8}}^{kj}\cdot\boldsymbol{m}_{\mathbf{8}}^{il}.
\end{align}
The bounds are 
\begin{align}
    |M_{\mathbf{6}}^{ijkl}|+|M_{\mathbf{6}}^{ilkj}|&\leq\sqrt{M^{ijij}_{\mathbf{6}}M^{klkl}_{\mathbf{6}}}+\sqrt{M^{ilil}_{\mathbf{6}}M^{kjkj}_{\mathbf{6}}}\\
    |M_{\mathbf{6}}^{ijkl}+M_{\overline{\mathbf{3}}}^{ijkl}|+|M_{\mathbf{6}}^{ilkj}+M_{\overline{\mathbf{3}}}^{ilkj}|&\leq \sqrt{\left(M_{\mathbf{6}}^{ijij}+M_{\overline{\mathbf{3}}}^{ijij}\right)\left(M_{\mathbf{6}}^{klkl}+M_{\overline{\mathbf{3}}}^{klkl}\right)}\nonumber\\
    &+\sqrt{\left(M_{\mathbf{6}}^{ilil}+M_{\overline{\mathbf{3}}}^{ilil}\right)\left(M_{\mathbf{6}}^{kj}+M_{\overline{\mathbf{3}}}^{kjkj}\right)}.
\end{align}
This applies regardless of which species of right-handed quarks are identified with the pairs $i,k$ and $j,l$ (all that is important is that the amplitudes are elastic). Similarly, these results also apply to right-handed quarks scattering off left-handed quarks. 

The bounds for left-handed quarks are more intricate because the convex cone describing the purely elastic, flavour-conserving amplitudes has a non-trivial (polyhedral) shape (in other words, more extremal rays than dimension). This issue also arises and is a general problem when there are multiple degenerate irreps of states in non-trivial symmetry representations. This was discussed in \cite{Yamashita:2020gtt} in the case of the hypercharge boson coupling to $W$ bosons with parity symmetry respected, where restriction to the latter of is described by a such a non-trivial cone. This will be elaborated upon further below, but will here be left as an open problem. It is nevertheless clear that (\ref{InelBound}) can be directly applied to provide necessary upper bounds. 

The general pattern described in the examples here is clear and would also apply to flavour-changing processes in which a fermion scatters off a boson. There are, of course, numerous other inelastic processes that can involve flavour violation that would likewise be bounded in more complicated ways (just as the underlying processes with flavour ignored). It should be again emphasised that (\ref{FlavInelBound}) is, by itself, also not complete, and further constraints on the general three-flavour systems remain to be precisely determined.

\section{Bounds with Helicity}\label{sec:SpinBounds}
In this section, the residual rotational invariance about the beam axis will be treated as a global symmetry in a similar way to the internal symmetries described above. This will allow bounds to be placed on theories with spinning particles in which there is a transfer of angular momentum.

\subsection{Rotational symmetry}\label{sec:RotSym}
In the limit of exactly forward scattering, the rotational symmetry about the beam axis is preserved. This is an additional symmetry that can be managed just as for the internal symmetries discussed above. In the center-of-mass frame, call the direction of particle $1$ the $z$-direction, with respect to which all spin projections will be quantised. It is however natural to label the external states by helicity, or stated equivalently, by their little group symmetry of rotations about their momenta. There are two equivalent options for describing this: states of definite helicity (as was done in Section \ref{sec:SMFermions}) and $SO(2)$ vectors, which were employed in \cite{Zhang:2020jyn}, \cite{Yamashita:2020gtt}. 
Helicity eigenstates will be predominantly used in the following examples, in which case the angular momentum along the beam axis is treated as a $U(1)$ charge in a similar way to the fermion and scalar examples previously. However, as part of the simple illustrative example below in Section \ref{sec:idparticles} of identical spinning particles, I will compare this with an analysis of the sum rule in $SO(2)$ form. This subsection will be devoted to deriving the relevant projectors and crossing relations required specifically for this case and addressing issues related to parity-violation. The more general results necessary for implementing the rotational symmetry in $SO(2)$ form will be given in the Appendix. The sum rules for the examples in Sections \ref{sec:FermionBounds} and \ref{Sec:MultiSpin} will also be given there in $SO(2)$ form for comparison. 

Call $h_i$ the magnitude of the helicity of particle $i$. The polarisation of particle $i$ may be represented equivalently by $\mathbf{2}_{\pm h_i}$ vectors, where a $\mathbf{2}_{z}$ vector responds to a spatial rotation of angle $\phi$ about the beam axis by a rotation by angle $z\phi$. Because the helicity quantisation axis of each particle is opposite, the rotational symmetry should act oppositely on each particle's little group indices, so it is natural to represent the polarisation of particle  $1$ by a $\mathbf{2}_{h_1}$ vector and particle $2$ by a $\mathbf{2}_{-h_2}$ vector (and likewise for the outgoing states). The general relationship between the states in tensors of this form and helicity eigenstates $\{|h\rangle,|-h\rangle\}$ is 
\begin{align}
    \mathbf{2}_{\pm h}\sim\frac{1}{\sqrt{2}}\begin{bmatrix}
    |h\rangle+|-h\rangle \\
    \mp i\left(|h\rangle-|-h\rangle\right)
    \end{bmatrix}.
\end{align}

While there are two components to a $SO(2)$ vector, the Wigner-Eckart theorem does not require transitions between the two component states to be related by a symmetry transformation. Instead, only the $U(1)$ charges (in this case, spin projection) need be conserved. That the positive and negative charge irreps can have different partial amplitudes is an expression of the possibility of charge conjugation or parity violation. Separate projectors $P_\pm$ are therefore needed for each distinct charge configuration. It is possible to define projectors $P_{P}=P_++P_-$ and $P_{\cancel{P}}=P_+-P_-$ corresponding to $P$ symmetric and violating transitions (these are normalised so that $(P_{P})^*P_{P}=(P_{\cancel{P}})^*P_{\cancel{P}}=2$). In many simple examples, such as those already discussed in Section \ref{sec:SMFermions}, $CPT$ is sufficient to accidentally rule-out $(C)P$-violating transitions. 

Now, consider the product of two states of helicities $h_1$ and $h_2$. The Clebsch-Gordan coefficients are 
\begin{align}\label{CGs}
    C^{ij}_{h_1+h_2}=\frac{1}{2}P^{ij}-\frac{i}{2}S^{ij}\qquad
    C^{ij}_{h_1-h_2}=\frac{1}{2}\delta^{ij}+\frac{i}{2}\epsilon^{ij}
\end{align}
and $C^{ij}_{-\left(h_1\pm h_2\right)}=\left(C^{ij}_{h_1\pm h_2}\right)^*$. The subscripts here denote $J_z$ eigenstate, while the superscript indices are $SO(2)$ $\mathbf{2}_{h_1}$ and $\mathbf{2}_{-h_2}$ components respectively. The symbol $P^{ij}$ is defined as having values $P^{11}=-P^{22}=1$ and $P^{12}=P^{21}=0$, while the symbol $S^{ij}$ is defined as having components $S^{12}=S^{21}=1$ and $S^{11}=S^{22}=0$. Note that here and throughout, as the four particle's little group indices are in altogether different representations, use of $\delta$ and $\epsilon$ is purely symbolic - these are not to be interpreted as invariant tensors. The helicities identified with the $i$ and $j$ indices are also necessary to uniquely specify the Clebsch-Gordan coefficients, but have been omitted from the notation here to avoid clutter, although there are several examples in the appendix where they must be kept track of.

For the special, yet prevalent case in which the particles have equal (non-zero) helicity $h_1=h_2=h$, the $\mathbf{2}_{h_1-h_2}$ representation instead decomposes into two degenerate singlets. A basis for these will be chosen here to be labelled $A$ and $B$, where 
\begin{align}
|0\rangle_A&=\frac{1}{\sqrt{2}}\left(|h\rangle|h\rangle+|-h\rangle|-h\rangle\right)\nonumber\\
|0\rangle_B&=\frac{1}{\sqrt{2}}\left(|h\rangle|h\rangle-|-h\rangle|-h\rangle\right).
\end{align}
These correspond to the standard $SO(2)$ components of the $\mathbf{2}_{h_1-h_2}$ vectors above. Note that the helicity labels denote spin numbers along opposite quantisation axes. Likewise, these states have Clebsch-Gordan coefficients
\begin{align}
    C^{ij}_{A}=\frac{1}{\sqrt{2}}\delta^{ij}\qquad
    C^{ij}_{B}=-\frac{i}{\sqrt{2}}\epsilon^{ij}.
\end{align}

Note that implicit in this discussion has been a particular phase convention in which eigenstates of helicity of particle $2$ are directly equated with eigenstates of $J_z$, the rotation generator about the beam axis. In the present context, this has the further simplifying implication that the polarisations can be all chosen to be real (or, more precisely, their spinorial representations) and that the action of $P$ on the states does not produce a momentum-dependent phase. See Appendices C and I of \cite{Dreiner:2008tw} for more details, as well as \cite{Weinberg:1964ev}. 

For identical particles, states $A$ and $B$ are therefore $P$ eigenstates with opposite eigenvalues, once (anti-)symmetrisation is accounted for. Because fermion pairs have an intrinsic odd P phase, state $A$ is $P$ even for both bosons and fermions and state $B$ is $P$ odd. Transitions between these states in the $S$-matrix are prohibited if $P$ is conserved.

The next step is to find the projectors into which each spinning amplitude decomposes. When $h_1=h_2=h_3=h_4=h\neq 0$, the projectors for $m_z=\pm 2h$ are
\begin{align}
    P_{\cancel{P}\,2h}^{ijkl}&=\frac{i}{2}\left(\delta^{ik}\epsilon^{jl}+\delta^{jl}\epsilon^{ik}\right)\\
    P_{P\,2h}^{ijkl}&=\frac{1}{2}\left(\delta^{ik}\delta^{jl}+\delta^{il}\delta^{jk}-\delta^{ij}\delta^{kl}\right),
\end{align}
while for singlet states they are
\begin{align}
    P_{AA}^{ijkl}&=\frac{1}{2}\delta^{ij}\delta^{kl}\\
    P_{BB}^{ijkl}&=\frac{1}{2}\left(\delta^{ik}\delta^{jl}-\delta^{il}\delta^{jk}\right)\\
    P_{AB}^{ijkl}&=\frac{i}{2}\delta^{ij}\epsilon^{kl}\\
    P_{BA}^{ijkl}&=\frac{-i}{2}\epsilon^{ij}\delta^{kl}.
\end{align}
The projectors in the $u$-channel may be decomposed as
\begin{align}
    P_{P2h}^{ilkj}&=P_{AA}^{ijkl}+P_{BB}^{ijkl}\\
    P_{AA}^{ilkj}&=\frac{1}{2}\left( P_{P2h}^{ijkl}+P_{AA}^{ijkl}-P_{BB}^{ijkl}\right)\\
    P_{BB}^{ilkj}&=\frac{1}{2}\left( P_{P2h}^{ijkl}-P_{AA}^{ijkl}+P_{BB}^{ijkl}\right)\\
    P_{\cancel{P}2h}^{ilkj}&=-P_{AB}^{ijkl}-P_{BA}^{ijkl}\\
    P_{AB}^{ilkj}&=\frac{1}{2}\left(P_{AB}^{ijkl}-P_{BA}^{ijkl}-P_{\cancel{P}2h}^{ijkl}\right)\label{PhotonCross1}\\
    P_{BA}^{ilkj}&=\frac{1}{2}\left(-P_{AB}^{ijkl}+P_{BA}^{ijkl}-P_{\cancel{P}2h}^{ijkl}\right)\label{PhotonCross2}.
\end{align}

\subsection{Simple example: identical spinning particles}\label{sec:idparticles}
The complete set of dimension $8$ positivity theorems for pure photon operators can be easily derived from requiring positivity of forward scattering of linearly polarised photons, with polarisations inclined by some relative angle tuned to give an optimal constraint that is a function of the Wilson-coefficients \cite{Remmen:2019cyz}. Here, I will present an alternative derivation directly from inspection of the sum rule. This is of particular educational value, as it provides a simple illustrates of several intricacies that can arise in the organisation of the symmetry structure of the sum rule. A discussion of these issues will also be both useful and necessary for further applications to theories of spinning particles. I will then present another derivation of the same results using the convex cone picture. While more complicated, this will again provide a simple archetypal example of a non-polyhedral cone. The arguments presented here applies generally for interactions of four identical particles of any non-zero helicity. While dimension $8$ level is assumed here (as everywhere else), a near identical argument applies to any mass dimension $4n$ as well, which would be applicable for analogous results for gravitons assuming that the results of Section \ref{sec:SumRule} continue to be valid. 

Most of the argument is already complete given the projectors above. The four-particle amplitude can be decomposed into terms given by the projectors. The helicity violating partial amplitude is, expressed in terms of amplitudes between helicity eigenstates,  $A_{\cancel{P}2h}=\frac{1}{2}(P^{ijkl}_{\cancel{P}2h})^*A^{ijkl}=\frac{1}{2}\left(A^{+-+-}-A^{-+-+}\right)$. In this form, it is clear that $CPT$ implies that this vanishes, as $CPT$ equates the two forward helicity amplitudes (as commented above, possible phases that may arise away from the forward limit are conventional and can be eliminated). The helicity-conserving amplitude is therefore time-reversal and parity-conserving. However, the crossing relations (\ref{PhotonCross1}) and (\ref{PhotonCross2}) would appear to generate this partial amplitude from crossing $A_{AB}$ and $A_{BA}$. This is avoided if the singlet partial amplitudes obey
\begin{align}\label{DegenCross}
    A_{BA}(s)=-A_{AB}(s)=(A_{AB}(s))^*,
\end{align}
where the second equality uses (\ref{reality}) and assumes that $s$ is real (that the amplitude is evaluated away from the threshold singularities). In other words, $A_{AB}$, and hence $M_{AB}$, is purely imaginary. This constraint also saves the parity-violating singlet amplitudes from the Wigner-Eckart theorem. Without $P^{ijkl}_{\cancel{P}2h}$, the tensors $P_{AB}^{ilkj}$ and $P_{BA}^{ilkj}$ cannot be individually decomposed into the $s$-channel projectors. However, 
\begin{align}
    P_{AB}^{ilkj}-P_{BA}^{ilkj}=P_{AB}^{ijkl}-P_{BA}^{ijkl}, 
\end{align}
as a result of the identity (\ref{2dId}). The relation (\ref{DegenCross}) thus ensures that it is only this crossing-consistent combination of parity-violating projectors that appear in the amplitude.

The $s$-channel term in the sum rule is
\begin{align}
    M^{ijkl}_s=|\boldsymbol{m}_2|^2P_{2}^{ijkl}+|\boldsymbol{m}_A|^2P_{AA}^{ijkl}+|\boldsymbol{m}_B|^2P_{BB}^{ijkl}+\boldsymbol{m}_A\cdot\boldsymbol{m}_BP_{AB}^{ijkl}+\boldsymbol{m}_B\cdot\boldsymbol{m}_AP_{BA}^{ijkl}.
\end{align}
Adding the $u$-channel crossed term gives 
\begin{align}\label{PhotonSum}
    M^{ijkl}=\left(|\boldsymbol{m}_2|^2+\frac{1}{2}\left(|\boldsymbol{m}_A|^2+|\boldsymbol{m}_C|^2\right)\right)P_{2}^{ijkl}+\frac{1}{2}\left(2|\boldsymbol{m}_2|^2+3|\boldsymbol{m}_A|^2-|\boldsymbol{m}_B|^2\right)P_{AA}^{ijkl}\nonumber\\
    +\frac{1}{2}\left(2|\boldsymbol{m}_2|^2-|\boldsymbol{m}_A|^2+3|\boldsymbol{m}_B|^2\right)P_{BB}^{ijkl}+2\boldsymbol{m}_A\cdot\boldsymbol{m}_B\left(P_{AB}^{ijkl}-P_{BA}^{ijkl}\right),
\end{align}
where the crossing relation (\ref{DegenCross}) implies that $\boldsymbol{m}_A\cdot\boldsymbol{m}_B$ is imaginary. This is to be expected because this transition amplitude is $P$-violating. The term parameterised by the vector $\boldsymbol{\alpha}$ is redundant, so can be ignored. 

Clearly $M_{2}\geq 0$ and implies that the coefficient of the helicity-conserving amplitudes is positive. A second inequality can then be determined by finding a relation between partial amplitudes in (\ref{PhotonSum}). The optimal constraint follows from $|\boldsymbol{m}_A\cdot\boldsymbol{m}_B|\leq |\boldsymbol{m}_A||\boldsymbol{m}_B|$ which, on the LHS, implies that 
\begin{align}\label{ConvexPhotonBound}
    |M_{AB}|\leq \frac{1}{2}\sqrt{16M_{2}^2-|M_{AA}-M_{BB}|^2},
\end{align}
where $M_{R}$ denotes twice differentiated forward partial amplitude corresponding to representation $R$. This represents an upper bound on the size of the $P$-violating transitions determined from the $P$-conserving ones.

The partial amplitudes can be converted into amplitudes between helicity eigenstates. These are
\begin{align}
    M_2&=M(+,-\rightarrow +,-)\label{HelDecomp1}\\
    M_{AA,BB}&=M(+,+\rightarrow +,+)\pm\frac{1}{2}\left((M(+,+\rightarrow -,-)+M(-,-\rightarrow +,+))\right)\label{HelDecomp2}\\
    M_{AB}&=-\frac{1}{2}\left(M(+,+\rightarrow -,-)-M(-,-\rightarrow +,+)\right)\label{HelDecomp3},
\end{align}
where the $(+)$ corresponds to $AA$ and $(-)$ to $BB$. $CPT$ has been invoked for simplification. After some rearrangement, the constraints can be re-expressed as 
\begin{align}\label{photonboundhel}
    M^{+-+-}>\frac{1}{4}\sqrt{|M^{++--}-M^{--++}|^2+|M^{++--}+M^{--++}|^2}.
\end{align}
As mentioned in Section \ref{sec:RotSym}, with the phase conventions chosen here, $A(+,+\rightarrow -,-)=A(-,-\rightarrow +,+)$ if $P$ is conserved. The $M_{AB}$ partial amplitudes (in the context of $W$-bosons) were not included in \cite{Yamashita:2020gtt}, where $P$-symmetry was assumed. In this case, the weaker bound on the $P$-conserving helicity violating interactions alone can be derived directly from positivity of the combinations of diagonal partial amplitudes $\frac{1}{2}\left(M_{AA}+M_{BB}\right)\pm\frac{1}{4}\left(M_{AA}-M_{BB}\right)$.

It is possible to perform the above analysis more directly with helicity eigenstates instead so that both helicity-violating interactions are treated symmetrically, as would be expected from the structure of the bound. In this case, the sum rule would have the same structure as the complex scalar example in Section \ref{sec:Scalars}, but with the single charge-violating amplitudes prohibited. This is given in Table \ref{Tab:Photons}. The helicity-violating amplitudes are immediately manifest and the bound in the form of (\ref{photonboundhel}) follows directly from bounding the off-diagonal entry, the real and imaginary parts of which correspond to $P$-conserving and violating interactions respectively. 

\begin{table}
\centering
\begin{tabular}{ |p{1cm}||p{2.9cm}|p{2.9cm}|p{2.9cm}|p{2.9cm}| }
 \hline
  \, & $+-$ & $++$ & $--$ & $-+$\\
 \hline
 \hline
 $+-$   & $|\boldsymbol{m}^{+-}|^2+|\boldsymbol{m}^{++}|^2$    & $0$ & $0$  & $0$\\
 $++$ & $0$ &  $|\boldsymbol{m}^{++}|^2+|\boldsymbol{m}^{+-}|^2$  & $2\boldsymbol{m}^{--}\cdot\boldsymbol{m}^{++}$  & $0$\\
 $--$ & $0$ & $2\boldsymbol{m}^{++}\cdot\boldsymbol{m}^{--}$ &  $|\boldsymbol{m}^{--}|^2+|\boldsymbol{m}^{+-}|^2$ & $0$\\
 $-+$    & $0$ & $0$ & $0$ & $|\boldsymbol{m}^{+-}|^2+|\boldsymbol{m}^{--}|^2$\\
 \hline
\end{tabular}
\caption{\label{Tab:Photons} Sum rule for photons.}
\end{table}

Specialising now to photons, the general effective action up to dim-$8$ is
\begin{align}\label{EHL}
    \mathcal{L}_{EFT_8}=\frac{c}{16\Lambda^4}\left(\left(F^2\right)^2+(F\tilde{F})^2\right)+\frac{d}{32\Lambda^4}\left(\left(F^2\right)^2-(F\tilde{F})^2\right)+\frac{e}{16\Lambda^4}F^2\left(F\tilde{F}\right).
\end{align}
The operator basis has been selected to match onto specific tree-level $4$-leg amplitudes between helicity eigenstates. The first operator is helicity preserving, the others are helicity-violating, with the coefficient $e$ being $P$ and $CP$ violating and providing the imaginary part of the coupling in the corresponding amplitudes. Evaluating the LHS entries at tree-level, the constraints reduce to \begin{align}
    c>\frac{1}{2}\sqrt{d^2+e^2}.
\end{align} 

However, loops of scalar particles mediated by dim-$6$ operators of the form $F^2\phi^2$ also contribute to the dim-$8$ order four photon amplitudes. If the scalar is complex, the possible operators are 
\begin{align}
    \mathcal{L}_{EFT_6}&=\frac{a}{\Lambda^2}F^2\phi^2+\frac{b}{\Lambda^2}F\tilde{F}\phi^2+\text{conj}\nonumber\\
    &\quad+\frac{\tilde{a}}{\Lambda^2}F^2\phi\phi^\dagger+\frac{\tilde{b}}{\Lambda^2}F\tilde{F}\phi\phi^\dagger.
\end{align}
where $a,b\in\mathbb{C}$ and $\tilde{a},\tilde{b}\in\mathbb{R}$. In a supersymmetric theory, discussed more below, $\tilde{a}$ and $\tilde{b}$ are prohibited if the scalars are the same, while $b=ia$. On-shell, this is the statement that the only permitted contact interactions induce amplitudes $\mathcal{A}(\gamma^+,\gamma^+,\overline{\phi},\overline{\phi})$ and $\mathcal{A}(\gamma^-,\gamma^-,\phi,\phi)$. In this case, helicity charge of the photons may be extended into a conserved charge also carried by the scalars, which is a statement of electric-magnetic duality \cite{Gaillard:1981rj}. Alternatively, $b=-ia$ is also compatible if the identification of particle and antiparticle is reversed. Including these contributions to the amplitudes, the constraints become:
\begin{align}
    &2c+\frac{8}{(4\pi)^2}\left(\tilde{a}^2+\tilde{b}^2+2|a-ib|^2+2|a+ib|^2\right)\left(2+\log\left(\frac{\mu^2}{\Lambda^2}\right)\right)\nonumber\\
    &\qquad>\left|d+ie+\frac{8}{(4\pi)^2}\left((\tilde{a}-i\tilde{b})^2+4(a-ib)(a^*-ib^*)\right)\left(2+\log\left(\frac{\mu^2}{\Lambda^2}\right)\right)\right|.
\end{align}
Just as in the scalar example discussed in Section \ref{sec:Scalars}, the dim-$6$ operators strengthen the lower bound on $c$ (absorbing into it the rational part of the loop correction), at least assuming that there is little change in the size of the expression on the RHS of the inequality. 

It is possibly enlightening to consider the effect of each term in isolation, with all others set to zero. In order to maintain consistency with positivity, $c$ will remain active so that the dim-$6$ corrections can be consistently negative, while all $d$ and $e$ will be chosen to cancel the rational terms generated from the dim-$6$ operators. Activating only $\tilde{a}$ and $\tilde{b}$, then both terms on the left and right side of the inequality are equal in magnitude. If both terms were positive, then these would simply account for each other on each side of the sum rule and saturate it to give no information. However, as the logarithm is negative, these terms actually reinforce each other and strengthen the lower bound on $c$ beyond $0$, the degree to which depending on the size of the hierarchy between IR and UV scales. Next, if only $a$ and $b$ are active, then the terms on the LHS of the inequality are always greater in magnitude than those on the RHS although, as before, both strengthen the tree-level bound. Interestingly, for a give coupling $a$, the weakest contribution to the constraint is made for the symmetry-enhanced choice $b=\pm ia$. This is because the on-shell amplitudes into which the cut scalar loop factorise would be prohibited for the helicity-violating configurations, as well as for one of the two possible contributions to the helicity-conserving case. This is typical of the suppression of RG evolution caused by the enhancement of symmetries. 

There are also other potential contributions to the dim-$8$ order four photon amplitudes, such as from a fermion box of three-particle dim-$5$ operators, that would appear on the LHS. A more thorough analysis of the way that lower-dim operators affect the constraints will be left for another work.

The bound (\ref{ConvexPhotonBound}) is therefore the statement that the helicity-violating amplitudes must be smaller than the helicity-conserving ones. For photons, this is a leading-order statement of the hierarchy in coupling strengths of symmetry preserving and symmetry violating interactions. Helicity conservation corresponds to electric-magnetic duality and is also selected by supersymmetry, as will be explored further below. This point of view also ``explains'' the observation of \cite{Remmen:2019cyz} of the consequential suppression of the $P$ and $T$ violation in the vector boson EFT. These discrete symmetries can only be violated by operators that mediate helicity-violating interactions, or, more generally, off-diagonal, inelastic $S$-matrix entries. Because the sum rule curbs the size of these interactions, it consequently also places fundamental restrictions on the size of $T$-violation. Note that, while it is possible to perform a field redefinition (or ``duality rotation'') in the effective action (\ref{EHL}) to remove the $T$-violating term (and transfer it into the coupling of the photon to sources), the combination $d^2+e^2$ remains invariant and the constraint is unchanged. This is the reason that the two couplings must necessarily appear added in quadrature. On-shell, this is reflected in the overall phase ambiguity of the amplitudes. This similarly applies to the complex $a$ coefficient in the case that $b=\pm ia$ and $\tilde{a}=\tilde{b}=0$.


The bound (\ref{ConvexPhotonBound}) can be alternatively derived from the convex cone picture. Beginning with the sum rule (\ref{PhotonSum}) (with $\boldsymbol{\alpha}=0$), the first step is to find the PERs. These may be determined as the independent contributions with the factorised form $\boldsymbol{m}\cdot\boldsymbol{m}$ in the $s$-channel before the crossed $u$-channel is added. Choose as independent coordinates the partial amplitudes $\boldsymbol{x}=(M_{AA},M_{BB},\Im\left(M_{AB}\right))$, where $M_{AB}$ is purely imaginary, so represents only one real dimension. Each PER corresponds to the contribution from a single UV state, so only a single component of the complex vectors in the sum rule need be chosen. These will be labelled as $\beta$ and $\gamma$. Because of the crossed-amplitude, there is only a single ray structure that combines all terms generated by both parameters. As the ray is only defined modulo positive real factors, the coordinates may be rescaled by a factor of $2/|\beta|^2$ to give $\boldsymbol{e}(r)=(3-r^2,-1+3r^2,4r)$, where $r=-i\gamma/\beta\in\mathbb{R}$. It can be verified that this is extremal for all $r$. This family of ERs is effectively a $2d$ surface in a $3d$ space parameterised by $2$ real parameters (including a positive real parameter rescaling the ray). The cone itself has parabolic sections. The face of the cone is defined by the normal $\boldsymbol{n}$ to the surface, which (up to an arbitrary scale) has components $n_i= \epsilon_{ijk}e^j(r)\frac{\partial e^k}{\partial r}=\left(-1-3r^2,-3-r^2,4r\right)$. The EFT must induce amplitudes that lie inside the cone, so this implies that $\boldsymbol{x}\cdot\boldsymbol{n}(r)<0$ for all $r\in\mathbb{R}$. Imposing this latter condition implies that $3M_{AA}+M_{BB}>0$, $M_{AA}+3M_{BB}>0$ and $|M_{AB}|<\frac{1}{2}\sqrt{\left(3M_{AA}+M_{BB}\right)\left(M_{AA}+3M_{BB}\right)}$, which is just a restatement of (\ref{ConvexPhotonBound}) given the relation $M_2=\frac{1}{2}\left(M_{AA}+M_{BB}\right)$.

\subsection{Non-Abelian vector bosons}\label{sec:NAVec}

I here make some comments about extending the above analysis to scattering of (massless) $W$-bosons in which $CP$-violation is permitted.

The $W$-bosons have helicity and adjoint $SU(2)$ indices. The projectors that span the amplitude are given by products of adjoint isospin and helicity projectors, of which there are $15$. Denoting by $\boldsymbol{\alpha}_{I,m}$ the vector of UV couplings in the RHS of the sum rule for isospin irrep $I$ and $J_z$ irrep $m$, then the sum rule can be expressed as
\begin{align}
    M^{ai,bj,ck,dl}=\sum_{I,m,n}\boldsymbol{\alpha}_{I,m}\cdot\boldsymbol{\alpha}_{I,n}\left(P^{abcd}_IP^{ijkl}_{mn}+P^{adcb}_IP^{ilkj}_{mn}\right).
\end{align}
The crossed projectors are then decomposed using the relations stated in the previous sections to derive the partial amplitudes as a function of the complex vectors $\boldsymbol{\alpha}_{I,m}$. Similarly to photons, $CPT$ implies that the helicity-conserving amplitudes are parity symmetric, so the relations (\ref{DegenCross}) hold for each isospin partial amplitude such that $\boldsymbol{\alpha}_{I,A}\cdot\boldsymbol{\alpha}_{I,B}=-\boldsymbol{\alpha}_{I,B}\cdot\boldsymbol{\alpha}_{I,A}$. The $\boldsymbol{\alpha}_{I,m}\cdot\boldsymbol{\alpha}_{I,n}$ coefficients determine the space of partial (forward) amplitudes. 

In the simple examples above, the vectors parameterising the sum rule were loosely in correspondance with the elastic partial amplitudes. An expression of the form $|\boldsymbol{\alpha}|^2$ could be simply translated into a partial amplitude in order to determine the bounds. However, in the $W$ theory (and any other with sufficiently many degrees of freedom), these vectors exceed the number of independent partial amplitudes. If only diagonal $S$-matrix transitions were permitted, this would correspond to a non-trivial polyhedral cone with more facets than dimension. Each point in the cone may admit multiple decompositions into positive sums of extremal rays and different subsets of rays span different regions of the cone, complicating the simple inspection arguments used to derive bounds in the examples above. Geometrically, the problem becomes that of performing vertex enumeration for non-polyhedral cones, or more simply finding the curved facets bounding the cone. Some ideas for addressing this were described in \cite{Yamashita:2020gtt}, but for now this will be left for future work. The solution to this problem represents the next step toward bootstrapping constraints on realistic EFTs such as Standard Model EFT.

\subsection{Two chiral fermions}\label{sec:FermionBounds}
A simple expansion of the previous example in Section \ref{sec:idparticles} is given by introducing a second distinct particle with the same helicity. I will commit to assuming that both particles are chiral fermions, $\psi$ and $\lambda$, because this will be of interest later. 
However, the conclusions are more general. It will be additionally assumed for simplicity that each is charged under its own $\mathbb{Z}_2$ symmetry so that they can only be destroyed or created in pairs.

The elastic amplitudes are affected by the same constraints derived above in Section \ref{sec:idparticles}. The sum rule for these are reproduced here in Table \ref{Tab:2fermions}. Entries below the main diagonal have been omitted as they are simply related by Hermiticity of the matrix. Some of the vectors in the last block are $Y$-rotated versions of counterparts in the third block and satisfy $|\boldsymbol{m}^{\psi^+\lambda^+}|=|\boldsymbol{m}^{\lambda^+\psi^+}|$, $|\boldsymbol{m}^{\psi^+\lambda^-}|=|\boldsymbol{m}^{\lambda^-\psi^+}|$.

\begin{table}
\centering
\begin{adjustwidth}{-1.5cm}{0cm}
\begin{tabular}{ |p{1cm}||p{3.7cm}|p{3.7cm}|p{3.7cm}|p{3.7cm}| }
 \hline
  \, & $\lambda^+\lambda^-$ & $\lambda^+\lambda^+$ & $\lambda^-\lambda^-$ & $\lambda^-\lambda^+$\\
 \hline
 \hline
 $\lambda^+\lambda^-$   & $|\boldsymbol{m}^{\lambda^+\lambda^-}|^2+|\boldsymbol{m}^{\lambda^+\lambda^+}|^2$    & $0$ & $0$  & $0$\\
 $\lambda^+\lambda^+$ & . &  $|\boldsymbol{m}^{\lambda^+\lambda^-}|^2+|\boldsymbol{m}^{\lambda^+\lambda^+}|^2$  & $2\boldsymbol{m}^{\lambda^-\lambda^-}\cdot\boldsymbol{m}^{\lambda^+\lambda^+}$  & $0$\\
 $\lambda^-\lambda^-$ & . & . &  $|\boldsymbol{m}^{\lambda^-\lambda^-}|^2+|\boldsymbol{m}^{\lambda^+\lambda^-}|^2$ & $0$\\
 $\lambda^-\lambda^+$    & . & . & . & $|\boldsymbol{m}^{\lambda^-\lambda^-}|^2+|\boldsymbol{m}^{\lambda^+\lambda^-}|^2$\\
 \hline
\end{tabular}

\begin{tabular}{ |p{1cm}||p{3.7cm}|p{3.7cm}|p{3.7cm}|p{3.7cm}| }
 \hline
  \, & $\psi^+\psi^-$ & $\psi^+\psi^+$ & $\psi^-\psi^-$ & $\psi^-\psi^+$\\
 \hline
 \hline
 $\psi^+\psi^-$   & $|\boldsymbol{m}^{\psi^+\psi^-}|^2+|\boldsymbol{m}^{\psi^+\psi^+}|^2$    & $0$ & $0$  & $0$\\
 $\psi^+\psi^+$ & . &  $|\boldsymbol{m}^{\psi^+\psi^-}|^2+|\boldsymbol{m}^{\psi^+\psi^+}|^2$  & $2\boldsymbol{m}^{\psi^-\psi^-}\cdot\boldsymbol{m}^{\psi^+\psi^+}$  & $0$\\
 $\psi^-\psi^-$ & . & . &  $|\boldsymbol{m}^{\psi^-\psi^-}|^2+|\boldsymbol{m}^{\psi^+\psi^-}|^2$ & $0$\\
 $\psi^-\psi^+$    & . & . & . & $|\boldsymbol{m}^{\psi^-\psi^-}|^2+|\boldsymbol{m}^{\psi^-\psi^-}|^2$\\
 \hline
\end{tabular}

\begin{tabular}{ |p{1cm}||p{3.7cm}|p{3.7cm}|p{3.7cm}|p{3.7cm}| }
 \hline
  \, & $\lambda^+\psi^-$ & $\lambda^+\psi^+$ & $\lambda^-\psi^-$ & $\lambda^-\psi^+$\\
 \hline
 \hline
 $\lambda^+\psi^-$   & $|\boldsymbol{m}^{\lambda^+\psi^-}|^2+|\boldsymbol{m}^{\lambda^+\psi^+}|^2$    & $0$ & $0$  & $0$\\
 $\lambda^+\psi^+$ & . &  $|\boldsymbol{m}^{\lambda^+\psi^-}|^2+|\boldsymbol{m}^{\lambda^+\psi^+}|^2$  & $2\boldsymbol{m}^{\lambda^-\psi^-}\cdot\boldsymbol{m}^{\lambda^+\psi^+}$  & $0$\\
 $\lambda^-\psi^-$ & . & . &  $|\boldsymbol{m}^{\lambda^-\psi^-}|^2+|\boldsymbol{m}^{\lambda^-\psi^+}|^2$ & $0$\\
 $\lambda^-\psi^+$    & . & . & . & $|\boldsymbol{m}^{\lambda^-\psi^-}|^2+|\boldsymbol{m}^{\lambda^-\psi^+}|^2$\\
 \hline
\end{tabular}

\begin{tabular}{ |p{1cm}||p{3.7cm}|p{3.7cm}|p{3.7cm}|p{3.7cm}| }
 \hline
  \, & $\psi^+\psi^-$ & $\psi^+\psi^+$ & $\psi^-\psi^-$ & $\psi^-\psi^+$\\
 \hline
 \hline
 $\lambda^+\lambda^-$   & $\boldsymbol{m}^{\psi^+\psi^-}\cdot\boldsymbol{m}^{\lambda^+\lambda^-}+\boldsymbol{m}^{\psi^+\lambda^+}\cdot\boldsymbol{m}^{\lambda^+\psi^+}$    & $0$ & $0$  & $0$\\
 $\lambda^+\lambda^+$ & $0$ &  $\boldsymbol{m}^{\psi^+\psi^+}\cdot\boldsymbol{m}^{\lambda^+\lambda^+}+\boldsymbol{m}^{\psi^+\lambda^-}\cdot\boldsymbol{m}^{\lambda^+\psi^-}$  & $\boldsymbol{m}^{\psi^-\psi^-}\cdot\boldsymbol{m}^{\lambda^+\lambda^+}+\boldsymbol{m}^{\lambda^-\psi^-}\cdot\boldsymbol{m}^{\psi^+\lambda^+}$  & $0$\\
 $\lambda^-\lambda^-$ & $0$ & $\boldsymbol{m}^{\psi^+\psi^+}\cdot\boldsymbol{m}^{\lambda^-\lambda^-}+\boldsymbol{m}^{\psi^+\lambda^+}\cdot\boldsymbol{m}^{\lambda^-\psi^-}$ &  $\boldsymbol{m}^{\psi^-\psi^-}\cdot\boldsymbol{m}^{\lambda^-\lambda^-}+\boldsymbol{m}^{\lambda^+\psi^-}\cdot\boldsymbol{m}^{\psi^+\lambda^-}$ & $0$\\
 $\lambda^-\lambda^+$  & $0$ & $0$ & $0$ & $\boldsymbol{m}^{\psi^+\psi^-}\cdot\boldsymbol{m}^{\lambda^+\lambda^-}+\boldsymbol{m}^{\lambda^+\psi^+}\cdot\boldsymbol{m}^{\psi^+\lambda^+}$\\
 \hline
\end{tabular}
\end{adjustwidth}
\caption{\label{Tab:2fermions} Sum rule for two fermion theory.}
\end{table}

Both $CPT$ and $Y$ can be invoked to reduce the number of independent vector magnitudes $|\boldsymbol{m}|$ to six. After applying the Schwarz and triangle inequalities to the off-diagonal amplitudes, these all have upper bounds of the form of either $|\boldsymbol{m}^{\lambda+\lambda-}||\boldsymbol{m}^{\psi+\psi-}|+|\boldsymbol{m}^{\lambda+\psi+}|^2$ or $|\boldsymbol{m}^{\psi^+\psi^+}||\boldsymbol{m}^{\lambda^+\lambda^+}|+|\boldsymbol{m}^{\lambda^+\psi^-}|^2$. Analogous bounds of the form of (\ref{InelBound}) are then given by 
\begin{align}\label{twofermionbounds}
    M_1+M_2\leq \sqrt{M^{\lambda^+\lambda^-\lambda^+\lambda^-}M^{\psi^+\psi^-\psi^+\psi^-}}+M^{\lambda^+\psi^-\lambda^+\psi^-},
\end{align}
where 
\begin{align}
    M_1 &\in\{|M^{\lambda^+\lambda^+\psi^+\psi^+}|,|M^{\lambda^-\lambda^-\psi^-\psi^-}|,|M^{\lambda^+\lambda^+\psi^-\psi^-}|,|M^{\lambda^-\lambda^-\psi^+\psi^+}|\}\\
    M_2 &\in\{|M^{\lambda^+\lambda^-\psi^+\psi^-}|,|M^{\lambda^-\lambda^+\psi^-\psi^+}|\}
\end{align}
are, respectively, any of the $m_z=0$ and $m_z=\pm 1$ amplitudes.

For reference, the general effective action for this theory has terms 
\begin{align}
    \mathcal{L}_{EFT_6}&= \frac{f}{\Lambda^2}\psi\psi\psi^\dagger\psi^\dagger+\frac{g}{\Lambda^2}\lambda\lambda\lambda^\dagger\lambda^\dagger+\frac{h}{\Lambda^2}\psi\lambda\psi^\dagger\lambda^\dagger\nonumber\\
    &\qquad +\frac{k}{\Lambda^2}\psi\psi\lambda^\dagger\lambda^\dagger+\frac{\tilde{k}}{\Lambda^2}\psi\psi\lambda\lambda+\text{conj.}
\end{align}
and 
\begin{align}
    &\mathcal{L}_{EFT_8}=\frac{a}{2\Lambda^4}\psi\psi\partial^2\left(\psi^\dagger\psi^\dagger\right)+\frac{b}{2\Lambda^4}\lambda\lambda\partial^2\left(\lambda^\dagger\lambda^\dagger\right)+\frac{c}{\Lambda^4}\psi\lambda\partial^2\left(\psi^\dagger\lambda^\dagger\right)+\frac{d}{\Lambda^4}\partial\psi\lambda\cdot\partial\psi^\dagger\lambda^\dagger\nonumber\\
    &\,+\frac{\tilde{a}}{2\Lambda^4}\psi\psi\partial^2\left(\psi\psi\right)+\frac{\tilde{b}}{2\Lambda^4}\lambda\lambda\partial^2\left(\lambda\lambda\right)+\frac{\tilde{c}}{\Lambda^4}\psi\lambda\partial^2\left(\psi\lambda\right)+\frac{\tilde{d}}{\Lambda^4}\psi\psi\partial^2\left(\lambda\lambda\right)+\frac{e}{\Lambda^4}\psi\psi\partial^2\left(\lambda^\dagger\lambda^\dagger\right)+\text{conj}.\label{dim8fermions}
\end{align}
The couplings $f,g,h,a,b,c,d\in\mathbb{R}$ and $k,\tilde{k},\tilde{a},\tilde{b},\tilde{c},\tilde{d},e\in\mathbb{C}$. The operators with Wilson coefficients denoted with a tilde mediate helicity-violating interactions. Consistency with supersymmetry will be elaborated on further below (see Section \ref{sec:SUSY}). The bounds on amplitudes are readily converted into bounds on dim-$8$ Wilson coefficients at tree-level after identifying them with the particular transitions between helicity eigenstates and channels.

\subsection{Bounds on simple theories of spinning particles}\label{Sec:MultiSpin}
Similar analyses can be performed on systems with particles of different helicities. Here, simple cases of mixed scalar, fermion and vector amplitudes will be discussed in order to derive new constraints.

To begin with, the system will be restricted to consisting of a real scalar, a photon and a chiral fermion. The full table has dimensions $25\times 25$ when external helicity eigenstates are chosen as a basis, but most of the entries vanish either altogether by conservation of angular momentum or, specifically at dimension $8$, by incompatibility of Lorentz invariance and dimensional analysis (see discussion below in Section \ref{sec:SUSYEFT}). For this reason, only the (few) important entries containing new information will be quoted here. The full sum rule in $SO(2)$ form is given in the Appendix. 

Restricting entirely to the bosons to begin with, the entries relevant for photons are given in Table \ref{Tab:Photons} in the section above, while the others of relevance are:
\begin{align}
    M^{\phi\phi\phi\phi}&=2|\boldsymbol{m}^{\phi\phi}|^2\\
    M^{\gamma^+\phi\gamma^+\phi}&=2|\boldsymbol{m}^{\gamma^+\phi}|^2\\
    M^{\gamma^-\phi\gamma^-\phi}&=2|\boldsymbol{m}^{\gamma^-\phi}|^2\\
    M^{\phi\phi\gamma^+\gamma^+}&=\boldsymbol{m}^{\gamma^+\gamma^+}\cdot\boldsymbol{m}^{\phi\phi}+\boldsymbol{m}^{\gamma^+\phi}\cdot\boldsymbol{m}^{\phi\gamma^-}\\
    M^{\phi\phi\gamma^-\gamma^-}&=\boldsymbol{m}^{\gamma^-\gamma^-}\cdot\boldsymbol{m}^{\phi\phi}+\boldsymbol{m}^{\phi\gamma^-}\cdot\boldsymbol{m}^{\gamma^+\phi}
\end{align}
(and others related by crossing). Simplification with $CPT$ and $Y$ (which also equate $|\boldsymbol{m}^{\gamma^+\phi}|=|\boldsymbol{m}^{\phi\gamma^-}|$ has been invoked. The bounds are identified as:
\begin{align}\label{MixedBosonBound}
    |M^{\phi\phi\gamma^+\gamma^+}|,|M^{\phi\phi\gamma^-\gamma^-}|\leq \frac{1}{2}M^{\phi\gamma^+\phi\gamma^+}+\sqrt{\frac{1}{2}M^{\phi\phi\phi\phi}M^{\gamma^+\gamma^-\gamma^+\gamma^-}}.
\end{align}
Unlike the helicity-violating four-vector operators, these bounds are not accessible by considering elastic forward scattering of a scalar with a linearly polarised vector. Superpositions of vectors and scalars are instead necessary. The bounds are also stronger by various factors of $2$ compared to what would be anticipated from direct application of (\ref{InelBound}). This is because the identity of the scalars is crossing symmetric, which simplifies the sum rule. The analogous bounds with complex scalars, used below in Section \ref{sec:PositivityCombined}, are weaker. 

The other constraint arises for a mixed spin amplitude. The relevant entries are
\begin{align}
    M^{\phi\psi^+\phi\psi^+}&=|\boldsymbol{m}^{\phi\psi^+}|^2+|\boldsymbol{m}^{\phi\psi^-}|^2\\
    M^{\gamma^+\psi^+\gamma^+\psi^+}&=|\boldsymbol{m}^{\gamma^+\psi^+}|^2+|\boldsymbol{m}^{\gamma^-\psi^-}|^2\\
    M^{\phi\psi^-\gamma^+\psi^+}&=2\boldsymbol{m}^{\gamma^+\psi^+}\cdot\boldsymbol{m}^{\phi\psi^-}\\
    M^{\phi\psi^+\gamma^-\psi^-}&=2\boldsymbol{m}^{\gamma^-\psi^-}\cdot\boldsymbol{m}^{\phi\psi^+},
\end{align}
where $|\boldsymbol{m}^{\phi\psi^+}|=|\boldsymbol{m}^{\phi\psi^-}|$. The resulting inelastic constraint is 
\begin{align}
    |M^{\phi\psi^-\gamma^+\psi^+}|,|M^{\phi\psi^+\gamma^-\psi^-}|\leq \sqrt{2M^{\phi\psi^+\phi\psi^+}M^{\psi^+\gamma^+\psi^+\gamma^+}}.
\end{align}

Along with the postivity constraints given in \cite{Bellazzini:2016xrt}, this completes the causality bounds for the simple minimal, toy theories of spinning particles. Similar arguments can be used to adapt these to more complicated theories with more states. This will be partly done in the supersymetric case below.

\section{Supersymmetry}\label{sec:SUSY}
Having addressed the management of spin in the sum rule, it is natural to now extend this to supermultiplets. Supersymmetry unifies states of different spin and likewise their interactions. Supersymmetry is a consistent extension of the spacetime symmetry algebra, so should not affect conclusions drawn from the (rigid) causal structure of background flat Minkowski space. It is therefore expected that causality constraints on scattering of a particular set of component states should be shared by the other interactions related by supersymmetry. Precisely these connections will be explored in this section. For simplicity, attention will be restricted to EFTs with minimal particle content.

\subsection{Superamplitudes}
Superspaces at the level of the effective action are generally arduous and cumbersome to work with. As is very well appreciated, on-shell scattering amplitudes cut-through the off-shell baggage of the effective action, not only making computations substantially easier, but also clarifying the presence and action of symmetries that are either not manifest or are convoluted in the Lagrangian field theory. The ``on-shell superspace'', to be employed here, makes the super-Ward identities (SWIs) manifest as relations between scattering amplitudes - see e.g. \cite{Elvang:2013cua} for review. Amplitudes between individual states in a multiplet are unified into superamplitudes. This makes transparent the relation between the unified effective interactions and their component operators without recourse to an off-shell superspace. The especially simple case of $2\rightarrow 2$ scattering amplitudes, under discussion here, are highly constrained by fundamental principles. See \cite{Elvang:2010jv} for numerous examples in supergravity.

The chiral superspace of \cite{Lal:2009gn}, later used by \cite{Elvang:2011fx}, will be employed here, where the highest helicity state is selected as the Clifford vacuum for the representation. This determines the little group representation of the entire ``superfield''. The massless multiplets for $\mathcal{N}=1$ theories are 
\begin{align}
\begin{split}
    \Phi^+&=\psi^++\eta \overline{\phi}\nonumber\\
    V^+&=v^++\eta\lambda^+
\end{split}
\begin{split}
    \Phi^-&=\phi+\eta\psi^-\nonumber\\
    V^-&=\lambda^-+\eta v^-.
\end{split}
\end{align}
For $\mathcal{N}=2$, they are 
\begin{align}
\begin{split}
    K &= \chi^++\eta^A\phi_A-\frac{1}{2}\epsilon_{AB}\eta^A\eta^B\chi^-\nonumber\\
    V^+&=v^++\eta^A\lambda^+_A-\frac{1}{2}\epsilon_{AB}\eta^A\eta^B\overline{\phi}
\end{split}
\begin{split}
    \overline{K} &=\overline{\chi}^++\eta^A\overline{\phi}_A-\frac{1}{2}\epsilon_{AB}\eta^A\eta^B\overline{\chi}^-\nonumber\\
    V^-&=\phi+\eta^A\lambda^-_A-\frac{1}{2}\epsilon_{AB}\eta^A\eta^Bv^-.
\end{split}
\end{align}
See \cite{Elvang:2011fx} for general explanation of notation. The multiplet $K$ is a half-hypermultiplet and is usually paired with a conjugate multiplet of antiparticles, $\overline{K}$. However, as both multiplets have identical helicity structure, it will not be important here to continue to distinguish between the two. The $\mathcal{N}=4$ vector is defined in \cite{Elvang:2013cua} and will not be reproduced here.

All external states defining the superamplitudes will be taken to be outgoing. This is the convention adopted in \cite{Elvang:2013cua}. However, it will be necessary to cross two of the states to be incoming. Crossing has been discussed in the present context in \cite{Bellazzini:2016xrt} and will be performed here on the component amplitudes. As mentioned in at the end of Section \ref{sec:SumRule}, the ordering of the superfields in the correlator from which the amplitude is derived is $\langle 0|4312|0\rangle$. This gives the order of the states and the Grassmann variables in the superamplitude, which determines the order in which the Grassmann derivatives should be applied to extract the components. Note that the Feynman rules for external antifermion legs include a factor of $-1$ that is frequently dropped \cite{Srednicki:2007qs}, but is required here. This implies that a single fermion leg must be accompanied by a factor of $-1$ when crossed, in addition to the usual rules of reversing the momentum and replacing the external polarisation.

\subsection{Effective operators}\label{sec:SUSYEFT}
It is of general interest to classify the effective contact interactions combined together under various degrees of supersymmetry. Here these will be systematically classified from dimension $5$ to dimension $8$ for operators inducing contact interactions between three or four particles. Again, the discussion will be restricted to helicities $h\leq 1$ (so no (super)gravity). Different species of multiplets with the same superspin will not be distinguished in order to emphasise the purely kinematical structure of the allowed interactions, but no assumptions will be made about permutation symmetries and internal quantum numbers (unrelated to the supersymmetry algebra). 

Most of the interactions considered here will be four-particle contact interactions. These are severely constrained by consistency with dimensional analysis, little group representation, Lorentz invariance, locality and supersymmetry. The last condition is the requirement that the superamplitude depend on the Grassmann variables exactly through $\delta^{(2)}(Q^\dagger)$, while the former conditions demand that the amplitudes be polynomials in spinor bilinears with the required mass dimension and total helicity charge for each leg. 

While supersymmetry unifies interactions, it can also prohibit them. A common reason for this is that the spectrum of effective interactions for higher spin particles is sparser than for lower spin particles, so not all lower spin interactions can be uniquely paired with a higher spin interaction. 

\subsubsection{Dimension $5$}
An anomalous magnetic dipole moment (MDM)-like operator for matter fermions is prohibited by supersymmetry (this is already prohibited by exchange antisymmetry if the fermions are identical). However, the axion/dilaton coupling is promoted to
\begin{align}\label{AxionCoup}
    \mathcal{A}(\Phi^+,V^+,V^+)\propto\frac{1}{\Lambda}\tilde{\delta}^{(1)}(Q)\ds{23},
\end{align}
which also contains a mixed MDM-like interaction between the matter fermion and the gaugino. This interaction can be further promoted to $\mathcal{N}=2$ in the superamplitude
\begin{align}
    \mathcal{A}(V^+,V^+,V^+)\propto\frac{1}{\Lambda}\tilde{\delta}^{(2)}(Q),
\end{align}
which contains no further interactions. There are likewise conjugate superamplitudes between the corresponding anti-multiplets. Notably, the axion/dilaton cannot belong to a hypermultiplet. 

The Weinberg operator (uniquely) supersymmetrises into itself:
\begin{align}
    \mathcal{A}(\Phi^+,\Phi^+,\Phi^+,\Phi^+)\propto \frac{1}{\Lambda}\delta^{(2)}(Q^\dagger)\frac{\ds{12}}{\da{34}}.
\end{align}
The conjugate amplitude  $\mathcal{A}(\Phi^-\Phi^-\Phi^-\Phi^-)$ is similar. Despite appearances, all spinor prefactors are equivalent. The operator is altogether incompatible with $\mathcal{N}\geq 2$.

\subsubsection{Dimension $6$}
Cubic vector interactions are altogether prohibited by supersymmetry, so there are no $3$-particle operators to consider. 

Permissible $\mathcal{N}=1$ superamplitudes are 
\begin{align}
    \mathcal{A}(\Phi^+,\Phi^+,\Phi^-,\Phi^-)&\propto\frac{1}{\Lambda^2}\delta^{(2)}(Q^\dagger)\ds{12}\label{Matter6N1}\\
    \mathcal{A}(V^+,V^+,\Phi^+,\Phi^+)&\propto\frac{1}{\Lambda^2}\delta^{(2)}(Q^\dagger)\frac{\ds{12}^2}{\da{34}}\label{Vector6N1}
\end{align}
and analogous conjugates. The first superamplitude combines dimension-$6$ four fermion operators with scalar and mixed fermion-scalar operators. The superamplitude must be helicity conserving, so helicity-violating matter interactions are forbidden (such as those induced by the operators of the form $\psi\psi\partial^2\psi\psi$). For the scalar interactions, ``helicity preserving'' becomes charge preserving (each $\phi$ must be paired with a $\phi^\dagger$ in the operator). The second superamplitude does allow for helicity violation, provided that it involves a gaugino and a matter fermion. It relates this to the bosonic operators of the form $F^2\phi^2$, where the scalar must be charge-violating, as well as the MDM-like operator in (\ref{AxionCoup}), dressed with an additional scalar. 

Each of these superamplitudes may be respectively further enhanced to $\mathcal{N}=2$, 
\begin{align}
    \mathcal{A}(K,K,K,K)\propto\frac{1}{\Lambda^2}\delta^{(4)}(Q^\dagger)\frac{\ds{12}}{\da{34}}\label{Matter6N2}\\
    \mathcal{A}(V^+,V^+,V^+,V^+)\propto\frac{1}{\Lambda^2}\delta^{(4)}(Q^\dagger)\frac{\ds{12}^2}{\da{34}^2}\label{Vector6N2}
\end{align}
(and conjugates), neither of which contains any new types of component interactions. The chiral multiplets of (\ref{Vector6N1}) must descend from the vector multiplets in (\ref{Vector6N2}), while the $\mathcal{N}=1$ matter interactions can only be promoted to $\mathcal{N}=2$ matter interactions. Interestingly, the $\mathcal{N}=2$ vectors still cannot couple to the hypermultiplets at this order.

\subsubsection{Dimension $7$}
The $\mathcal{N}=1$ possibilities are
\begin{align}
    \mathcal{A}(\Phi^+,\Phi^+,\Phi^+,\Phi^+)&\propto\frac{1}{\Lambda^3}\delta^{(2)}(Q^\dagger)\{\ds{12}\ds{34},\ds{13}\ds{24}\}\label{7dN1Matter}\\
    \mathcal{A}(V^+,V^+,\Phi^-,\Phi^-)&\propto\frac{1}{\Lambda^3}\delta^{(2)}(Q^\dagger)\ds{12}^2\label{7dN1Vector}
\end{align}
The first superamplitude is the first example that admits multiple possible independent terms, a particular basis for which is given inside the brackets. These correspond to helicity-violating operators of the schematic form $\psi\psi\partial^2\phi\phi$, where each term corresponds to a particular distribution of the derivatives. Like the Weinberg operator, these supersymmetrise into themselves. The second superamplitude describes operators of the form $F^2\psi^2$ and its superpartners: the gaugino-scalar coupling, similar to that in (\ref{7dN1Matter}), but restricted to the term proportional to the Mandelstam invariant of both scalars' momenta, and a coupling of the schematic form $F\lambda\psi^\dagger\cancel{\partial}\phi^\dagger$.

The only $\mathcal{N}=2$ possibility is
\begin{align}
    \mathcal{A}(V^+,V^+,K,K)&\propto\frac{1}{\Lambda^3}\delta^{(4)}(Q^\dagger)\frac{\ds{12}^2}{\da{34}}.
\end{align}
This superamplitude unifies both of the $\mathcal{N}=1$ superamplitudes listed above (although selecting-out only one of the terms (\ref{7dN1Matter}) determined by which chiral multiplets are embedded in the $\mathcal{N}=2$ vectors).

\subsubsection{Dimension $8$}
$\mathcal{N}=4$ compatible interactions become admissible at dimension $8$. The only possible superamplitude consistent with dimensional analysis is 
\begin{align}
    \mathcal{A}(V,V,V,V)\propto \frac{1}{\Lambda^4}\delta^{(8)}(Q^\dagger)\frac{\ds{12}^2}{\da{34}^2}.
\end{align}
This is the supersymmetrisation of the helicity-preserving $F^4$ operator and is (kinematically) unique.

For $\mathcal{N}<4$, more possibilities arise than for lower dimension, as dimensional analysis permits more derivatives and therefore more ways that they can be distributed, as well as new Lorentz-invariant combinations of fermion chirality. However, supersymmetry still places stringent constraints on the possible component interactions. 

For $\mathcal{N}=1$, the possible superamplitudes are 
\begin{align}
    \mathcal{A}(\Phi^+,\Phi^+,\Phi^-,\Phi^-)&=\frac{1}{\Lambda^4}\delta^{(2)}(Q^\dagger)\ds{12}\{c_{\Phi^4s}s,c_{\Phi^4t}t\}\label{8dN1Matter}\\
    \mathcal{A}(V^+,V^-,\Phi^+,\Phi^-)&=\frac{c_{V^2\Phi^2}}{\Lambda^4}\delta^{(2)}(Q^\dagger)\ds{13}\ds{14}\da{24}\\
    \mathcal{A}(V^+,V^+,\Phi^+,\Phi^+)&=\frac{1}{\Lambda^4}\delta^{(2)}(Q^\dagger)\ds{12}\{d_{V^2\Phi^2s}\ds{12}\ds{34},d_{V^2\Phi^2t}\ds{31}\ds{24}\}\label{8dN1HV}\\
    \mathcal{A}(V^+,V^+,V^-,V^-)&=\frac{c_{V^4}}{\Lambda^4}\delta^{(2)}(Q^\dagger)\ds{12}^2\da{34}.
\end{align}
The numerical Wilson coefficients are retained in these expressions to match them with the operators to be given below.

Extending to $\mathcal{N}=2$, the permitted superamplitudes are 
\begin{align}
    \mathcal{A}(K,K,K,K)&\propto\frac{1}{\Lambda^4}\delta^{(4)}(Q^\dagger)\{\ds{12}\ds{34},\ds{13}\ds{24}\}\\
    \mathcal{A}(V^+,V^-,K,K,)&\propto\frac{1}{\Lambda^4}\delta^{(4)}(Q^\dagger)\ds{13}\ds{14}\label{8dn2Mix1}\\
    \mathcal{A}(V^+,V^+,V^+,V^+)&\propto\frac{1}{\Lambda^4}\delta^{(4)}(Q^\dagger)\frac{\ds{12}}{\da{34}}\{\ds{12}\ds{34},\ds{13}\ds{24}\}\\
    \mathcal{A}(V^+,V^+,V^-,V^-)&\propto\frac{1}{\Lambda^4}\delta^{(4)}(Q^\dagger)\ds{12}^2.\label{8dn2Vec}
\end{align}
These mostly just describe promotions of the respective $\mathcal{N}=1$ superamplitudes into $\mathcal{N}=2$. The superamplitudes (\ref{8dn2Mix1}) and (\ref{8dn2Vec}) also decompose into $\mathcal{N}=1$ components that include the $c_{\Phi^4s}$ term in (\ref{8dN1Matter}).

Note that the $\mathcal{N}=1,2$ helicity-violating interactions (which have been singled-out in (\ref{8dN1HV}) by having coupling labelled as $d$ rather than $c$), which are the only examples of inelastic superamplitudes listed above, are also the only type that do not appear when that $\mathcal{N}=4$ superamplitude is decomposed into lower $\mathcal{N}$ components and are therefore not $\mathcal{N}=4$ compatible. 

For minimal field theories, it is interesting to consider the terms in an effective action that would generate the superamplitudes listed above and identify the Wilson coefficients united by supersymmetry. For a $\mathcal{N}=1$ chiral multiplet, the dimension $8$ operators are 
\begin{align}\label{Phi4Op}
    \mathcal{L}_{EFT_8}&\propto\frac{c_{\Phi^4_s}}{\Lambda^4}\left(\frac{1}{4}\phi\phi(\partial^2)^2(\phi^*\phi^*)+\frac{1}{2}\psi\psi\partial^2(\psi^\dagger\psi^\dagger)+2i\partial_\mu\phi^\dagger\partial^\nu\phi\partial_\nu\psi\sigma^\mu\psi^\dagger\right)
\end{align}
(identical particle exchange symmetry rules-out the other possible operator displayed above in (\ref{8dN1Matter})). For the vector multiplet, 
\begin{align}\label{Vec4Op}
    \mathcal{L}_{EFT_8}\propto \frac{c_{V^4}}{\Lambda^4}\left(\frac{1}{16}\left((F^2)^2+(F\tilde{F})^2\right)+\frac{1}{2}\lambda\lambda\partial^2(\lambda^\dagger\lambda^\dagger)+2i\lambda F_L\sigma^\mu F_R\partial_\mu\lambda^\dagger\right).
\end{align}
Here, $F_L=F_{\mu\nu}S_L^{\mu\nu}$ and $F_R=F_{\mu\nu}S_R^{\mu\nu}$. The mixed interactions are:
\begin{align}
    \mathcal{L}_{EFT_8}&\propto \frac{c_{V^2\Phi^2}}{\Lambda^4}\Big(2\text{tr}\left(F_L\sigma^\mu F_R\overline{\sigma}^\nu\right)\partial_\mu\phi^\dagger\partial_\nu\phi-\partial_\mu\psi\lambda\psi^\dagger\partial^\mu\lambda^\dagger\nonumber\\
    &\qquad\qquad\qquad-2i \lambda^\dagger\overline{\sigma}^\mu\partial_\nu\lambda\partial_\mu\phi^\dagger\partial^\nu\phi+2i\psi 
    F_L\sigma^\mu F_R\partial_\mu\psi^\dagger\nonumber\\
    &\qquad\qquad\qquad\qquad\qquad\qquad\qquad\qquad-\sqrt{2}i\partial_\mu\psi\lambda F^{\mu\nu}\partial_\nu\phi^\dagger+conj.\Big)\label{MixedDim8ActN1}\\
    &+\frac{d_{V^2\Phi^2s}}{\Lambda^4}\left(\frac{1}{8}\left(F^2+iF\tilde{F}\right)\partial^2\phi^2+\frac{1}{4}\psi\psi\partial^2\left(\lambda\lambda\right)+2\sqrt{2}i\partial_\mu\phi\lambda F_L\partial^\mu\psi\right)+\text{conj}.\label{MixedDim8ActN2}
\end{align}

The terms in the effective action above, or equivalently, the amplitudes that they correspond to, can be compared to those listed in Section \ref{sec:FermionBounds} for the two fermion system. The notable differences are that supersymmetry forbids the helicity-violating interactions involving only a single species $\psi$ or $\lambda$ (i.e. $\psi^2\partial^2\psi^2$, $\lambda^2\partial^2\lambda^2$ and their conjugates). For interactions between two different species of fermion, helicity-violating interactions are permitted. However, only the two interactions listed in (\ref{MixedDim8ActN1}) and (\ref{MixedDim8ActN2}) are allowed. The others listed in (\ref{dim8fermions}) are forbidden by supersymmetry. Note that $\partial\psi\lambda\cdot\partial\psi^\dagger\lambda^\dagger+\frac{1}{2}\psi\lambda\partial^2\left(\psi^\dagger\lambda^\dagger\right)=-\partial_\mu\psi\lambda\psi^\dagger\partial^\mu\lambda^\dagger$, so supersymmetry only permits this single combination of mixed helicity conserving interaction.

Before advancing on to the superymmetrised postivity constraints, I first digress to make a comment on the application to supersymmetry breaking given in \cite{Dine:2009sw}. The general low-energy EFT of a goldstino and $R$-axion was constructed in \cite{Dine:2009sw} to describe the breaking of $\mathcal{N}=1$ supersymmetry and its $R$-symmetry. This action included an interaction of the form given by the mixed interaction in (\ref{Phi4Op}) (but with a real scalar), as well as dimension $8$ helicity preserving and violating pure goldstino operators. By demanding positivity of the mixed interaction, an upper bound on the vev of the superpotential was derived from the product of the $R$-axion and goldstino decay constants (all parameters determining the low energy constants in the effective action). This same bound could be equivalently obtained by instead applying the conclusions of Section \ref{sec:idparticles} directly to the goldstino interactions.

\subsection{Unity of the positivity theorems and new bounds}\label{sec:PositivityCombined}

To begin with, the positivity constraints on the scalar \cite{Adams:2006sv} and the fermion \cite{Bellazzini:2016xrt} in (\ref{Phi4Op}) are unified when the fields are combined in a supermultiplet. Note that these interactions are consistent with both a Goldstone shift-symmetry for the scalar and a goldstino non-linear supersymmetry for the fermion. These interactions are expected for a Goldstone multiplet in a theory with extended SUSY spontaneously broken to $\mathcal{N}=1$. Likewise, the interactions in (\ref{Vec4Op}) must each have positive coefficients \cite{Bellazzini:2016xrt}, which is consistent with their unification under supersymmetry. 

The mixed interactions are similar to those discussed in the previous Section \ref{Sec:MultiSpin}. The first four terms of (\ref{MixedDim8ActN1}) induce elastic scattering of different species off each other, so the positivity of the coefficient $c_{V^2\Phi^2}>0$ is again expected. Interestingly however, the inelastic partner operators of the form  $\partial_\mu\psi\lambda F^{\mu\nu}\partial_\nu\phi^\dagger$ seem to inherit this condition. It is unclear how these operators would be constrained in the absence of supersymmetry. Their on-shell contact amplitudes that vanish in the forward limit in all channels, so a departure away from the forward scattering would seem necessary to access them. This operator will remain a puzzle here.

Promoting to $\mathcal{N}=2$, the $V^4$ and $V^2\Phi^2$ (super)-operators unify further and the positivity of their Wilson coefficients is combined into that of the ($\mathcal{N}=2$) $V^4$ operator. Similarly, positivity of the $\mathcal{N}=1$ chiral multiplet interactions (\ref{Phi4Op}) becomes positivity of the analogous $\mathcal{N}=2$ hypermultiplet interactions. Promoting further to $\mathcal{N}=4$, all of these are unified into the single positivity constraint on the dim-$8$ vector multiplet interaction.

Finally, the new inelastic constraints derived in the previous section also unify. With supersymmetry, the fermion constraints from Section \ref{sec:FermionBounds} simplify, as the amplitudes 
\begin{align}
    A\left(\psi^\pm\psi^\pm\rightarrow\psi^\mp\psi^\mp\right)=A\left(\lambda^\pm\lambda^\pm\rightarrow\lambda^\mp\lambda^\mp\right)=A\left(\lambda^\pm\lambda^\mp\rightarrow\psi^\pm\psi^\mp\right)=A\left(\lambda^\pm\lambda^\pm\rightarrow\psi^\pm\psi^\pm\right)=0,
\end{align}
while specifically for identical particles, $M\left(\psi^\pm\lambda^\pm\rightarrow\psi^\mp\lambda^\mp\right)=0$. This leaves only one type of inelastic amplitude and its parity conjugate and the constraints can be stated as
\begin{align}
    \frac{1}{2}|M^{\lambda^-\lambda^-\psi^+\psi^+}\pm M^{\lambda^+\lambda^+\psi^-\psi^-}|<M^{\psi^+\lambda^-\psi^+\lambda^-}+\sqrt{M^{\psi^+\psi^-\psi^+\psi^-}M^{\lambda^+\lambda^-\lambda^+\lambda^-}}.\label{SUSYfermions}
\end{align}
The combinations appearing on the LHS correspond to the $P$ conserving and violating interactions in the inelastic transitions (corresponding to, at tree-level, the real and imaginary parts of the coupling $d_{V^2\Phi^2s}$ above). Notably, the amplitudes forbidden by the SWIs would all appear as additional contributions to the left hand side of (\ref{SUSYfermions}), strengthening the lower bound on the elastic amplitudes.

The analysis of Section \ref{Sec:MultiSpin} can be easily extended to a complex scalar and two fermion species. With the scalar complex, the bound (\ref{MixedBosonBound}) generalises to 
\begin{align}
    \frac{1}{2}|M^{\phi\phi\gamma^+\gamma^+}\pm M^{\overline{\phi}\overline{\phi}\gamma^-\gamma^-}|<M^{\phi\gamma^+\phi\gamma^+}+\sqrt{M^{\phi\phi\phi\phi}M^{\gamma^+\gamma^-\gamma^+\gamma^-}}
\end{align}
(after simplifying with $CPT$ and $Y$), which has the expected form resembling (\ref{SUSYfermions}). The last partner relation, involving the mixed fermion-boson amplitudes, can also be found to be 
\begin{align}
    \frac{1}{2}|M^{\phi\psi^-\gamma^+\lambda^+}\pm M^{\overline{\phi}\psi^+\gamma^-\lambda^-}|<\sqrt{M^{\phi\lambda^+\phi\lambda^+}M^{\psi^+\gamma^-\psi^+\gamma^-}}+\sqrt{M^{\phi\psi^+\phi\psi^+}M^{\lambda^+\gamma^-\lambda^+\gamma^-}}.
\end{align}
SWIs imply that $M^{\phi\lambda^+\phi\lambda^+}=M^{\psi^+\gamma^-\psi^+\gamma^-}$, so this bound has identical structure to the previous two, completing the full super-positivity constraint. Again, the inelastic components of (\ref{MixedDim8ActN1}) do not appear in any of these bounds and seem only to be dragged into participation by supersymmetry. For Wilson coefficients at tree-level, these bounds are encapsulated by 
\begin{align}
    |\Re(d_{V^2\Phi^2s})|,|\Im(d_{V^2\Phi^2s})|<c_{V^2\Phi^2}+\sqrt{c_{\Phi^4s}c_{V^4}}.
\end{align}
In contrast to the case from Section \ref{sec:idparticles}, the space of consistent $P$-violating inelastic couplings is a square rather than a disc. 

As mentioned above, the inelastic amplitudes are also the only type not consistent with $\mathcal{N}=4$ supersymmetry. This indicates that the lower bounds in these inequalities are minimised by requiring increasingly more supersymmetry (where simple $\mathcal{N}=1$ is sufficient to rule-out many possible inelastic amplitudes that may potentially appear, such as the other fermionic helicity configurations in Section \ref{sec:FermionBounds}). This is the (expected) consequence of symmetries imposing selection rules that prohibit inelastic processes, but also illustrates how extreme symmetry breaking can be prohibited by the positivity bounds.

\section{Conclusion}

The general implication of unitarity for the causality sum rule (\ref{SumRule}) is to bound the size of inelastic scattering amplitudes by the size of the elastic ones (\ref{InelBound}). This places fundamental limits on the extent to which hypothetical symmetries, which manifest themselves in the $S$-matrix as selection rules, can be broken by effective interactions. These constraints appear naively invisible in the construction of a general effective action of local contact interactions. 

Employing the convex cone picture of \cite{Zhang:2020jyn}, the general set of positivity bounds for fundamental $SU(2)$ and $SU(3)$ fermions were derived, including some that cannot be obtained from considering scattering of factorised states. Separately, general constraints on flavour violation were also derived for the cases in which the fermions are only non-singlets under one symmetry group. Simple inelastic bounds for spinning particles were also derived, in particular for two scalars and two vectors and the mixed case of two fermions, a vector and a scalar, where each particle has the same-sign helicity in the all outgoing convention. It was then shown that all of the standard bounds for particles of different spin unify under supersymmetry.

In the examples discussed in \ref{sec:SMFermions}, all states under consideration are related by symmetries. When the particles transform under multiple symmetry groups, the convex cone picture is needed for a complete characterisation of the information in the sum rule. However, if transitions between multiple distinct states are permitted by the symmetries, then the cone is non-polyhedral and the standard results for polyhedral cones cannot be so simply applied. This was analysed in \cite{Yamashita:2020gtt} for parity-symmetric weak boson operators, where a set of necessary constraints were derived using analytic and numerical methods. It is easy to apply (\ref{InelBound}) directly to obtain necessary conditions on the couplings. However, finding the complete set of sufficient bounds remains the most immediate open problem in applying the constraints from the sum rule to EFTs of multiple species, in particular the SMEFT. Subsequent to the release of this work, \cite{Li:2021cjv} were able to reformulate the the sum rule (\ref{SumRule}) as a positive semi-definite statement in a dual space to the space of external scattering states. Once appropriately crossing-symmetrised, a tensor in external particle labels may be contracted with (\ref{SumRule}) to obtain an expression of positive-definiteness. This dual space of positive-definite matrices was identified as a ``spectrahedron'', the geometry of which has been studied in \cite{Ramana:1995xxx}, and, as a space of positive-definite matrices, enabled methods from semi-definite programming to be applied (see \cite{Simmons-Duffin:2016gjk} for introduction of recent applications of these ideas to solving the CFT bootstrap equations). See also \cite{Hebbar:2020ukp} for recent progress in applying the $S$-matrix bootstrap to constraining Wilson coefficients in EFTs directly from the full crossing constraints. This leverages positive semi-definiteness of the $S$-matrix to utilise semi-definite programming. 

Generalisation to higher mass dimension, departure from the forward limit and massive particles are obvious future directions \cite{deRham:2017zjm}, \cite{Bellazzini:2020cot}, \cite{Tolley:2020gtv}, \cite{Caron-Huot:2020cmc}. See e.g. \cite{Green:2019tpt}, \cite{Huang:2020nqy} for recent discussions in string theory.

The entire discussion of this work has been concentrated at the level of dimension-$8$ order effective interactions. These have the minimum energy scaling to ensure that the integral over the contour deformed to infinity converges to zero and does not introduce an additional unknown UV ingredient into the sum rule with an unknown impact on the structure. The results described here at dimension $8$ readily extends to higher dimension $4n$ for $n>2$. However, no statement about lower dimensional amplitudes has been made, so it would seem that the remarks about symmetry violation would not extend to them. However, lower dimensional interactions will typically contribute to dimension-$8$ level interactions through multiple insertions and still appear in the sum rule. The impact of loops in higher order constraints was recently discussed in \cite{Bellazzini:2020cot}. In agreement with the observations here, these strengthen the positivity bounds that would naively apply at tree-level. It would be of interest to further investigate the implications of these constraints for RG flow, both in the context for the SMEFT and more generally. As mentioned in \cite{Bellazzini:2020cot}, this would require a treatment of IR divergences and inclusive observables. Furthermore, as recent works have shown, even operators appearing at higher dimension $4n+2$, for which the standard sum rule does not imply positivity, are highly constrained away from the forward limit. It would be interesting to establish precise points of distinction between the rigid set of constraints applicable at dimension $8$ and above and freedom for lower dimensional (including renormalisable) operators. Examination of UV completions may provide insight into this. Some discussion of dimension $6$ operators has recently been given in \cite{Remmen:2020uze}, \cite{Gu:2020thj}, \cite{Bellazzini:2014waa}, \cite{Falkowski:2012vh}, although conclusions drawn for the IR interactions have been predicated on assumptions about the amplitudes having extra-friendly high-energy scaling. 

While most of the applications presented here have had an eye toward the SMEFT, no analysis of the impacts of these constraints on tests of the SM 
has been attempted here. See \cite{Bellazzini:2018paj}, \cite{Gu:2020ldn}, \cite{Bi:2019phv}, \cite{Zhang:2018shp}, \cite{Fuks:2020ujk} for some recent discussion of this. The bounds discussed here activate at dimension $8$ order, which is expected to be typically sub-leading to the (many) dimension $6$ interactions that pervade the SMEFT. However, it may still be possible to access the affected dimension $8$ interactions through non-interference effects and angular distributions \cite{Alioli:2020kez}, \cite{Azatov:2016sqh}. While amplitudes vanishing in the forward limit (possibly because of angular momentum conservation) appear naively unaffected, this is not necessarily true in a crossed, inelastic channel described by the same amplitude, and as result, such a process would not escape constraint. 

This entire work has relied upon the $S$-matrix formulation of causality in order to derive constraints on effective interactions. However, it would also be interesting to construct background solutions (such as was done in \cite{Adams:2006sv}) in order to see precisely how such a breakdown arises if the constraints are violated, especially for the fermionic and mixed spin interactions. The validity and scope of the $S$-matrix formulation is also dependent upon the analytic structure being established and the standard derivation of the dispersion relations in Section \ref{sec:SumRule} relied upon this. These have been established for Wightman theories (at least for the analogous correlation functions) and theories loosely satisfying the requirements of LSZ reduction. However, this does not include applicability to perturbative gauge theories. Clarification over this issue would be informative.
Some possibly related foundational questions that have practical implications are the interpretation and use of the sum rule in the presence of IR divergences, as well as possible extension away from the forward limit where some understanding of the singularity structure may become necessary. The assumption that all particles can be given a small mass and that the results will apply to the exactly massless theory also needs to be validated (in particular, the assumption that the forward and massless limits commute), although no effort has been made in this direction here. This is least clear for possible applications to gravitational systems, where other foundational assumptions about locality in the UV completion are also uncertain.

It would also be interesting to find applications of these bounds to model building. As mentioned above, the fact that dimension $8$ operators usually contribute subleading effects naively poses an obstruction to the constraints having widespread, leading-order consequences. Continuing the analysis of \cite{Bellazzini:2018paj} for constraining massive higher spin particles and their hypothetical coupling to the SM is another possible application with direct consequence for the constraining the space of possible particle models of dark matter or other hypothetically fields associated with other cosmological mysteries. See e.g. \cite{Falkowski:2020mjq} for discussion of coupling of massive gravitons to (regular) matter, or \cite{deRham:2018qqo} for various other theories related to modified gravity. Alternatively, it would be of interest if these results can be used to make general statements about the nature and scope of symmetry breaking (such as of time-reversal) that is possible at low energies.

\acknowledgments 

Thank you to Grant Remmen for discussion, Isabel Garcia Garcia for comments on a draft, Nathaniel Craig for assistance with funding and Cen Zhang for some questions and comments. This work is supported by the US Department of Energy under the grant DE-SC0011702.

\appendix

\section{Sum Rules for Spinning Particles with $SO(2)$}

\subsection{Angular momentum projectors in $SO(2)$ form}\label{app:so2proj}

The Clebsch-Gordan coefficients for general spinning particles were given in (\ref{CGs}). If one of the helicities is zero, then the Clebsch-Gordan coefficients are
\begin{align}
    C^{i0}_{h}=C^{0i}_{h}=\frac{1}{\sqrt{2}}\begin{cases}1,\qquad  i=1\\
    -i,\qquad  i=2
    \end{cases}.
\end{align}
The $0$ superscript denotes the scalar state. The parity-conjugate coefficients are $C^{i0}_{-h}=C^{0i}_{-h}=(C^{i0}_{h})^*$.

The next step is to find the projectors. There are several special cases. For the case in which one of the incoming and outgoing particles are scalars, the projectors are simply
\begin{align}
    P_{Ph}^{i0k0}&=\delta^{ik}\qquad P_{Ph}^{0j0l}=\delta^{jl}\qquad
    P_{Ph}^{i00l}=\delta^{il}\qquad P_{Ph}^{0jk0}=\delta^{jk}\\
    P_{\cancel{P}h}^{i0k0}&=i\epsilon^{ik}\qquad P_{\cancel{P}h}^{0j0l}=i\epsilon^{jl}\qquad
    P_{\cancel{P}h}^{i00l}=i\epsilon^{il}\qquad P_{\cancel{P}h}^{0jk0}=i\epsilon^{jk}.
\end{align}

More generally, if none of the particles are scalars and $h_1\neq h_2$, $h_3\neq h_4$ (so that singlets do not appear in the product irreps), then the possible projectors are 
\begin{align}
    P_{\cancel{P}\,h_1+h_2}^{ijkl}&=\frac{i}{2}\left(\delta^{ik}\epsilon^{jl}+\delta^{jl}\epsilon^{ik}\right)\\
    P_{P\,h_1+h_2}^{ijkl}&=\frac{1}{2}\left(\delta^{ik}\delta^{jl}+\delta^{il}\delta^{jk}-\delta^{ij}\delta^{kl}\right), \qquad&\text{if}\, h_1+h_2=h_3+h_4,\\
    P_{\cancel{P}\,|h_1-h_2|}^{ijkl}&=\frac{i}{2}\left(\epsilon^{ij}\delta^{kl}-\delta^{ij}\epsilon^{kl}\right)\\
    P_{P\,|h_1-h_2|}^{ijkl}&=\frac{1}{2}\left(\delta^{ij}\delta^{kl}+\delta^{ik}\delta^{jl}-\delta^{il}\delta^{jk}\right), \qquad&\text{if}\,\left(h_1-h_2\right)=\left(h_3-h_4\right),\\
    P_{\cancel{P}\,|h_1-h_2|}^{ijkl}&=\frac{i}{2}\left(\delta^{ij}\epsilon^{kl}+\epsilon^{ij}\delta^{kl}\right)\\
    P_{P\,|h_1-h_2|}^{ijkl}&=\frac{1}{2}\left(\delta^{ij}\delta^{kl}-\delta^{ik}\delta^{jl}+\delta^{il}\delta^{jk}\right), \qquad&\text{if}\,\left(h_1-h_2\right)=-\left(h_3-h_4\right),\\
    P_{\cancel{P}\,h_1+h_2}^{ijkl}&=\frac{-i}{2}\left(S^{ij}\delta^{kl}\pm P^{ij}\epsilon^{kl}\right)\\
    P_{P\,h_1+h_2}^{ijkl}&=\frac{1}{2}\left(P^{ij}\delta^{kl}\mp S^{ij}\epsilon^{kl}\right), \qquad&\text{if}\, h_1+h_2=\pm\left(h_3-h_4\right),\\
    P_{\cancel{P}\,h_3+h_4}^{ijkl}&=\frac{i}{2}\left(\delta^{ij}S^{kl}\pm \epsilon^{ij}P^{kl}\right)\\
    P_{P\,h_3+h_4}^{ijkl}&=\frac{1}{2}\left(\delta^{ij}P^{kl}\pm \epsilon^{ij}S^{kl}\right), \qquad&\text{if}\, \pm\left(h_1-h_2\right)=h_3+h_4.
\end{align}
The $u$-channel projectors all agree with the required $s$-channel projectors that they should be related to under crossing. For example, if $h_1+h_2=h_3+h_4$, it is easily verified that $P_{{h_1-h_4}}^{ilkj}=P_{{h_1+h_2}}^{ijkl}$ (in either parity-symmetric or violating cases). The identity 
\begin{align}\label{2dId}
    \delta^{ij}\epsilon^{kl}+\delta^{kl}\epsilon^{ij}=\delta^{il}\epsilon^{kj}+\delta^{kl}\epsilon^{il}
\end{align}
is useful in handling the parity-violating projectors.

For the special case in which exactly one of the particles is a scalar, the projectors are 
\begin{align}
    P_{\cancel{P}\,h_3+h_4}^{0jkl}&=\frac{i}{\sqrt{2}}\left(\delta^{j1}S^{kl}-\delta^{j2}P^{kl}\right)\\
    P_{P\,h_3+h_4}^{0jkl}&=\frac{1}{\sqrt{2}}\left(\delta^{j1}P^{kl}+\delta^{j2}S^{kl}\right)\\
    P_{\cancel{P}\,h_3-h_4}^{0jkl}&=\frac{-i}{\sqrt{2}}\left(\delta^{j1}\epsilon^{kl}+\delta^{j2}\delta^{kl}\right)\\
    P_{P\,h_3-h_4}^{0jkl}&=\frac{1}{\sqrt{2}}\left(\delta^{j1}\delta^{kl}-\delta^{j2}\epsilon^{kl}\right)\\
    P_{\cancel{P}\,-h_3+h_4}^{0jkl}&=\frac{i}{\sqrt{2}}\left(\delta^{j1}\epsilon^{kl}-\delta^{j2}\delta^{kl}\right)\\
    P_{P\,-h_3+h_4}^{0jkl}&=\frac{1}{\sqrt{2}}\left(\delta^{j1}\delta^{kl}+\delta^{j2}\epsilon^{kl}\right).
\end{align}
Then $P^{i0kl}=P^{0ikl}$ and $P^{ij0l}=P^{ijl0}=(P^{l0ij})^*$ for each helicity configuration. These expressions encapsulate each of the possible relations between the  helicities of the scattered particles. Again, it can be easily checked that these are consistent with crossing e.g. for the case $h_3=0$ and $h_1+h_2=h_4$, $P_{P\,h_4-h_1}^{il0j}=P_{P\,h_1+h_2}^{ij0l}$, while the parity-violating projectors pick-up a negative sign $P_{\cancel{P}\,h_4-h_1}^{il0j}=-P_{\cancel{P}\,h_1+h_2}^{ij0l}$.

For the special case in which the outgoing states are both scalars, the projectors are simply the Clebsch-Gordan coefficients (or if the scalars are incoming, their conjugates)
\begin{align}
    P^{ij00}_A=\frac{1}{\sqrt{2}}\delta^{ij}\qquad P^{00kl}_A=\frac{1}{\sqrt{2}}\delta^{kl}\qquad P^{ij00}_B=\frac{-i}{\sqrt{2}}\epsilon^{ij}\qquad P^{00kl}_B=\frac{i}{\sqrt{2}}\epsilon^{kl}.
\end{align}
These have crossing relations
\begin{align}
    P^{0lk0}_{Ph}=\sqrt{2}P_A^{00kl}\qquad\qquad P^{0lk0}_{\cancel{P}h}=-\sqrt{2}P_B^{00kl}.
\end{align}

When the particles are not scalars, then if $h_1=h_4$ and $h_2=h_3$, the $u$-channel projectors decompose as 
\begin{align}
    P_{\cancel{P}\,h_1+h_2}^{ilkj}&=-P_{AB}^{ijkl}-P_{BA}^{ijkl}\\
    P_{P\,h_1+h_2}^{ilkj}&=P_{AA}^{ijkl}+P_{BB}^{ijkl}\\
    P_{\cancel{P}\,h_1-h_2}^{ilkj}&=P_{AB}^{ijkl}-P_{BA}^{ijkl}\\
    P_{P\,h_1-h_2}^{ilkj}&=P_{AA}^{ijkl}-P_{BB}^{ijkl},
\end{align}
while in the opposite channel,
\begin{align}
    P_{AA}^{ilkj}&=\frac{1}{2}\left(P_{P\,h_1+h_2}^{ijkl}+P_{P\,h_1-h_2}^{ijkl}\right)\\
    P_{BB}^{ilkj}&=\frac{1}{2}\left(P_{P\,h_1+h_2}^{ijkl}-P_{P\,h_1-h_2}^{ijkl}\right)\\
    P_{AB}^{ilkj}&=\frac{1}{2}\left(P_{\cancel{P}\,h_1-h_2}^{ijkl}-P_{\cancel{P}\,h_1+h_2}^{ijkl}\right)\\
    P_{BA}^{ilkj}&=-\frac{1}{2}\left(P_{\cancel{P}\,h_1-h_2}^{ijkl}+P_{\cancel{P}\,h_1+h_2}^{ijkl}\right).
\end{align}

The results given in Section \ref{sec:RotSym} also apply to the case $h_1=h_2=h\neq h_3=h_4$, except that only the transitions between singlet irreps are possible.

\subsection{Sum rules for two fermion and multispin theories}\label{app:so2sumrules}
The sum rule entries for scattering in the two fermion theory of Section \ref{sec:FermionBounds} are, in $SO(2)$ form:

\hspace*{-1.5cm}\vbox{
\begin{align}
    M\left(\lambda,\lambda\rightarrow\lambda,\lambda\right)&=\left(|\boldsymbol{m}_{1}^{\lambda\lambda}|^2+\frac{1}{2}\left(|\boldsymbol{m}_{A}^{\lambda\lambda}|^2+|\boldsymbol{m}_{B}^{\lambda\lambda}|^2\right)\right)P_{P1}^{ijkl}+\frac{1}{2}\left(2|\boldsymbol{m}_{1}^{\lambda\lambda}|^2+3|\boldsymbol{m}_{A}^{\lambda\lambda}|^2-|\boldsymbol{m}_{B}^{\lambda\lambda}|^2\right)P_{AA}^{ijkl}\nonumber\\
    &\quad+\frac{1}{2}\left(2|\boldsymbol{m}_{1}^{\lambda\lambda}|^2-|\boldsymbol{m}_{A}^{\lambda\lambda}|^2+3|\boldsymbol{m}_{B}^{\lambda\lambda}|^2\right)P_{BB}^{ijkl}+2\boldsymbol{m}_{A}^{\lambda\lambda}\cdot\boldsymbol{m}_{B}^{\lambda\lambda}\left(P_{AB}^{ijkl}-P_{BA}^{ijkl}\right)\\
    M\left(\psi,\psi\rightarrow\psi,\psi\right)&=\left(|\boldsymbol{m}_{1}^{\psi\psi}|^2+\frac{1}{2}\left(|\boldsymbol{m}_{A}^{\psi\psi}|^2+|\boldsymbol{m}_{B}^{\psi\psi}|^2\right)\right)P_{P1}^{ijkl}+\frac{1}{2}\left(2|\boldsymbol{m}_{1}^{\psi\psi}|^2+3|\boldsymbol{m}_{A}^{\psi\psi}|^2-|\boldsymbol{m}_{B}^{\psi\psi}|^2\right)P_{AA}^{ijkl}\nonumber\\
    &\quad+\frac{1}{2}\left(2|\boldsymbol{m}_{1}^{\psi\psi}|^2-|\boldsymbol{m}_{A}^{\psi\psi}|^2+3|\boldsymbol{m}_{B}^{\psi\psi}|^2\right)P_{BB}^{ijkl}+2\boldsymbol{m}_{A}^{\psi\psi}\cdot\boldsymbol{m}_{B}^{\psi\psi}\left(P_{AB}^{ijkl}-P_{BA}^{ijkl}\right)\\
    M\left(\psi,\lambda\rightarrow\psi,\lambda\right)&=\left(|\boldsymbol{m}_{1}^{\psi\lambda}|^2+\frac{1}{2}\left(|\boldsymbol{m}_{A}^{\psi\lambda}|^2+|\boldsymbol{m}_{B}^{\psi\lambda}|^2\right)\right)P_{P1}^{ijkl}+\frac{1}{2}\left(2|\boldsymbol{m}_{1}^{\psi\lambda}|^2+3|\boldsymbol{m}_{B}^{\psi\lambda}|^2-|\boldsymbol{m}_{A}^{\psi\lambda}|^2\right)P_{AA}^{ijkl}\nonumber\\
    &\quad+\frac{1}{2}\left(2|\boldsymbol{m}_{1}^{\psi\lambda}|^2-|\boldsymbol{m}_{B}^{\psi\lambda}|^2+3|\boldsymbol{\lambda}|^2\right)P_{BB}^{ijkl}+2\boldsymbol{m}_{A}^{\psi\lambda}\cdot\boldsymbol{m}_{B}^{\psi\lambda}\left(P_{AB}^{ijkl}-P_{BA}^{ijkl}\right)\\
    M\left(\lambda,\psi\rightarrow\lambda,\psi\right)&=\left(|\boldsymbol{m}_{1}^{\lambda\psi}|^2+\frac{1}{2}\left(|\boldsymbol{m}_{A}^{\lambda\psi}|^2+|\boldsymbol{m}_{B}^{\lambda\psi}|^2\right)\right)P_{P1}^{ijkl}+\frac{1}{2}\left(2|\boldsymbol{m}_{1}^{\lambda\psi}|^2+3|\boldsymbol{m}_{B}^{\lambda\psi}|^2-|\boldsymbol{m}_{A}^{\lambda\psi}|^2\right)P_{AA}^{ijkl}\nonumber\\
    &\quad+\frac{1}{2}\left(2|\boldsymbol{m}_{1}^{\lambda\psi}|^2-|\boldsymbol{m}_{B}^{\lambda\psi}|^2+3|\boldsymbol{m}_{A}^{\lambda\psi}|^2\right)P_{BB}^{ijkl}+2\boldsymbol{m}_{A}^{\lambda\psi}\cdot\boldsymbol{m}_{B}^{\lambda\psi}\left(P_{AB}^{ijkl}-P_{BA}^{ijkl}\right)\\
    M\left(\lambda,\lambda\rightarrow\psi,\psi\right)&=\frac{1}{2}\left(\boldsymbol{m}_{1}^{\psi\psi}\cdot\boldsymbol{m}_{1}^{\lambda\lambda}+\boldsymbol{m}_{-1}^{\psi\psi}\cdot\boldsymbol{m}_{-1}^{\lambda\lambda}+\boldsymbol{m}_{B}^{\psi\lambda}\cdot\boldsymbol{m}_{B}^{\lambda\psi}+\boldsymbol{m}_{A}^{\psi\lambda}\cdot\boldsymbol{m}_{A}^{\lambda\psi}\right)P_{P1}^{ijkl}\nonumber\\
    &\quad+\frac{1}{2}\left(2\boldsymbol{m}_{A}^{\psi\psi}\cdot\boldsymbol{m}_{A}^{\lambda\lambda}+\boldsymbol{m}_{1}^{\psi\lambda}\cdot\boldsymbol{m}_{1}^{\lambda\psi}+\boldsymbol{m}_{-1}^{\psi\lambda}\cdot\boldsymbol{m}_{-1}^{\lambda\psi}+\boldsymbol{m}_{B}^{\psi\lambda}\cdot\boldsymbol{m}_{B}^{\lambda\psi}-\boldsymbol{m}_{A}^{\psi\lambda}\cdot\boldsymbol{m}_{A}^{\lambda\psi}\right)P_{AA}^{ijkl}\nonumber\\
    &\quad+\frac{1}{2}\left(2\boldsymbol{m}_{B}^{\psi\psi}\cdot\boldsymbol{m}_{B}^{\lambda\lambda}+\boldsymbol{m}_{1}^{\psi\lambda}\cdot\boldsymbol{m}_{1}^{\lambda\psi}+\boldsymbol{m}_{-1}^{\psi\lambda}\cdot\boldsymbol{m}_{-1}^{\lambda\psi}-\boldsymbol{m}_{B}^{\psi\lambda}\cdot\boldsymbol{m}_{B}^{\lambda\psi}+\boldsymbol{m}_{A}^{\psi\lambda}\cdot\boldsymbol{m}_{A}^{\lambda\psi}\right)P_{BB}^{ijkl}\nonumber\\
    &\quad+\frac{1}{2}\left(2\boldsymbol{m}_{A}^{\psi\psi}\cdot\boldsymbol{m}_{B}^{\lambda\lambda}-\boldsymbol{m}_{1}^{\psi\lambda}\cdot\boldsymbol{m}_{1}^{\lambda\psi}+\boldsymbol{m}_{-1}^{\psi\lambda}\cdot\boldsymbol{m}_{-1}^{\lambda\psi}+\boldsymbol{m}_{A}^{\psi\lambda}\cdot\boldsymbol{m}_{B}^{\lambda\psi}-\boldsymbol{m}_{B}^{\psi\lambda}\cdot\boldsymbol{m}_{A}^{\lambda\psi}\right)P_{AB}^{ijkl}\nonumber\\
    &\quad+\frac{1}{2}\left(2\boldsymbol{m}_{B}^{\psi\psi}\cdot\boldsymbol{m}_{A}^{\lambda\lambda}-\boldsymbol{m}_{1}^{\psi\lambda}\cdot\boldsymbol{m}_{1}^{\lambda\psi}+\boldsymbol{m}_{-1}^{\psi\lambda}\cdot\boldsymbol{m}_{-1}^{\lambda\psi}-\boldsymbol{m}_{A}^{\psi\lambda}\cdot\boldsymbol{m}_{B}^{\lambda\psi}+\boldsymbol{m}_{B}^{\psi\lambda}\cdot\boldsymbol{m}_{A}^{\lambda\psi}\right)P_{BA}^{ijkl}.
\end{align}
}
Here $\boldsymbol{m}_{-1}^{ff'}$ and $\boldsymbol{m}_{1}^{ff'}$ for example represent the UV couplings of the $f^+f'^-$ and $f^-f'^+$ helicity configurations respectively for any fermions $f$ and $f'$. For brevity, the amplitude $M\left(\lambda,\psi\rightarrow\psi,\lambda\right)$ has not been stated as it is entirely determined from crossing $M\left(\lambda,\lambda\rightarrow\psi,\psi\right)$. The $Y$-symmetry implies that $|\boldsymbol{m}_{-1}^{\lambda\lambda}|=|\boldsymbol{m}_{1}^{\lambda\lambda}|$, $|\boldsymbol{m}_{-1}^{\psi\psi}|=|\boldsymbol{m}_{1}^{\psi\psi}|$, $|\boldsymbol{m}_{B}^{\lambda\psi}|=|\boldsymbol{m}_{B}^{\psi\lambda}|$, $|\boldsymbol{m}_{A}^{\lambda\psi}|=|\boldsymbol{m}_{A}^{\psi\lambda}|$, $|\boldsymbol{m}_{1}^{\psi\lambda}|=|\boldsymbol{m}_{-1}^{\psi\lambda}|=|\boldsymbol{m}_{1}^{\lambda\psi}|=|\boldsymbol{m}_{-1}^{\lambda\psi}|$, which has been used to (partially) simplify the elastic amplitudes. As for the example of Section \ref{sec:idparticles}, fundamental principles rule-out the existence of the parity-violating, spinning component amplitudes in all configurations above (although it is present in the omitted $M\left(\lambda,\psi\rightarrow\psi,\lambda\right)$).

The sum rule entries for the simple multispin theory of Section \ref{Sec:MultiSpin} in $SO(2)$ form are given next. The terms can be classified by spin projection $m_z=0,\frac{1}{2},1,\frac{3}{2},2$. They are given in Table \ref{Tab:Multispin}.

\begin{table}
\centering
\begin{adjustwidth}{-2.5cm}{0cm}
\begin{tabular}{ |p{1.3cm}||p{1.5cm}|p{6.8cm}|p{8.8cm}| }
 \hline
  $m_z=0$ & $\phi\phi$ & $\psi\psi$ & $\gamma\gamma$ \\
 \hline
 \hline
 $\phi\phi$   & $2|\boldsymbol{m}^{\phi\phi}|^2$    & 0
 & $\left(\boldsymbol{m}_A^{\gamma\gamma}\cdot\boldsymbol{m}^{\phi\phi}+\sqrt{2}\boldsymbol{m}_1^{\phi\gamma}\cdot\boldsymbol{m}_1^{\gamma\phi}+\sqrt{2}\boldsymbol{m}_1^{\gamma\phi}\cdot\boldsymbol{m}_1^{\phi\gamma}\right)P_A+\left(\boldsymbol{m}_B^{\gamma\gamma}\cdot\boldsymbol{m}^{\phi\phi}-\sqrt{2}\boldsymbol{m}_1^{\phi\gamma}\cdot\boldsymbol{m}_1^{\gamma\phi}+\sqrt{2}\boldsymbol{m}_1^{\gamma\phi}\cdot\boldsymbol{m}_1^{\phi\gamma}\right)P_B$ \\
 $\psi\psi$ & . &  $\frac{1}{2}\left(2|\boldsymbol{m}_1^{\psi\psi}|^2+3|\boldsymbol{m}_A^{\psi\psi}|^2-|\boldsymbol{m}_B^{\psi\psi}|^2\right)P_{AA}+\frac{1}{2}\left(2|\boldsymbol{m}_1^{\psi\psi}|^2-|\boldsymbol{m}_A^{\psi\psi}|^2+3|\boldsymbol{m}_B^{\psi\psi}|^2\right)P_{BB}+2\boldsymbol{m}_A^{\psi\psi}\cdot\boldsymbol{m}_B^{\psi\psi}\left(P_{AB}-P_{BA}\right)$  & 0
 \\
 $\gamma\gamma$ & . & . &  $\frac{1}{2}\left(2|\boldsymbol{m}_2^{\gamma\gamma}|^2+3|\boldsymbol{m}_A^{\gamma\gamma}|^2-|\boldsymbol{m}_B^{\gamma\gamma}|^2\right)P_{AA}+\frac{1}{2}\left(2|\boldsymbol{m}_2^{\gamma\gamma}|^2-|\boldsymbol{m}_A^{\gamma\gamma}|^2+3|\boldsymbol{m}_B^{\gamma\gamma}|^2\right)P_{BB}+2\boldsymbol{m}_A^{\gamma\gamma}\cdot\boldsymbol{m}_B^{\gamma\gamma}\left(P_{AB}-P_{BA}\right)$ \\
 \hline
\end{tabular}
\end{adjustwidth}

\begin{adjustwidth}{-2.5cm}{0cm}
\begin{tabular}{ |p{1.3cm}||p{2.0cm}|p{2.0cm}|p{5.8cm}|p{5.8cm}| }
 \hline
  $m_z=\frac{1}{2}$ & $\phi\psi$ & $\psi\phi$ & $\psi\gamma$ & $\gamma\psi$ \\
 \hline
 \hline
 $\phi\psi$   & $|\boldsymbol{m}_{\frac{1}{2}}^{\phi\psi}|^2P_{P\frac{1}{2}}$    & 0 
 & 0
 & $\left(\boldsymbol{m}_{\frac{1}{2}}^{\psi\gamma}\cdot\boldsymbol{m}_{\frac{1}{2}}^{\phi\psi}+\boldsymbol{m}_{\frac{1}{2}}^{\phi\psi}\cdot\boldsymbol{m}_{\frac{1}{2}}^{\psi\gamma}\right)P_{P\frac{1}{2}}+\left(\boldsymbol{m}_{\frac{1}{2}}^{\psi\gamma}\cdot\boldsymbol{m}_{\frac{1}{2}}^{\phi\psi}-\boldsymbol{m}_{\frac{1}{2}}^{\phi\psi}\cdot\boldsymbol{m}_{\frac{1}{2}}^{\psi\gamma}\right)P_{\cancel{P}\frac{1}{2}}$
 \\
 $\psi\phi$ & . &  $|\boldsymbol{m}_{\frac{1}{2}}^{\phi\psi}|^2P_{P\frac{1}{2}}$  & $\left(\boldsymbol{m}_{\frac{1}{2}}^{\psi\gamma}\cdot\boldsymbol{m}_{\frac{1}{2}}^{\phi\psi}+\boldsymbol{m}_{\frac{1}{2}}^{\phi\psi}\cdot\boldsymbol{m}_{\frac{1}{2}}^{\psi\gamma}\right)P_{P\frac{1}{2}}-\left(\boldsymbol{m}_{\frac{1}{2}}^{\psi\gamma}\cdot\boldsymbol{m}_{\frac{1}{2}}^{\phi\psi}-\boldsymbol{m}_{\frac{1}{2}}^{\phi\psi}\cdot\boldsymbol{m}_{\frac{1}{2}}^{\psi\gamma}\right)P_{\cancel{P}\frac{1}{2}}$ 
 & 0
 \\
 $\psi\gamma$ & . & . &  $\left(|\boldsymbol{m}_{\frac{3}{2}}^{\psi\gamma}|^2+|\boldsymbol{m}_{\frac{1}{2}}^{\psi\gamma}|^2\right)P_{P\frac{1}{2}}$ & 0
 \\
 $\gamma\psi$ & . & . &  . & $\left(|\boldsymbol{m}_{\frac{3}{2}}^{\psi\gamma}|^2+|\boldsymbol{m}_{\frac{1}{2}}^{\psi\gamma}|^2\right)P_{P\frac{1}{2}}$\\
 \hline
\end{tabular}
\end{adjustwidth}

\begin{adjustwidth}{-2.5cm}{0cm}
\begin{tabular}{ |p{1.3cm}||p{6.5cm}|p{2.0cm}|p{8.1cm}| }
 \hline
  $m_z=1$ & $\psi\psi$ & $\phi\gamma$ & $\gamma\phi$ \\
 \hline
 \hline
 $\psi\psi$   & $\left(|\boldsymbol{m}_1^{\psi\psi}|^2+\frac{1}{2}\left(|\boldsymbol{m}_A^{\psi\psi}|^2+|\boldsymbol{m}_B^{\psi\psi}|^2\right)\right)P_{P1}$    & 0
 & 0
 \\
 $\phi\gamma$ & . &  $|\boldsymbol{m}_1^{\phi\gamma}|^2P_{P1}$  & $\left(\boldsymbol{m}_1^{\gamma\phi}\cdot\boldsymbol{m}_1^{\phi\gamma}+\boldsymbol{m}_1^{\phi\gamma}\cdot\boldsymbol{m}_1^{\gamma\phi}+\frac{1}{\sqrt{2}}\boldsymbol{m}_A^{\gamma\gamma}\cdot\boldsymbol{m}^{\phi\phi}\right)P_{P1}+\left(\boldsymbol{m}_1^{\gamma\phi}\cdot\boldsymbol{m}_1^{\phi\gamma}-\boldsymbol{m}_1^{\phi\gamma}\cdot\boldsymbol{m}_1^{\gamma\phi}-\frac{1}{\sqrt{2}}\boldsymbol{m}_B^{\gamma\gamma}\cdot\boldsymbol{m}^{\phi\phi}\right)P_{\cancel{P}1}$ \\
 $\gamma\phi$ & . & . &  $|\boldsymbol{m}_1^{\gamma\phi}|^2P_{P1}$ \\
 \hline
\end{tabular}
\end{adjustwidth}

\begin{adjustwidth}{-2.5cm}{0cm}
\begin{tabular}{ |p{1.3cm}||p{4cm}|p{4cm}| }
 \hline
  $m_z=\frac{3}{2}$ & $\psi\gamma$ & $\gamma\psi$ \\
 \hline
 \hline
 $\psi\gamma$   & $\left(|\boldsymbol{m}_{\frac{3}{2}}^{\psi\gamma}|^2+|\boldsymbol{m}_{\frac{1}{2}}^{\psi\gamma}|^2\right)P_{P\frac{3}{2}}$    & 0
 \\
 $\gamma\psi$ & . &  $\left(|\boldsymbol{m}_{\frac{3}{2}}^{\psi\gamma}|^2+|\boldsymbol{m}_{\frac{1}{2}}^{\psi\gamma}|^2\right)P_{P\frac{3}{2}}$ \\
 \hline
\end{tabular}
\end{adjustwidth}

\begin{adjustwidth}{-2.5cm}{0cm}
\begin{tabular}{ |p{1.3cm}||p{6.0cm}|}
 \hline
  $m_z=2$ & $\gamma\gamma$ \\
 \hline
 \hline
 $\gamma\gamma$   & $\left(|\boldsymbol{m}_2^{\gamma\gamma}|^2+\frac{1}{2}\left(|\boldsymbol{m}_A^{\gamma\gamma}|^2+|\boldsymbol{m}_B^{\gamma\gamma}|^2\right)\right)P_{P2}$     \\
 \hline
\end{tabular}
\end{adjustwidth}
\caption{\label{Tab:Multispin} Sum rule for theory of spinning particles.}
\end{table}

The vanishing entries correspond to amplitudes that do not have dimension $8$ order contributions. Amplitudes with the required mass dimension cannot be constructed consistently respecting Lorentz invariance and possessing the required little group scaling. Many of the processes, most notably the elastic ones, are also accidentally parity symmetric as a consequence of $CPT$ (and the fact that the particles are assumed to be self-conjugate in this theory). Both $Y$ and $CPT$ can be invoked to simplify the entries. 
The $Y$ symmetry further relates $|\boldsymbol{m}_1^{\phi\gamma}|=|\boldsymbol{m}_1^{\gamma\phi}|$. 
The couplings $\boldsymbol{m}_2^{\gamma\gamma}$, $\boldsymbol{m}_1^{\psi\psi}$ and $\boldsymbol{m}_{\frac{3}{2}}^{\psi\gamma}$ are redundant.

\bibliography{superspace}{}
\bibliographystyle{JHEP}

\end{document}